%% file: ms.tex
\documentclass[12pt,preprint]{aastex}

\usepackage{graphics}
\usepackage{natbib}

\newcommand{\eps}[1]{\mbox{log~$\epsilon$(#1)}}
\newcommand\iso[2]{$^{\rm #1}$#2}

\def\eg{\mbox{e.g.}}
\def\etal{\mbox{et al.}}
\def\ie{\mbox{i.e.}}
\def\kmsec{\mbox{km~s$^{\rm -1}$}}
\def\logg{\mbox{log~{\it g}}}
\def\Msun{\mbox{$M_{\odot}$}}
\def\teff{\mbox{$T_{\rm eff}$}}
\def\vturb{\mbox{$v_{\rm t}$}}

\slugcomment{Accepted to AJ}

\shorttitle{The Chemical Compositions of Non-Variable Red and Blue Field Horizontal Branch Stars}
\shortauthors{B.-Q. For and C. Sneden}

\begin{document}

\title{The Chemical Compositions of Non-Variable Red and Blue Field 
Horizontal Branch Stars}

\author{Bi-Qing For\altaffilmark{1} and Christopher Sneden}
\affil{Department of Astronomy, University of Texas, Austin, TX 78712, USA}

\altaffiltext{1}{biqing@astro.as.utexas.edu}

\begin{abstract}
We present a new detailed abundance study of field red horizontal 
branch (RHB) and blue horizontal branch (BHB) non-variable stars. 
High resolution and high S/N echelle spectra of 11 RHB and 12 BHB were 
obtained with the McDonald 2.7~m telescope, and the RHB sample was
augmented by reanalysis of spectra of 25 stars from a recent survey.
We derived stellar atmospheric parameters based on spectroscopic 
constraints, and computed relative abundance ratios for 24 species of 
19 elements.  
The species include \ion{Si}{2} and \ion{Ca}{2}, which have not been 
previously studied in RHB and BHB (\teff~$<$~9000~K) stars. 
The abundance ratios are generally consistent with those of 
similar-metallicity field stars in different evolutionary stages.
We estimated the masses of the RHB and BHB stars by comparing their 
\teff$-$\logg\ positions with HB model evolutionary tracks.
The mass distribution suggests that our program stars possess 
masses of $\sim0.5~M_{\sun}$.
Finally, we compared the temperature distributions of field 
RHB and BHB stars with field RR Lyraes in the metallicity range 
$-0.8 \gtrsim$ [Fe/H] $\gtrsim -2.5$.
This yielded effective temperatures estimates of 5900~K and
7400~K for the red and blue edges of the RR Lyrae 
instability strip.
\end{abstract}


\section{Introduction}

Horizontal branch (HB) stars are evolved objects that are fusing helium
in their cores \citep{HS55}. 
As low-mass main sequence stars age, they first ascend the red giant 
branch (RGB), undergo internal helium-flash (losing some of their mass 
somewhere along the RGB), and finally take up residence on the HB 
while they complete their helium consumption. 
The helium core mass is relatively constant in all types of HB stars 
($\sim0.5~M_{\sun}$), but they have a large hydrogen envelope mass range. 

HB stars are commonly found in globular clusters (GCs), as well as
in field disk and halo populations of our Milky Way. 
They exhibit a range of photometric 
colors (or temperatures) which is known as the HB morphology.
The distribution can be divided into several groups:
\begin{itemize} 
\item Red horizontal branch (RHB) stars, which are all HBs cooler 
than the instability strip (IS). 
\item RR Lyraes (RR Lyr), named after their prototype. 
These are variable stars with intermediate temperature and color, 
located in the IS.
\item Blue horizontal branch (BHB) stars, which are hotter than
the RR Lyr IS. 
Their temperatures ranges from 8000--20,000~K, which is also 
subdivided into HBA (\teff~$<$~10,000~K) and HBB stars 
(\teff~$>$~10,000~K) \citep{Mohler04}. 
This division corresponds roughly to A and B spectral type. 
In this paper, we analyze only HBA stars, referring to them
collectively as BHB stars. 
\item Extreme horizontal branch (EHB) stars, which are hotter extension of 
HB (20,000--40,000~K). 
These stars often lie below the main sequence in the Hertzsprung-Russell 
diagram, and thus they are also referred to as hot subdwarfs 
(see review by \citealp{Heber09}). 
\end{itemize}

The assignment of a star to a particular HB group is based on color 
(or temperature), but the physical cause that determines the
position could be affected by multiple parameters. 
Metallicity, also referred to as the first parameter, was suggested by 
\citet{SW60} to explain the HB morphology as seen in the GCs. 
Metal-rich clusters have mostly RHB stars and metal-poor 
clusters have mostly BHB and/or EHB stars. 

However, this is not the full story of the HB morphology. 
Globular clusters that possess similar metallicity often
exhibit different HB types. For example, compare the color-magnitude 
diagrams of M3 vs M13 (see \citealp{Rosenberg00}), which clearly
indicates that HB morphology is influenced by other parameter(s). 

The early study of \citet{SZ78} suggested that the cluster age 
could be the second parameter, but later investigation by, e.g, 
\citet{Peterson95} and \citet{BBehr03} argued that stellar rotation could 
also be a significant contributor.
Alternative explanations, such as CNO abundance \citep{RS81}, mixing 
and helium abundance \citep{Sweigart97}, central concentration of 
the cluster \citep{FusiPecci93}, and Na--O anticorrelation \citep{Gratton07} 
also have been proposed. 
\citet{Lee94} demonstrated that various second parameters can produce 
different HB morphologies. 
To what extent these potential second parameters influence the 
variety of observed HB distributions in GCs remains an open question. 

Chemical abundance studies of GCs provide ideal laboratories for 
testing predictions of stellar evolution and nucleosynthesis.  
Horizontal branch stars are particularly useful for probing several aspects
of post-main sequence evolution because they are sensitive to the 
composition and structure of main sequence stars prior to the exhaustion
of their hydrogen fuel \citep{Behr03}. 
Unfortunately, HBs in GCs and stellar streams are faint and as such, 
hard to observe at high spectral resolution. 
On the other hand, field horizontal branch (FHB) stars are 
significantly brighter than cluster stars, and could be useful 
in many aspects. 
For example, FHB stars have been used as tracers of 
Galactic structure (see \citealp{Wilhelm96}; \citealp{Altmann00}). 
In addition, field RR Lyrae stars (easy to identify from their variability) 
yield important information on stellar evolution and pulsation. 
Their absolute magnitudes and metallicities provide powerful constraints
on synthetic HB models (see \citealp{Cassisi04}; \citealp{Demarque00}). 

While FHB kinematics have been widely used to study Galactic structure, 
their chemical compositions have received scant attention.
There are only a handful of detailed abundance studies of FHB stars to date 
(see \citealp{Adelman87}; \citealp{Adelman90}; \citealp{Lambert96}). 
\citet{Behr03} conducted a rotational velocity study of FHB stars, 
with only the derivation of Mg abundances for all HB stars. 
He performed a more extensive chemical abundance study for BHB stars 
in GCs \citep{BBehr03}.  
A recent large survey of FHB stars was carried out by \citet{Preston06}, 
but their sample was limited to very metal-poor RHB stars 
([Fe/H]~$<$~$-$2) that were selected from the HK objective-prism survey. 
Their primary objectives were to investigate any abundance anomalies in 
these stars, and to derive the fundamental \teff\ red edge of 
the metal-poor RR Lyr IS. 
They concluded that: (a) FRHB stars generally possess normal enhancements 
of $\alpha$-elements; (b) there is a [Si/Fe] dependence on \teff\ which is 
unrelated to nucleosynthesis issues; (c) [Mn/Fe] is subsolar; and (d) the 
$n$-capture elements have large star-to-star relative abundance scatter.
They also derived the temperature of the red edge of the metal-poor 
RR Lyr IS, by interfacing the temperature distributions of field 
metal-poor RHB and RR Lyr stars with stars of similar metallicities in 
globular clusters.  

In this paper, we present the first detailed abundance study of field 
RHB and BHB stars that spans an effective temperature range of 4000~K. 
We explore possible abundance anomalies and their implications on 
HB evolution.
This work potentially can provide a different point of views toward 
understanding HB morphology, and results should aid in application
of HB chemical compositions to stellar stream investigations.
\S 2 describes the target selection and interstellar reddening. 
The observations and reduction are given 
in \S3. In \S4 and 5, we present the line list compilation, equivalent 
width measurements and analysis methods. 
The results of individual elemental 
abundances and evolutionary states of HB stars are given in \S6 and \S7.
We discuss the implication of several elemental abundances of our HB samples in \S8. 
Lastly, we summarize the results of this work in \S9.

\section{Target Selection and Reddening}

The observed targets for this program were selected from \citet{Behr03}. 
That paper contains a compilation of known FHB stars that he used
for his rotational velocity study. 
We selected the FHB stars that have $V < 11$, [Fe/H] $\leq -1.2$ and 
\teff~$<$~9000~K. 
The temperature restriction was chosen to avoid abundance anomalies 
due to gravitational settling and diffusion processes that are
observed in the hotter BHB stars (e.g, \citealp{BBehr03}). 
RR Lyr stars were 
deliberately excluded in this program; a companion study of their 
chemical compositions will be presented in paper II.

We also included metal-poor field red horizontal branch 
(MPFRHB) stars studied by \citet{Preston06} in our program. 
We did not re-observe the MPFRHB stars, but we analyzed them in a 
manner consistent with that of the newly observed targets. 
We refer the reader to the description of target selection and 
observational details in \citet{Preston06}. 
Table~\ref{targets} gives basic information for our program stars. 

Reddening estimates $E(B-V)$ of individual stars were obtained from 
the NASA/IPAC Extragalactic 
Database\footnote{ http://nedwww.ipac.caltech.edu/forms/calculator.html} 
(NED) extinction calculator. 
This technique is based on the \textit{Infrared Astronomical Satellite} 
(IRAS) and \textit{Diffuse Infrared Background Experiment} (DIRBE) 
measurements of dust IR emission maps of \citet{DS98} (hereafter SFD). 
We chose this method in preference to the older \citet{BH82} maps, which are   
based on \ion{H}{1} 21-cm column density and galaxy counts, because 
the \ion{H}{1} maps suffer from the general problem of saturation in 
the 21-cm line in high extinction regions and have lower 
spatial resolution than the SFD maps. 

Some uncertainties in $E(B-V)$ values estimated from the SFD maps might 
arise from missing cold dust emission that is not detected by IRAS. 
In fact, $E(B-V)$ values determined from SFD are probably systematically 
larger by $\sim$0.02 mag as compared to those of \citet{BH82} 
(e.g., see comments in \citealp{MSVR06} and references therein). 
\citet{BH82} maps are not error free. In fact, their maps 
contain systematic effect that arises from 
fluctuations in galaxy count and variation in gas-to-dust ratio. To 
be consistent and to reduce the degree of systematic effect in our analysis, 
we only adopted extinctions from SFD maps. 
To correct these systematic effects of SFD maps, we used 
a 10~\% correction factor as suggested by \citeauthor{MSVR06}:
\begin{equation}
cE(B-V) = 0.9E(B-V) - 0.01,
\end{equation} 
\noindent where $cE(B-V)$ is the corrected $E(B-V)$.
We employed the corrected $E(B-V)$ for calculating the photometric \teff,
which we used to compare with our independent spectroscopic \teff\ values.
The details will be given in \S 5.1.

\section{Observations and Reductions}

The observations were made with the McDonald 2.7~m Smith telescope,
using the ``2dcoud\'{e}'' cross-dispersed echelle spectrograph.
We used this instrument with a $1.2\arcsec$ slit and in its ``cs23-e2'' 
configuration;
it gives a 2-pixel resolving power of 
$R \equiv \lambda/\Delta\lambda \sim 60,000$ with spectra projected
onto a Tektronix $2048\times2048$ CCD chip with no binning. 
The total wavelength range is $\sim 3700-8200$ \AA\, 
with complete spectral coverage for $\lambda < 5900$ \AA, and
with gaps in coverage increasing toward the red. 
We usually integrated on the target stars for 1.5~hr, yielding
S/N per resolution element of $\sim 70$ near 4000~\AA, $\sim 140$ near
5000~\AA, and $\sim 240$ near 7000~\AA. The typical seeing 
for our observing runs varied from $1.5\arcsec$ to $2.2\arcsec$. 
Our observations in 2007--2008 were taken in conjunction 
with another project, 
for which we positioned the grating so that more red portion of the 
spectrum was projected onto the CCD. 
This resulted in sacrificing some useful blue-spectral echelle 
orders, which meant that there were fewer lines available for analysis. 
Optimal spectral coverage was obtained for observing run in 2009.

ThAr comparison lamp exposures were taken at the beginning and the end 
of each night. 
We also took the spectra of hot, rapidly rotating, relatively featureless 
stars throughout the night at different airmasses. 
These spectra were used to aid in removing telluric features from the 
spectra of our program stars.  
Table~\ref{obslog} summarizes the observations and stars that are listed 
but lack sufficient
numbers of detected \ion{Fe}{1} $\&$ \ion{Fe}{2} lines for stellar parameter
estimations were excluded from abundance analysis. 
   
We performed reductions of the spectra with the 
IRAF\footnote{The Image Reduction and Analysis Facility, a general purpose 
software package for astronomical data, is written and supported by the 
IRAF programming group of the National Optical Astronomy Observatories 
(NOAO) in Tucson, AZ.} ECHELLE package. 
The raw data were bias, flat-field, and scattered-light corrected, then
extracted to one-dimensional spectra and wavelength-calibrated in
standard fashion.
The wavelength calibration arc identification was based on the line list 
in the IRAF package data file (thar.dat) and the Th-Ar wavelength table 
for the 2dcoud\'{e} spectrograph \citep{Carlos01}. 
The individual wavelength-corrected spectra were then average combined into a single spectrum. 

\begin{figure}
\plotone{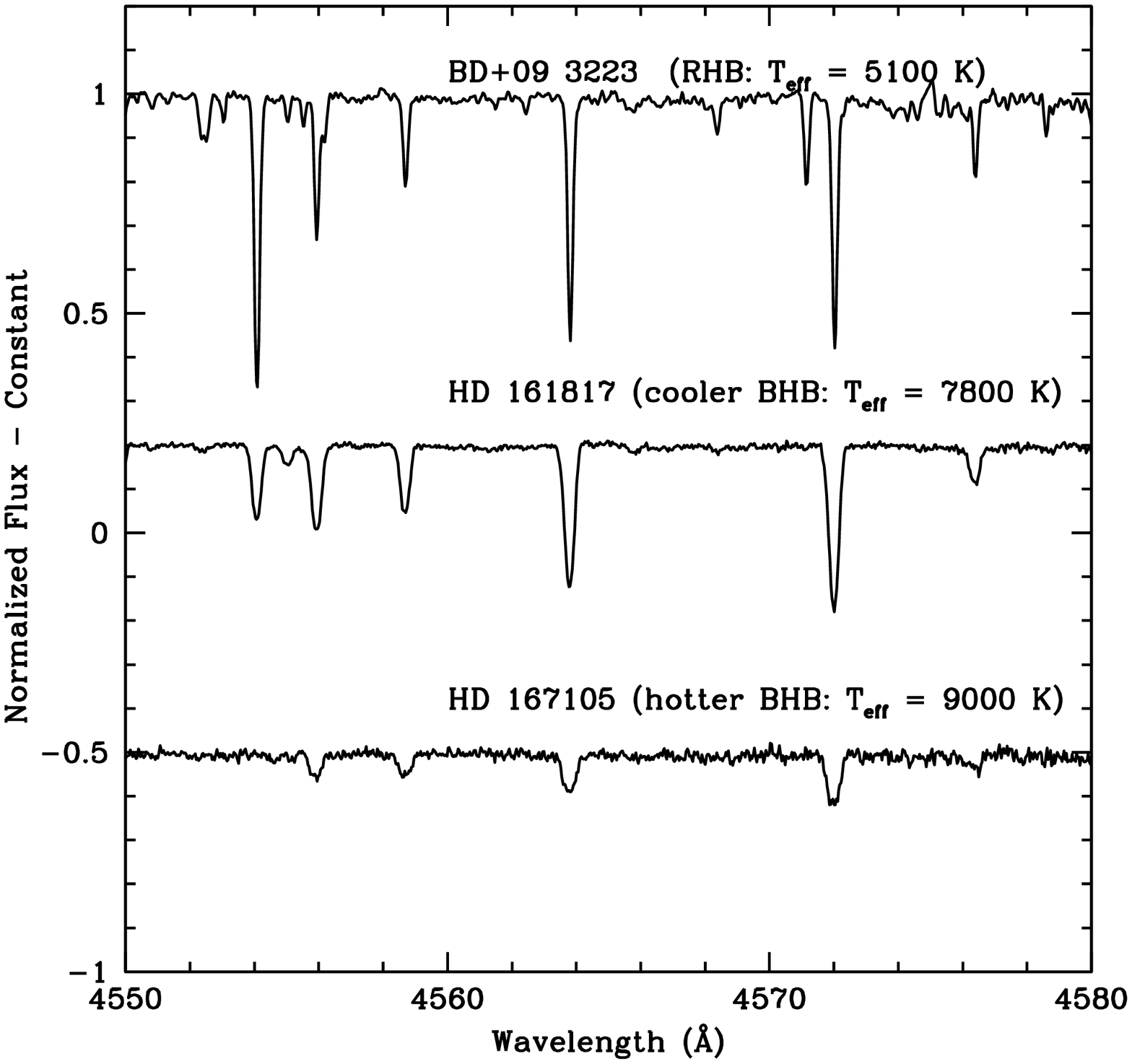}
\caption{Typical reduced, normalized spectra of RHB and BHB stars obtained at McDonald 2.7~m 
telescope. Large rotational velocity is seen in hotter BHB stars.\label{comp_spec}}
\end{figure}

Subsequently, we used the 
SPECTRE\footnote{An interactive spectrum measurement package,
available at http://verdi.as.utexas.edu/spectre.html} \citep{Fitz87} 
code to normalize the spectra and to remove cosmic rays contamination 
from the spectral lines. 
Figure~\ref{comp_spec} shows typical normalized spectra of 
RHB and BHB stars. 
Several of the hotter BHB stars exhibit significant 
rotational broadening.

\section{Line List and Equivalent Width Measurements}

We compiled an input line list of various elements from previous studies 
on HB stars (\ie, \citealp{Preston06,PTSSS06}; \citealp{Hubrig09}; 
\citealp{Khalack07,Khalack08}; \citealp{Clementini95} \& \citealp{Lambert96}). 
Species such as \ion{Si}{2} and \ion{Ca}{2} have been included in
past HBB studies, but to our knowledge this is the first use of these 
species for RHB and BHB analysis.
Excitation potentials (E.P.) and laboratory oscillator strengths ($\log gf$) 
are extracted from various sources, which we cite in 
Table~\ref{EW}. 

For each star, we measured the equivalent widths (EWs) of unblended atomic 
absorption lines interactively with SPECTRE. 
We either adopted the EW value given by fitting a Gaussian to the line 
profile or by integrating over the relative absorption across a line profile. 
If a particular line was contaminated by cosmic rays or had an
obviously distorted profile (especially lines in BHB stars can be 
blended with nearby lines due to rotational broadening), we excluded it. 
Very strong lines on the damping portion of the curve-of-growth 
(defined as those with reduced widths  
$\log \rm RW  \equiv \log EW/\lambda \gtrsim -4.0$) are relatively 
insensitive to abundance, and thus were not measured here.
After initial trials, we also excluded very weak lines (EW $< 5$ m\AA)
because the EW measurement errors were too large.
Since our program stars have a wide range of \teff\ and metallicity, 
the number of lines measured varied considerably. 
The lines used for each star, along with species, E.P., $\log gf$, 
its associated references, and measured EWs are listed in Table~\ref{EW}. 

\begin{figure}
\plottwo{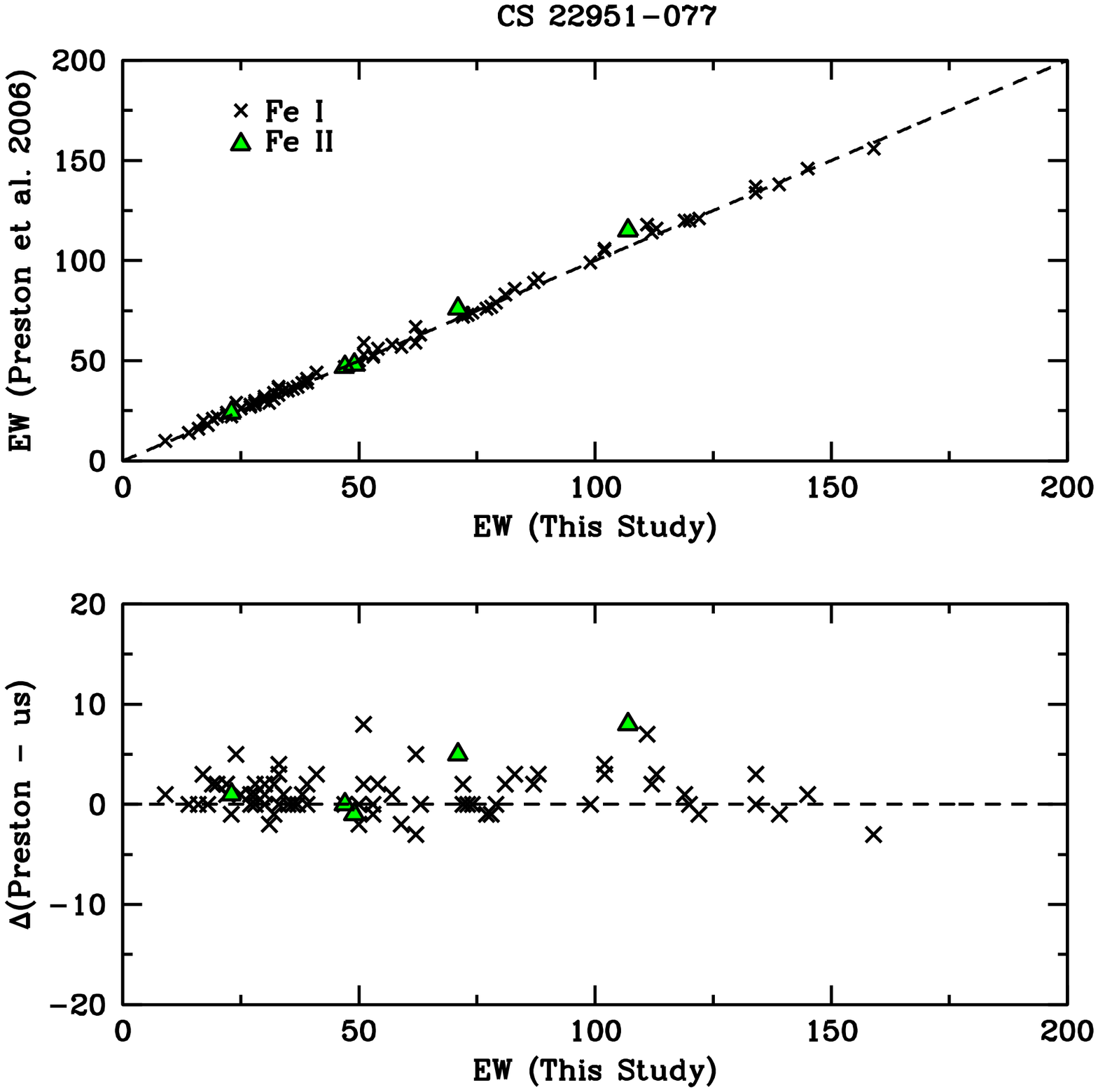}{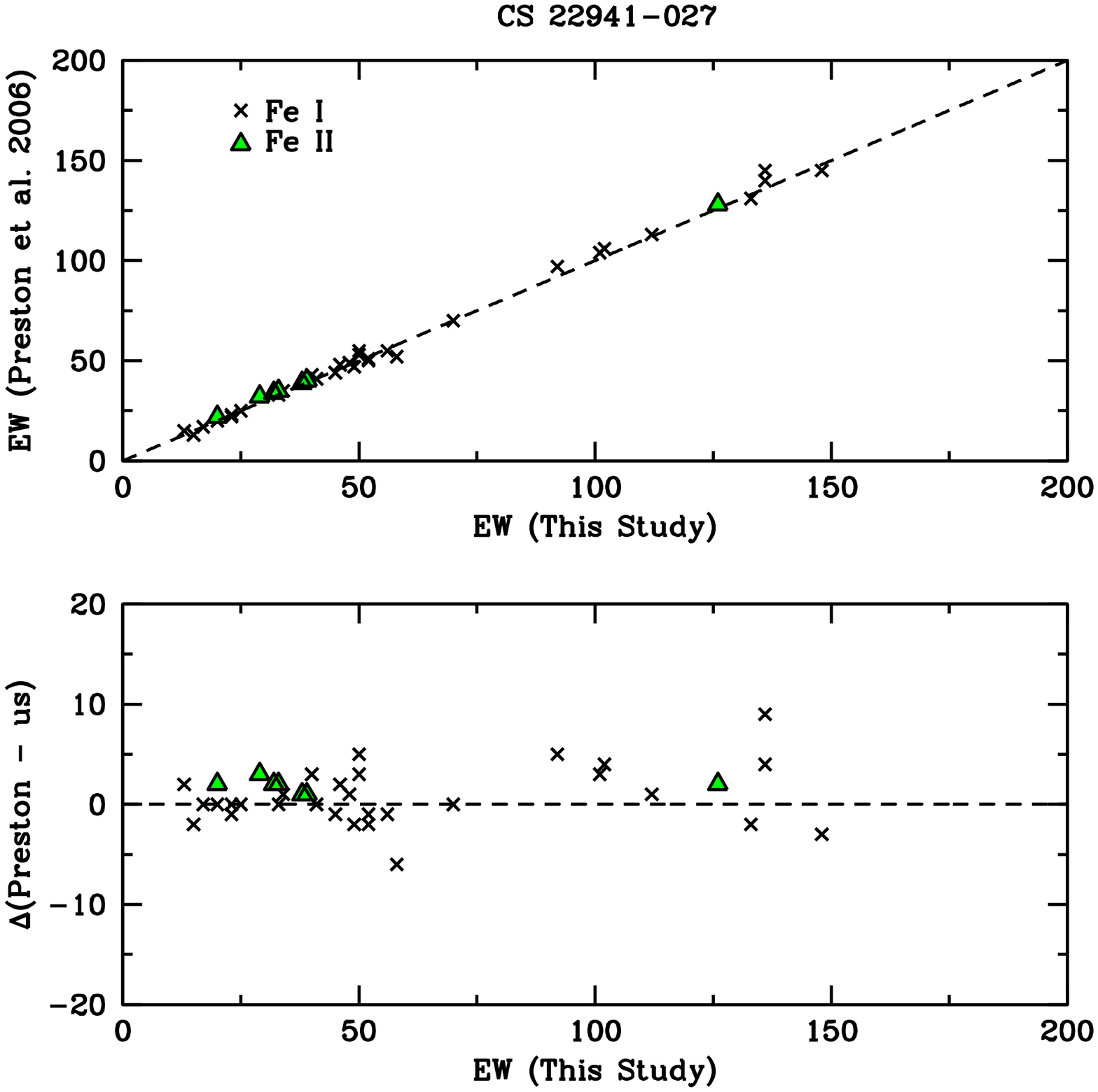}
\caption{Comparisons of our measured Fe I $\&$ II EWs of cooler (CS\,22951$-$077) and 
hotter (CS\,22941$-$027) MPFRHB stars with \citet{Preston06}. 
The top panels show 1:1 comparison of EW measurements. The bottom panels show 
the difference between our EW measurements and \citet{Preston06}. The crosses and triangles represent Fe I 
and Fe II lines, respectively. \label{ew_comp1}}
\end{figure}

\begin{figure}
\plotone{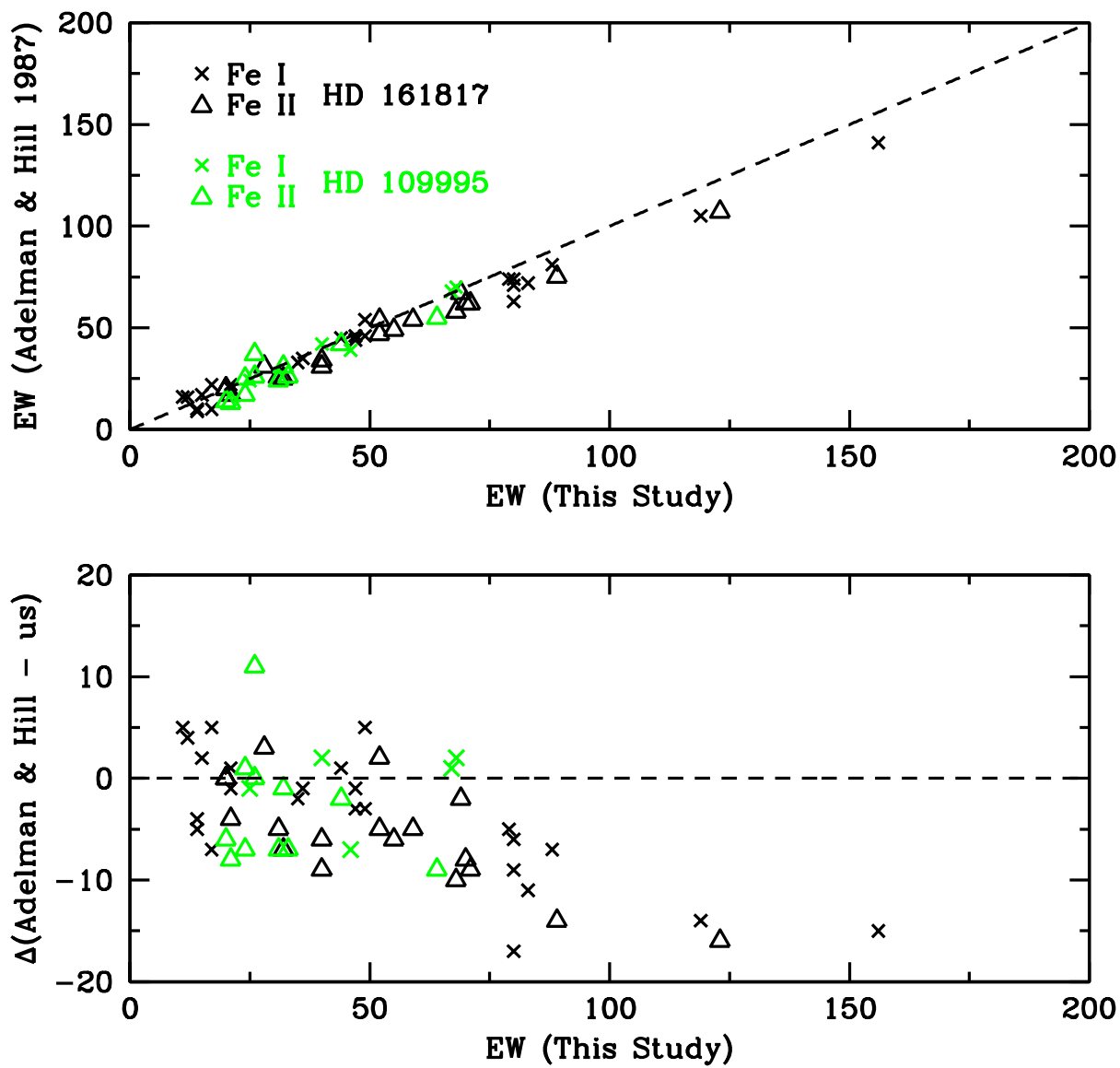}
\caption{Comparisons of our measured Fe I $\&$ II EWs of HD\,161817 and 
HD\,109995 with \citet{Adelman87}. 
The top panel shows 1:1 comparison of EW measurement. The bottom panel shows 
the difference between our EW measurements and \citet{Adelman87}. See text for 
explanation on the large deviation between ours and \citet{Adelman87} measurements. 
The crosses and triangles 
represent Fe I and Fe II lines. The green and black correspond to lines measured in 
HD\,109995 and HD\,161817, respectively. \label{ew_comp2}}
\end{figure}

We may compare our EW measurements of stars with existing previous studies. 
Only a few high-resolution, detailed chemical abundance investigations
of field 
BHB stars have been conducted to date.  
The only published iron EW measurements are from \citet{Adelman87} and 
\citet{Adelman90}, which were measured on coud\'{e} spectrograms 
recorded with photographic plates.  
Figure~\ref{ew_comp1} $\&$ \ref{ew_comp2} show the comparison of \ion{Fe}{1} 
\& \ion{Fe}{2} EW measurements in four stars. 
The literature data for the cooler (CS\,22951$-$077) and hotter 
(CS\,22941$-$027) MPFRHB stars are from \citet{Preston06} and those
for the two BHB stars (HD\,161817 $\&$ HD\,109995) are from 
\citet{Adelman87}. 
Taking the EW measurements difference between \citet{Preston06}, 
\citet{Adelman87} and this study (as shown in 
Figures~\ref{ew_comp1} $\&$ \ref{ew_comp2}), we find:
for CS\,22951$-$077, $\Delta$EW $=1.3\pm0.3$ m\AA, $\sigma = 2.7$ m\AA, 82 lines; 
for CS\,22941$-$027, $\Delta$EW $=1.0\pm0.4$ m\AA, $\sigma = 2.7$ m\AA, 37 lines;
for HD\,161817, $\Delta$EW $=-2.3\pm0.8$ m\AA, $\sigma = 4.4$ m\AA, 32 lines; and
for HD\,109995, $\Delta$EW $=-2.4\pm1.3$ m\AA, $\sigma = 5.3$ m\AA, 16 lines.
We only compute the EW difference of lines with EW $< 75$ m\AA\,in BHB 
stars because the larger EW difference in strong lines of HD\,161817 is 
probably due to the different measurement techniques of the two studies. 
In our case, strong lines were treated by either fitting the damping 
wing or integrating over the line profile. Since the deviations 
($\Delta$EW) are small, we conclude that our EW measurements are in 
excellence agreement with others.

\section{Analysis}

Our analysis is based on equivalent width matching and spectrum synthesis. 
Both methods require a stellar atmosphere model that is characterized by 
four parameters: effective temperature (\teff), 
surface gravity (\logg), metallicity ([M/H]) and 
microturbulence (\vturb). 
We constructed models by interpolation\footnote{
The interpolation code was kindly provided by Andrew McWilliam and
Inese Ivans.}
in Kurucz's non-convective-overshooting atmosphere model
grid \citep{Castelli97}. 
The elemental abundances were derived using the current version of 
the local thermodynamic equilibrium (LTE) spectral line synthesis code 
MOOG\footnote{Available at http://verdi.as.utexas.edu/moog.html .} 
\citep{Sneden73}.
With the exception of iron ($\log_{\epsilon} (\rm Fe) = 7.52)$, 
this code adopted the solar and meteoritic abundances of \citet{AG89}. 
The details on determining the stellar parameters and methodologies are 
given in the following subsections.

\subsection{Stellar Parameters}

An initial stellar atmosphere model was created based on the 
stellar parameters of \citet{Preston06} and \citet{Behr03}. 
Final model atmosphere parameters were determined by iteration, 
through spectroscopic constraints: 
(1) for  \teff, that the abundances of individual \ion{Fe}{1} lines 
show no trend with E.P.; 
(2) for \vturb, that the abundances of individual \ion{Fe}{1} lines show 
no trend with reduced width ($\log \rm RW$); 
(3) for \logg, that ionization equilibrium be achieved between the 
abundances derived from the \ion{Fe}{1} and \ion{Fe}{2} species; and
(4) for metallicity [M/H], that its value is consistent with
the [Fe/H] determination. 
In the case of [Fe/H] $< -2.5$, we adopted [M/H]$=-2.5$ for the stellar 
atmosphere model due to no available models in our grid below this 
metallicity.
Table~\ref{stellarpars} presents the derived stellar atmosphere 
model parameters and Fe metallicities of our program stars.

The standard spectroscopic constraints method has drawbacks. 
In particular, ``spectroscopic'' gravities derived from ionization balance 
may be lower than ``trigonometric'' gravities derived from stellar 
parallaxes ($\pi$) or ``evolutionary'' gravities inferred from
HR-diagram positions (see e.g., \citealp{AP99}). 
Such mismatches may arise from statistical equilibria that are
not well described by LTE.
These so-called NLTE effects are mainly due to the additional ionization 
of neutral-species beyond collisions by UV photons.
The problem can increase with decreasing metallicity due to smaller
UV line opacities in metal-poor stars.
Discrepancies in derived [\ion{Fe}{1}/H] and [\ion{Fe}{2}/H] are the 
result: \ion{Fe}{1} lines yield lower abundances than do \ion{Fe}{2} lines,
which are then ``corrected'' by decreasing assumed gravities in LTE
analysis \citep{TI99}. 
A full discussion of NLTE effects is beyond the scope of this paper. 
In the following section, we consider the effects of \logg\
uncertainties on our derived abundances.

We have compared our spectroscopic \teff's to those based purely
on photometry.
We computed photometric temperatures using the metallicity-dependent 
\teff-color formula of giants developed by \citet{Alonso99}.  
These relationships are based on the infrared flux method (IRFM) \citep{BS77}. 
We employed only $V-K$ colors for this exercise.
In contrast to $B-V$ colors, where blue continua are severely affected 
by line blanketing, $V-K$ colors are largely insensitive to the choice 
of metallicity and gravity. 

The $(V-K)$ values of our stars, as listed in Table~\ref{targets}, are based
on $V_{\rm Johnson}$ and 2MASS $J$ and $K_{s}$ magnitudes. The calibration 
curve of \citet{Alonso99} is based on $(V-K)_{\rm TCS}$. 
Therefore, several color transformations were required.
We converted these colors to the TCS system in two ways.
First, we simply shifted the 2MASS $K_{s}$ magnitudes to the 
$K_{\rm TCS}$\footnote{$K_{\rm TCS}$ is the broad-band $K$ magnitude
in the photometric system developed for the Observatorio del Teide
(Tenerife) 1.5m telescope \citep{Alonso94}.}
using Eq.~5(c) of \citet{RM05}: 
$K_{TCS}=K_{\rm 2MASS}-0.014+0.027(J-K)_{\rm 2MASS}$. 
The $V_{\rm TCS}$ magnitudes are essentially equal to $V_{\rm Johnson}$, 
thus the $K$ transformation should be sufficient to convert our $V-K$ 
values to $(V-K)_{\rm TCS}$. 
Second, a better method is to shift ($V_{\rm Johnson}-K_{s}$) into 
$(V-K)_{\rm TCS}$ by two corrections as described in \citet{Johnson05}; 
we computed the $(V-K)_{\rm TCS}$ using their Eq.~6:
$(V-K)_{\rm TCS}=0.050+0.993(V_{\rm Johnson}-K_{s})$.
For each of these conversion attempts, we then applied extinction corrections 
to the colors, adopting an extinction ratio of $k=E(V-K)/E(B-V)$, 
where $k=2.74$ for $(V-K)_{\rm TCS}$ \citep{RM05}. 
Photometric $T_{\rm eff}$ were subsequently calculated using polynomial
relation described in Eq.~8 of \citet{Alonso99}.
There are two BHB stars that possess $V-K$ colors that are smaller 
than $V-K$ range ($< 0.2$) of this equation's calibration.
For these stars we simply assumed that the polynomial fit could be 
extrapolated to $V-K$~$\simeq$~0. 

We compared the calculated photometric \teff\ of both methods 
and found that the difference is small 
($\Delta$\teff$ = 54\pm1$~K, $\sigma = 6$~K, $N_{\rm star} = 34$) for 
RHB stars and somewhat larger ($\Delta$\teff$ = 109\pm3$~K, $\sigma = 11$~K, $N_{\rm star} = 11$) for BHB stars. 
The larger difference for BHB stars is most likely due to the 
color-\teff\ transformation, because it is based mostly on cooler stars.
The error of calculated photometric $T_{\rm eff}$ depends on the slope of the polynomial fit, 
$\Delta$\teff/$\Delta X$, where $\Delta X$ is a function of extinction 
ratio ($k$) and error in reddening ($\Delta E(B-V)$). 
The error is represented by 17~K per 0.01 mag for $V-K < 2.2$ \citep{Alonso99}. 
 
\begin{figure}
\plotone{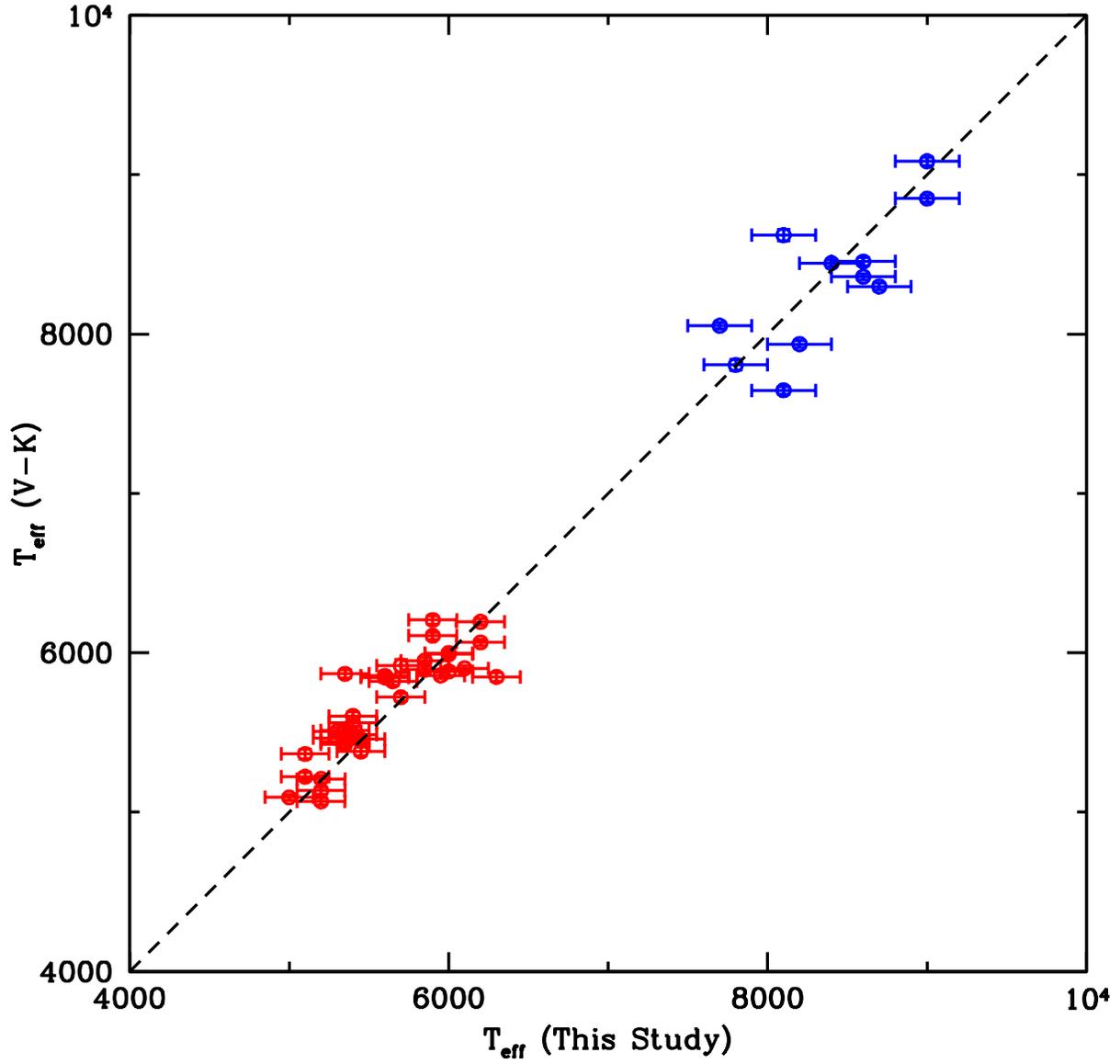}
\caption{Comparison of spectroscopic $T_{\rm eff}$ with photometric $T_{\rm eff}$ 
derived from $(V-K)_{\rm TCS}$ metallicity--dependent \teff--color formula 
of \citet{Alonso99}. The error of photometric $T_{\rm eff}$ is equal to or smaller 
than the size of the dots.\label{teff_vmk_comp}}
\end{figure}

We show the comparison of the calculated photometric \teff\ values that are adopted 
from the first color-transformation method to the derived spectroscopic \teff\ 
values in Figure~\ref{teff_vmk_comp}. 
Taking the difference (our spectroscopic \teff\ minus photometric \teff), 
we show that both \teff\ values of both RHB 
($\Delta$\teff$ = -73\pm30$~K, $\sigma = 177$~K, $N_{\rm star} = 34$) 
and BHB stars ($\Delta$\teff$ = 59\pm91$~K, $\sigma = 300$~K, $N_{\rm star} = 11$) 
are in good agreement.

Ideally our spectroscopic gravities should be compared with trigonometric 
or physical gravities, but such an exercise is not possible here.
Our stars have no reliable parallax data from $Hipparcos$ 
\citep{Perryman97}; they are too distant. 
Most stars selected from the \citet{BBehr03} catalog have large 
errors in their parallaxes, and no parallaxes have been reported for stars 
selected from \citet{Preston06}.

\subsection{Parameter Uncertainties} 
   
To estimate the effects of uncertainties in our spectroscopically-based
\teff\ on derived abundances, we varied the 
assumed \teff's of HD\,119516 (RHB) and HD\,161817 (BHB).  
For HD\,119516, raising \teff\ by 150~K from the derived 5400~K produced
an unacceptably large trend of derived \eps{Fe} with 
excitation potential.
For the BHB star, HD\,161817, \teff\ can be raised to 200~K before the 
trend of \eps{Fe} with E.P. becomes too large.
Repeating these trials for other stars suggested that 150~K and 200~K are
typical uncertainties for the RHB and BHB stars, respectively.
The difference between the two groups is due to the lesser number of 
available \ion{Fe}{1} lines in BHB spectra, which causes larger error in 
\teff\ derivation. 

We estimated \vturb\ uncertainties in a similar manner, assessing the
trends of \ion{Fe}{1} abundances with $\log\,(\rm{RW})$.
This yielded $v_{t}$ errors of 0.2 km s$^{-1}$ 
and 0.3 km s$^{-1}$ for RHB and BHB stars, respectively. 
Finally, (assuming that $\log g$ based on the neutral/ion ionization 
balance of Fe abundance is correct) from the dependence \ion{Fe}{2}
abundances with $\log g$, we estimated the error of $\log g$ to be $2\sigma$ 
of \ion{Fe}{2} abundance error. The mean error of \logg\ to be $\sim 0.16$ dex. 
We adopted the internal error ($\sigma$) of \ion{Fe}{1} abundances
as the model [M/H] error.

\subsection{Comparisons with Previous Studies}

\begin{figure}
\plotone{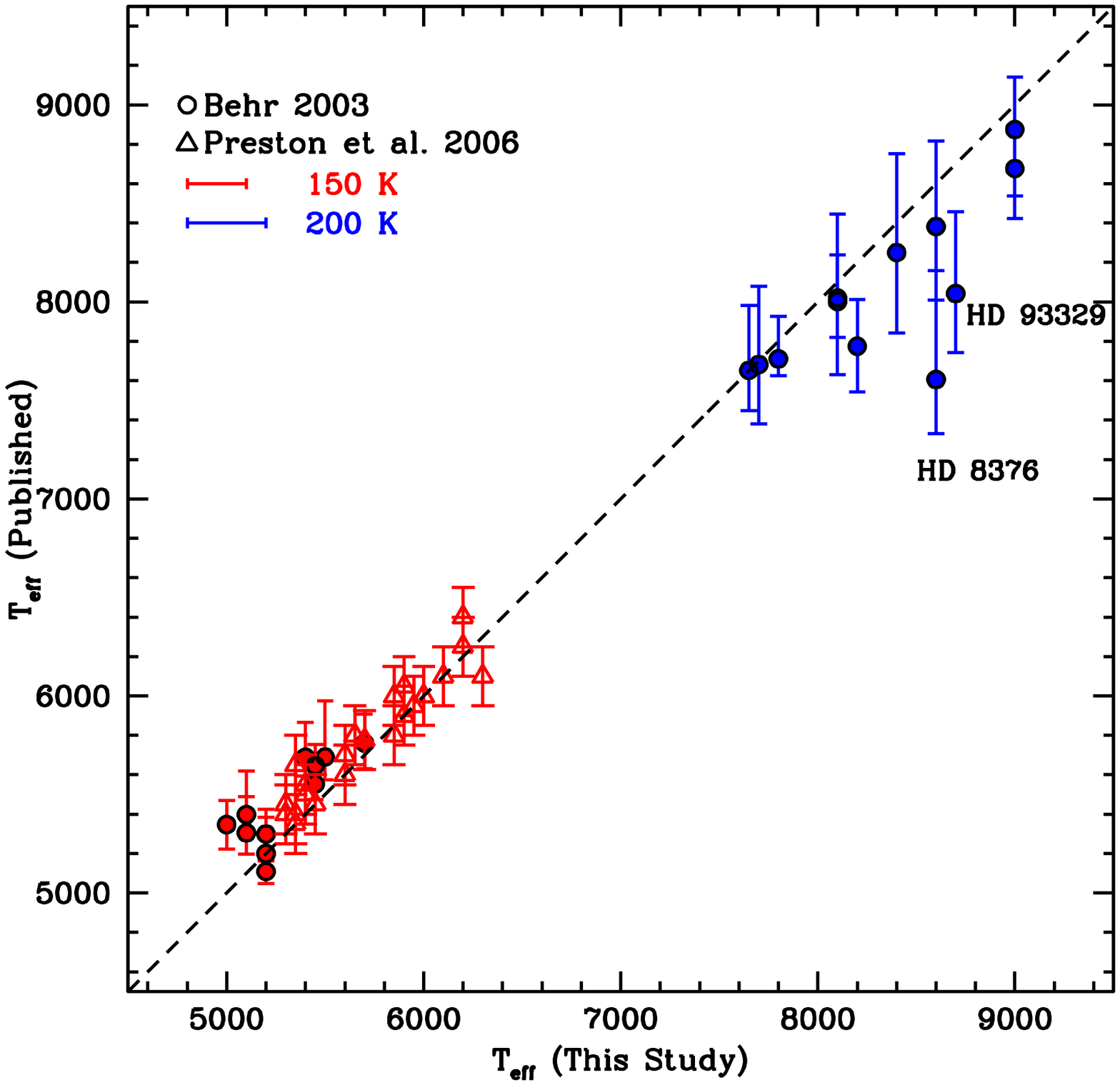}
\caption{Comparison of spectroscopic \teff\ derived from this 
study with \teff\ values from \citet{Preston06} and \citet{Behr03}. 
The triangles and circles represent \citet{Preston06} and \citet{Behr03} study, 
respectively. The red and blue colors correspond to RHB and BHB stars. 
For clarity in the figure, we do not plot error bars from our work
for each star, but instead indicate typical \teff\ uncertainties for 
this study, 150~K and 200~K for RHB and BHB stars. 
Comparison of BHB stars can only be made with \citet{Behr03}. 
\label{comp_teff}}
\end{figure}

\begin{figure}
\plotone{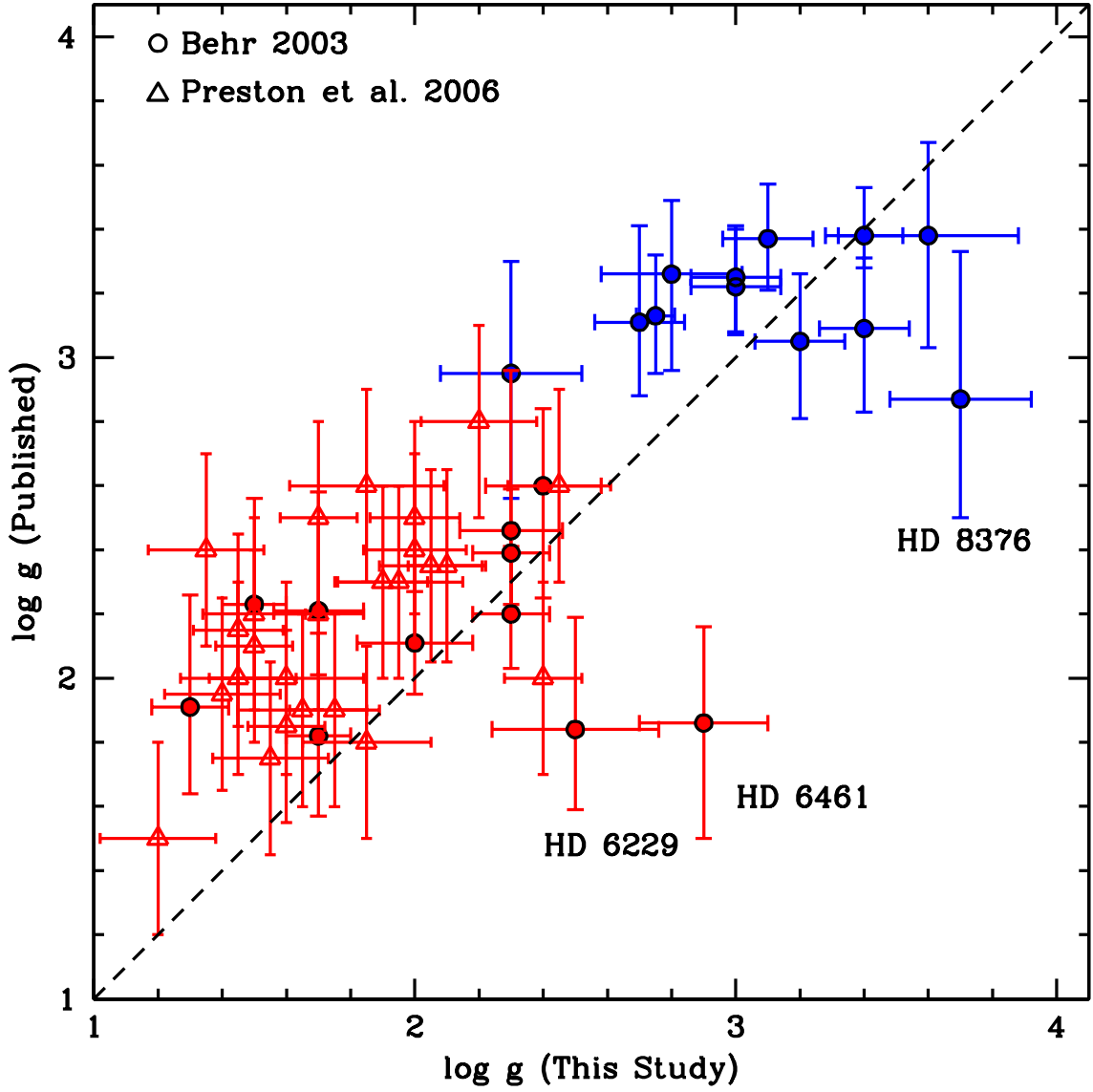}
\caption{Comparison of spectroscopic $\log g$ derived from this study with $\log g$ derived 
by \citet{Preston06} and \citet{Behr03}. The triangles and circles represent \citet{Preston06} and 
\citet{Behr03} study, respectively. The red and blue colors correspond to RHB and BHB stars.
\label{comp_logg}}
\end{figure}

We compared our derived $\log g$ and $T_{\rm eff}$ values with those of
\citet{Preston06} and \citet{Behr03}, as shown in Figures~\ref{comp_teff} 
$\&$ \ref{comp_logg}. 
\citet{Behr03} derived these quantities by comparing the synthetic 
photometric color and the observed color over a grid 
of $T_{\rm eff}-\log g$ values. 
\citet{Preston06} employed the same method as we do, \ie, from
spectroscopic constraints, but they used both Fe and Ti abundances
for determining \logg\ from ionization-balance considerations.
We decided here not to use Ti in the \logg\ estimation, because the 
\ion{Ti}{1} $\log gf$ values from the NIST atomic transition 
database\footnote{National Institute of Standards 
and Technology (NIST): http://www.nist.gov/physlab/data/asd.cfm .}
are of relatively high uncertainty and there are not many measurable 
\ion{Ti}{1} lines ($N < 6$) in most cases for our RHB stars. 
Using small number of lines would cause larger error in \logg\ estimation 
and could yield systematic error (see below).   
Additionally, we have no detections of \ion{Ti}{1} lines in our BHB sample.
Therefore to be consistent in our RHB and BHB star analyses, we decided 
to only use \ion{Fe}{1} and \ion{Fe}{2} abundances in estimations
of \logg.
   
Our \teff's for RHB stars are 
$\Delta$\teff (Preston$-$us) $= 59\pm20$~K ($\sigma = 100$~K, $N = 25$) 
and $\Delta$\teff (Behr$-$us) $= 154\pm40$~K ($\sigma = 134$~K, $N = 11$), which 
are in good agreement.
Comparison of BHB stars can only be made with Behr.
Our \teff\ values generally agree with his, $\Delta$\teff(Behr-us) $= -152\pm43$~K 
($\sigma = 134$~K, $N = 10$) except for HD\,8376 and possibly HD\,93329.
Our derived RHB \logg\ values are systematically lower 
($\Delta \log g$ (Preston$-$us) $= 0.41\pm0.06$ dex, $\sigma = 0.3$ dex, $N = 25$) 
than those of Preston \etal, which is due to different derivation methods. 
To demonstrate such systematic effect, we performed tests using both Fe and Ti lines. 
Abundances of neutral species of Titanium is generally larger than ionized species 
by 0.12--0.2 dex. As such, this requies a larger \logg, which is 0.2--0.5 dex, 
to achieve the ionization equilibrium for Ti. 
   
Our derived \logg\ values show no correlation with Behr's, and 
we note significant deviations for HD\,8376, HD\,6461 and HD\,6229.
For HD\,6461 our derived [\ion{Fe}{1}/H] is $+0.6$ dex higher than Behr's, 
which in turn forces a larger \logg\ to achieve the ionization equilibrium.
Our \teff\ for HD\,8376 is about 500~K larger than Behr's estimate, which
forces a much larger \logg\ value in our analysis.
We do not have an explanation for the $\log g$ deviation of HD\,6229.

\subsection{Microturbulence vs Effective Temperature}

\begin{figure}
\plotone{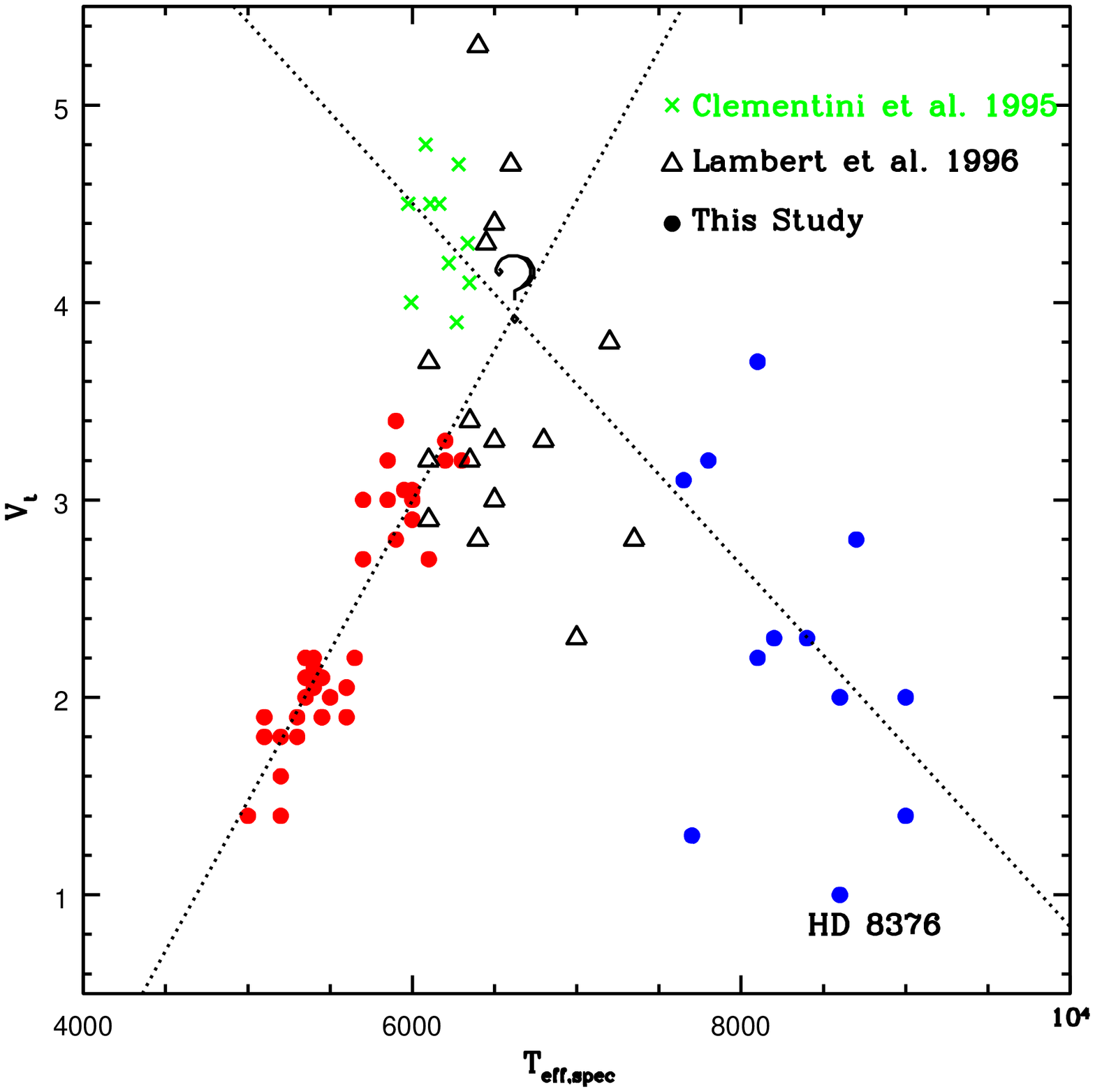}
\caption{The correlation and anti-correlation between \vturb\ and \teff\ for RHB and BHB stars. Linear least square 
equations were fitted to all the RHB stars and BHB stars, excluding HD~8376. The crosses 
and open triangles represent the \vturb\ and \teff\ of RR Lyrs studies by \citet{Clementini95} and \citet{Lambert96}, 
respectively. The readers are warned that there is no correlation in the RR~Lyr IS region and beyond the 
intersection of dashed lines, where question mark is marked.\label{vt_teff}}
\end{figure}

We plot our \vturb\ values against \teff\ in Figure~\ref{vt_teff}, where the 
correlations (dashed lines) were derived by fitting linear least squares regression lines to the RHB and BHB data.
The clear positive correlation of microturbulent velocity with 
temperature in RHB stars has been found by others (see \citealp{Preston06} 
and references therein).  
It is possible that the BHB stars have an anti-correlation between
these two quantities. The star-to-star scatter is large, but if we exclude HD\,8376\footnote{
Our derived \vturb\ for HD\,8376 is rather uncertain because 
no \vturb\ choice can eliminate the trend of \eps{Fe} with 
$\log (\rm{EW/\lambda})$ for this star.
This is the only program star for which we have trouble in finding an 
acceptable \vturb\ value.}, the anti-correlation remains. We have extended the 
dashed lines beyond their intersection in the figure; comparison of these lines with the RR Lyr 
data indicates that there is no \vturb\ correlation with \teff\ in this doman. 
This issue will be revisited in paper II.   

These trends in derived \vturb\ with \teff\ undoubtedly are related to the
envelope/atmosphere instabilities of RR~Lyr stars.
The evolutionary track of a HB star indicates that it evolves from the 
hot end, crosses the RR Lyr IS into the cool HB region, 
before ascending to the AGB. 
As an HB star evolves toward the RR Lyr IS blue edge, its atmosphere 
begins to be unstable, which results in increasing line widths that 
we model as increasing microturbulence.
And as the HB star evolves away from the RR Lyr IS red edge, 
the line widths decrease as the stability is regained.
We caution here that our microturbulence values are simple compensations
for complex physical changes that are occurring in HB stars near
the instability strip, and thus should be interpreted with caution.

\section{Chemical Abundances}

With the model atmosphere parameters listed in Table~\ref{stellarpars}, 
we derived the abundances of most elements from their EW measurements. 
In the cases of \ion{Ca}{2}, \ion{Mn}{1}, \ion{Ni}{2}, \ion{Sr}{2},
\ion{Zr}{2}, \ion{Ba}{2}, \ion{La}{2}, and \ion{Eu}{2}, the detectable
transitions are complex: they are either partially blended, or have 
significant hyperfine and/or isotopic substructure, or all of these
things. 
We employed spectrum synthesis to determine abundances for these species.
That is, for each line we computed theoretical spectra of a wavelength
region within $\pm 10$\AA\ of the line for a variety of assumed abundances, 
then broadened the computed spectrum with Gaussian line profile (or a 
combination of Gaussian plus rotational velocity line profile), and finally
compared these spectra to the observed ones.
The assumed abundances were changed iteratively to obtain acceptable 
synthetic/observed spectrum matches.
For stars with detectable rotational line broadening, we began with the 
$v$sin~$i$ estimates of \citet{Behr03} and derived the final $v$sin~$i$ 
based on the fit to observed line profile. 
Our final numbers were always in good agreement 
($\Delta v$sin~$i \simeq 1-2$~km s$^{-1}$) 
with initial values. The damping constant of \citet{BO98} was adopted whenever possible in 
both EW analyses and spectrum syntheses method. 

We present the derived abundances ratio [X/Fe] in 
Tables~\ref{abund1}--\ref{abund4}, and plot these as functions of
metallicity in Figure~\ref{xfe}--\ref{xfe_2} and $T_{\rm eff}$ in Figures~\ref{xfe_teff}--\ref{xfe_teff2}. Non-LTE corrections have been applied to the data 
on these figures and tables wherever applicable. 
The mean [X/Fe] values of RHB and BHB stars are summarized in 
Table~\ref{mabund}.
In the following subsections we comment on individual elements.

The total error in the abundances is a combination of internal error 
(line-to-line scatter), and external errors (induced by stellar model 
atmosphere parameter uncertainties). 
The line-to-line scatter is given by the abundance standard deviation 
($\sigma$) from individual spectral lines. 
To estimate the errors caused by model parameter uncertainties, we 
performed numerical experiments for four stars, in which we varied the 
model parameter errors as estimated in \S{5.2}. 
These stars are 
CS\,22898$-$043 (very metal-poor),
HD\,25532 (moderately metal-poor),
HD\,93329 (BHB) and 
BD+18$\degr$\,2890 (RHB). 
They were selected because they are representative of our whole sample. 
The results of [X/Fe] sensitivity to different stellar model atmosphere 
parameter variations are shown in Table~\ref{sensitivity1} $\&$ 
\ref{sensitivity2}. 
In most cases $\Delta\rm{[X/Fe]}$~$\lesssim$~0.05 in response to 
changes in \logg, [M/H] and \vturb. 
On the other hand, varying \teff\ by 150~K has a larger effect on the 
abundance ratios of cool, metal-poor RHB star BD+18$\degr$\,2890, 
especially on the neutral species. 
The overall average variations in [X/Fe] are small, $\simeq$0.05. 
Thus, in general external error from stellar model atmosphere parameters 
do not greatly influence the derived abundance ratios. 
For abundances derived from one spectral 
line, we adopted an error of 0.2 dex, 
judging from the statistical source of error (ie., 
sensitivity of $\Delta [X/Fe]$ with stellar parameters error, uncertainties in 
measuring the EW or matching a synthetic spectrum etc).

\subsection{The Light Alpha Elements: Magnesium, Calcium and Titanium} 

It has been known for decades that metal-poor stars are generally 
overabundant in $\alpha$-elements (\eg, \citealp{Wall63}). 
Our HB stars show standard enhancements in these elements, with  
neutral species $<$[Mg,Ca,Si,Ti/Fe]$> \simeq +0.3$ (see Figure~\ref{xfe}). 

Two RHB stars, BD+18$\degr$\,2890 and CS\,22883$-$037, exhibit relatively low [Mg/Fe], 
but not in other $\alpha$-elements. 
Only a single \ion{Mg}{1} line was analyzed in both of these cases, 
which resulted in larger abundance uncertainties.
Caution is advised in interpreting the Mg abundances of BD+18$\degr$\,2890 and 
CS\,22883$-$037.

The Calcium abundances of BHB stars have a larger scatter than the 
RHB stars. There is also an offset, $\sim 0.3$ dex of 
mean [Ca/Fe] of RHB and BHB stars.
We investigated this offset by synthesizing the \ion{Ca}{2} 3933\AA\,K-line of BHB stars.
This line is rarely used in abundance analyses, as it is extremely strong 
in cool stars. 
In our case, the K-line could be analyzed in BHB stars, in which
the line is not very strong and uncontaminated in most cases. 
There are weak interstellar contamination for HD\,2857 and BD+25$\degr$\,2602.  
However, it does not affect our abundances derivation, which is based on a Gaussian 
line profile fitting to the line. 
The abundances in BHB stars for \ion{Ca}{1} and \ion{Ca}{2}
are approximately consistent with each other. 
The presence of the BHB/RHB offset is currently unknown. 
We also note that there is an unexplained trend of decreasing [Ca/Fe] 
with increasing \teff\ for BHB stars (see Figure~\ref{xfe_teff}). 
Investigation of larger sample of BHB stars might resolve this puzzle.

There are no \ion{Ti}{1} lines detectable in our BHB stars. 
Additionally, our $\log gf$ values for the \ion{Ti}{1} lines are 
taken from the NIST compilation, but their estimated uncertainties 
are large.
In the RHB stars, \ion{Ti}{1} lines are visible, but not many measurable lines. 
The analysis yields a trend of increasing [\ion{Ti}{1}/Fe] with increasing \teff\ 
(see Figure~\ref{xfe_teff}).
This trend is opposite the sense of Si (discussed below) and 
has been noted by others (see \citealp{Lai08} and references 
therein). 
The abundance ratios derived from \ion{Ti}{2}, unlike those 
of the other 
$\alpha$-elements, do not decline as the metallicity increases. 
The mean value is flat, with small scatter, across the entire 
metallicity range. 
The \ion{Ti}{2}-based titanium abundances should be trustworthy as many 
\ion{Ti}{2} lines were used to determine the abundances.

\subsection{The Alpha Element Silicon: A Special Case}

Substantial dependence of [\ion{Si}{1}/Fe] with temperature has been 
found in previous studies of metal-poor field stars (see \citealp{Cayrel04}, 
\citealp{Cohen04}, \citealp{Preston06}, \citealp{SL08} $\&$ \citealp{Lai08}). 
This effect seems to depend entirely on \teff; there is no
apparent trend with \logg.
To address this puzzle, \citet{Shi09} investigated NLTE effects in warm 
metal-poor stars. 
They showed that the \ion{Si}{1} 3905.53 \AA\,lines and \ion{Si}{2} 
6347 \AA, 6371 \AA\,lines exhibit significant NLTE departures in warm 
metal-poor stars. 
Their study was limited to a sample of metal-poor dwarfs and a
single cool giant. 
Observationally however, warmer FRHB stars 
(6000~K~$\lesssim$ \teff\ $\lesssim$~6400~K) have similar Si abundances to 
those of metal-poor main sequence turnoff stars, [Si/Fe]~$\simeq$~0 
(see Figure~10 of \citeauthor{Preston06} or Figure~8 of 
\citeauthor{SL08}), in spite of their
large gravity differences ($<$$\Delta$\logg$>$~$\sim$~2).
Thus, the effect seems to be most dependent on \teff, so we assume that
the predicted NLTE effects for main sequence stars will also affect our 
low gravity, metal-poor, warm RHB and BHB stars. 
Taking the offsets of $+0.1$ dex 
and $-0.1$ dex to the \ion{Si}{1} and \ion{Si}{2} abundances from these lines, 
as suggested by \citeauthor{Shi09}, 
we corrected the abundances of these two species in our program stars
with \teff$\geq6000$~K.
Note that there is a large star-to-star scatter 
for RHB and BHB stars even after this adjustment (see Figure~\ref{xfe_teff}).
This suggests, in agreement with the conclusions of \citeauthor{Shi09}, 
that addition of an offset is inadequate to produce abundance consistency 
for this species.

The \ion{Si}{1} abundances of all the BHB stars and the CS stars, with 
the exception of CS\,22940$-$070, were exclusively derived 
from the 3905.53 \AA\,line. 
As always, the reader is cautioned about the abundances derived from a
single line.  
The blue-spectral region of hot stars are not overcrowded with lines, 
so blending is not an issue in this case. 
For cool stars, 3905.53 \AA\,might be blended with a weak CH transition 
\citep{Cohen04} which would become stronger with decreasing temperature. 
However, \citet{Preston06} argue that the CH contamination in metal-poor 
RHB stars is very weak, and will not seriously affect the derived
Si abundance.
The line is thus essentially ublended and weak enough for abundance
determinations in all BHB stars, and in RHB stars 
with $T_{\rm eff} \geq 5400$~K and [Fe/H]~$\leq -2$.\footnote{
We could not determine a Si abundance for HD\,119516 because our spectrum
of the 3905\AA\ line was corrupted by cosmic rays.}
Lines of \ion{Si}{1} in the red-spectral region ($> 5600$\AA) were used to 
derive abundances for the rest of the RHB stars. 
There are eight stars that we used at least four lines for 
determining the abundances. 
For these stars, we derived $<$[\ion{Si}{1}/Fe]$>$~=~$+0.42$, which is 
consistent with the mean of typical $\alpha$-enhancement in metal-poor stars. 

In Figure~\ref{si_1_teff}, we summarize the \ion{Si}{1} abundances found 
in large-sample studies and the spectral regions that were used to derive the \ion{Si}{1} abundances.  
All investigators agree on the declining trend of [\ion{Si}{1}/Fe] with
increasing $T_{\rm eff}$ among cooler metal-poor stars, and we have shown 
that the abundances reach a (low) plateau in BHB stars. 
Resolution of this unsatisfactory situation is beyond the scope of this
study.

An important check on the Si abundances is provided by our detection
of \ion{Si}{2}, which has mainly been studied in stars with 
\teff~$>$~10,000~K. 
Only a handful of dwarfs have reported \ion{Si}{2} abundances 
(see \citealp{SB02}), and no prior investigation has been done for RHB stars. 
In general, \ion{Si}{2} lines are very weak for RHB stars, only becoming
strong (EW $> 30$~m\AA) in BHB stars. We caution that weak lines and 1--3 \ion{Si}{2} 
lines were used for deriving the \ion{Si}{2} abundances. 

In Figure~\ref{si_2_teff}, we illustrate the mixture of lines that have been 
used to derive \ion{Si}{2} abundances for both RHB and BHB stars. 
The scatter of [\ion{Si}{2}/Fe] is large but the mean abundances agree 
with the general $\alpha$-enhancement indicated by Mg and Ca for our HB stars. 
We find unusually large \ion{Si}{2} abundances for CS\,22955$-$174 
and CS\,22937$-$072. 
However, they show normal enhancement in \ion{Si}{1} (\ie, +0.3 and +0.5 
dex, respectively). 
Unfortunately, in both cases, only 1--2 \ion{Si}{1} or \ion{Si}{2} lines 
were used to derive their abundances, so these abnormally large abundances
should be viewed with caution.

\subsection{Light Odd-Z Elements Sodium and Aluminum} 

\begin{figure}
\plotone{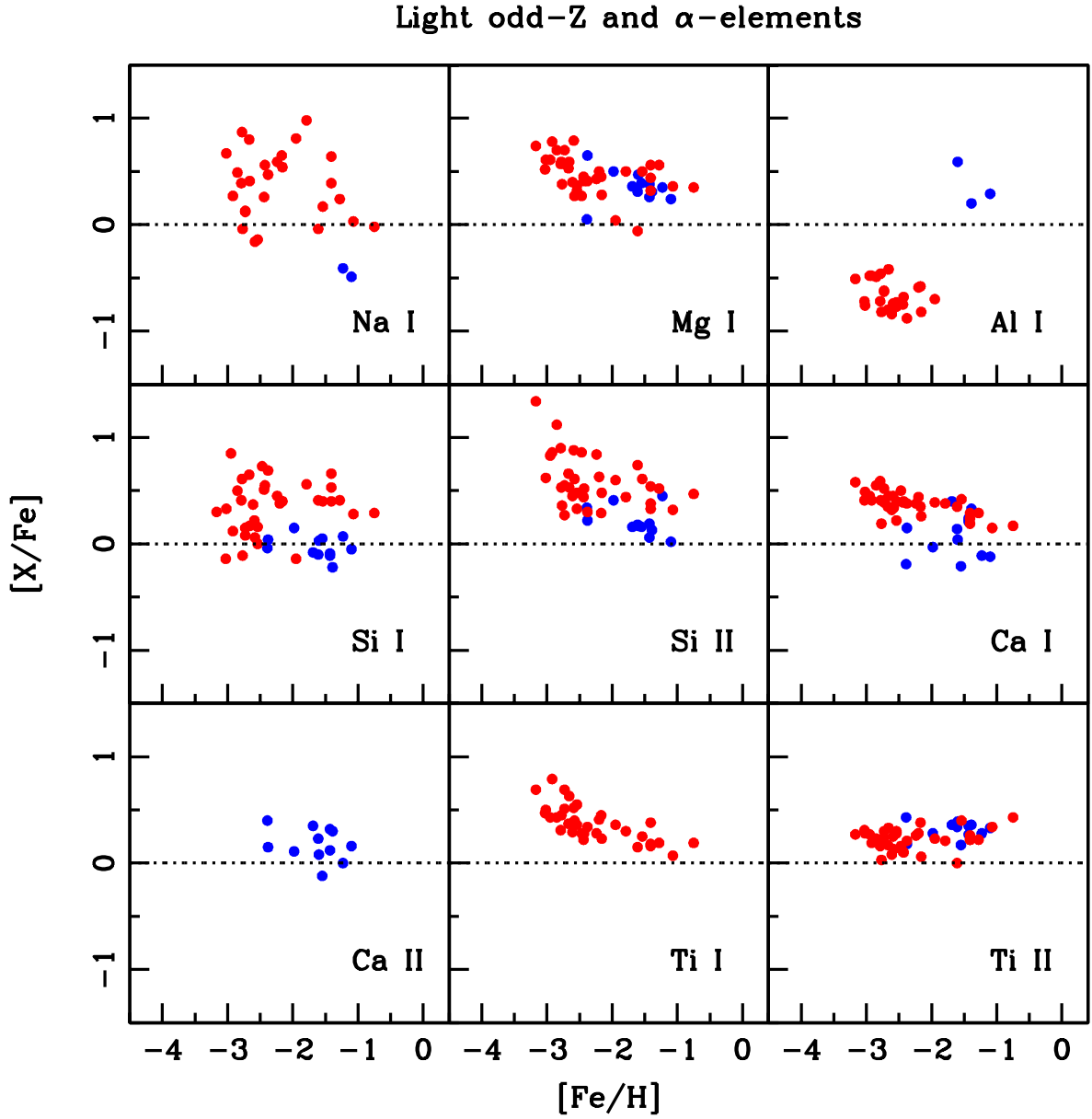}
\caption{Abundance ratios of odd-Z and $\alpha$-elements as a function of metallicity. 
NLTE corrections applied to \ion{Na}{1}, \ion{Al}{1}, \ion{Si}{1} $\&$ \ion{Si}{2} as described in text. 
The red and blue dots represent RHB and BHB stars.\label{xfe}}
\end{figure}

For sodium abundances, we used mainly the \ion{Na}{1} resonance D-lines 
(5889.9 \AA, 5895.9 \AA). 
Only a few of the cooler RHB stars have detectable, albeit weak, 
higher excitation \ion{Na}{1} lines (the 5682.6 \AA, 5688.2 \AA\,and the
6154.2 \AA, 6160.7\AA\,doublets). 
We visually inspected the D-line spectral region to search for ISM 
contamination of the stellar lines.
Any suspected line blending resulted in dropping the D-line measures
for a star.
The derived [Na/Fe] values exhibit a large star-to-star scatter (see Figure~\ref{xfe}). 
We warn the reader that the \ion{Na}{1} D-lines are relatively strong in 
the RHB stars as compared to the BHB stars. 
Unfortunately, there are only two BHB stars in our samples that have 
measurable, clean D-lines. 
Therefore, we could not make direct comparison with the star-to-star 
scatters in BHB and RHB stars. 
Nevertheless, the large variations derived here are consistent with
those seen in previous field metal-poor star studies 
(see \citealp{Pilachowski96}; \citealp{Venn04} and references therein). 

Aluminum is underabundant in RHB stars, $<$[Al/Fe]$> \simeq-0.64$, 
and overabundant in BHB stars, $<$[Al/Fe]$> \simeq+0.36$ (see Figure~\ref{xfe}). 
There are only two \ion{Al}{1} lines, the resonance transitions
3944 \AA\ and 3961 \AA\ in the blue spectral region, which we can 
employ for this study. The 3944 \AA\ line 
can be contaminated by CH transition \citep{AM83}. However, it is not an issue in 
our very warm BHB stars and it is even undetectable in our metal-poor RHB stars. Additionally, 
the 3961 \AA\ line can only be a reliable abundance indicator in metal-poor 
stars, as it is affected by the strong wing of \ion{Ca}{2} H and 
$\rm H_{\epsilon}$ features in higher metallicity stars \citep{SL08}. 
Higher excitation Al lines in the red spectral region, \eg, the 6696 \AA, 
6698 \AA\ pair, generally result in higher [Al/Fe] (see 
discussion of \citealp{Francois84}). 
The discrepancy of [Al/Fe] between the transitions of red and the 
blue spectral region is currently not completely understood.
Unfortunately we could not detect the red \ion{Al}{1} lines in our stars.

As noted by others, Na D lines and the \ion{Al}{1} red and blue resonance 
spectral region can be significantly altered from NLTE effects. 
These corrections are important for warm, metal-poor turnoff stars 
with \teff $\gtrsim$ 6000~K \citep{Baumueller98}.  
The suggested NLTE corrections are $-0.5$ dex for Na \citep{Baumueller98} 
and $+0.65$ dex for Al \citep{BG97}. 
Since the majority of our RHB stars are below this \teff\, 
we only applied NLTE corrections of suggested values to Na and Al 
abundance ratios of our BHB stars.

\subsection{The iron-peak elements: Scandium through Zinc}. 

\begin{figure}
\plotone{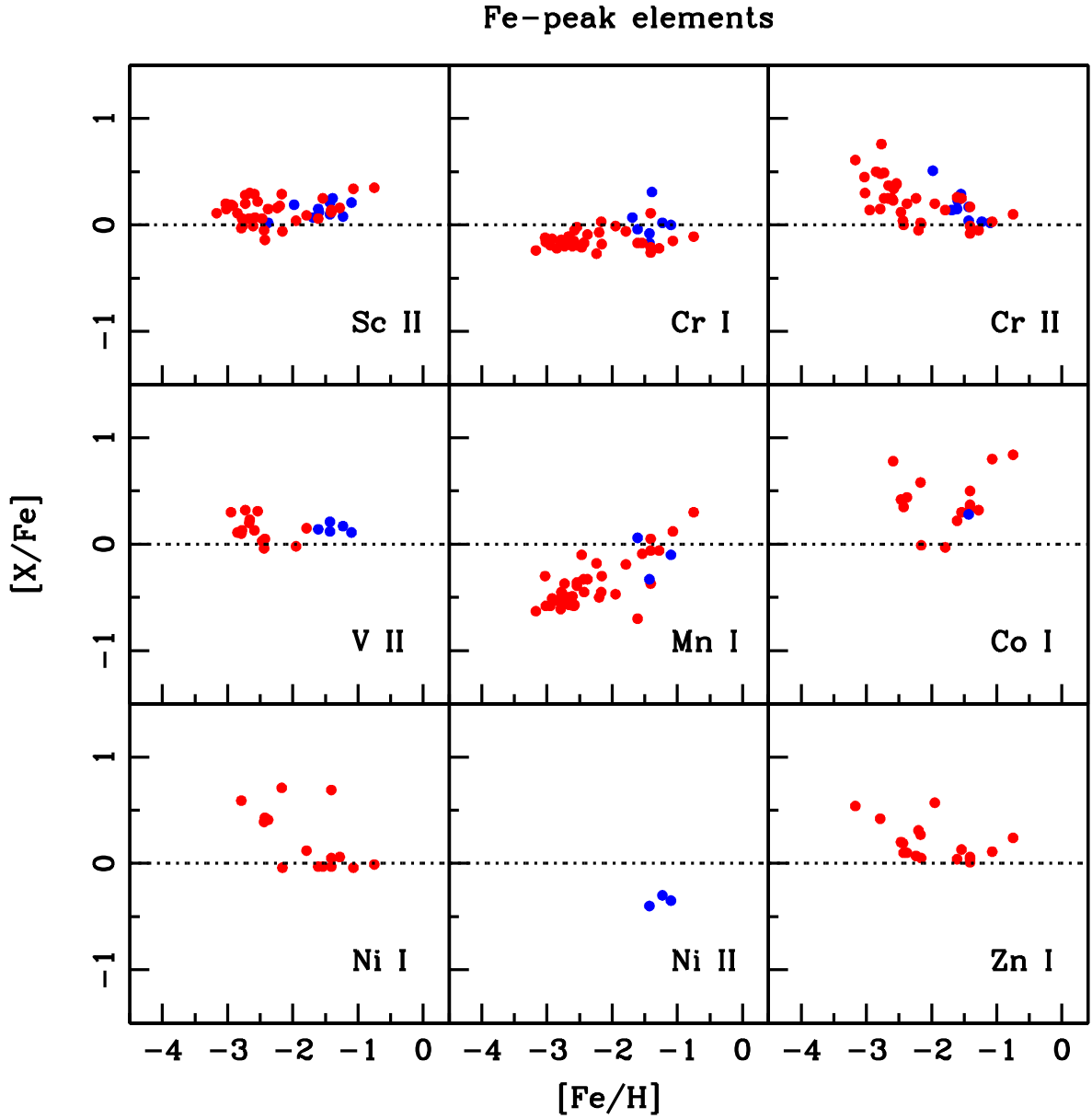}
\caption{Abundance ratios of Fe-peak elements as a function of metallicity. 
The red and blue dots represent RHB and BHB stars. \label{xfe_1}}
\end{figure}

Scandium lines can have substantial hypefine substructure. 
We synthesized a few \ion{Sc}{2} lines with their full substructure,
and found that the abundances derived from synthesis do not differ by more than 0.05 dex from 
those derived by the single-line EW method. 
Thus, we used the EW method for deriving all final \ion{Sc}{2} abundances. 
A study by \citet{Cohen04} showed that there are 
discrepancies of [\ion{Sc}{2}/Fe] among
different evolutionary groups of metal-poor stars, in which 
they are generally enhanced in main sequence stars while RGB stars
exhibit deficiencies.
Our results are more in accord with those of main-sequence stars, 
$<$[Sc II/Fe]$> \simeq +0.13$ (see Figure~\ref{xfe_1}). 

Our vanadium abundances come exclusively from \ion{V}{2} lines, which
were detectable in both RHB and BHB stars.
We find no trends of [V/Fe] with either [Fe/H] or \teff.

Chromium abundances derived from \ion{Cr}{1} transitions generally
yield smaller abundances than those from \ion{Cr}{2} lines in 
metal-poor stars (e.g, \citealp{Preston06}, \citealp{Sobeck07}, 
and references therein). 
Ideally, we would have preferred to use recent laboratory transition
probabilities for both \ion{Cr}{1} \citep{Sobeck07} and 
\ion{Cr}{2} \citep{Nilsson06} for our study. 
However, there are no \ion{Cr}{2} lines studied by \citet{Nilsson06} 
that are routinely detectable in our spectra. 
Therefore, we employed the transition probabilites of detectable 
\ion{Cr}{1} and \ion{Cr}{2} lines from 
\citet{Sobeck07} and NIST, respectively. 
The offset between \ion{Cr}{1} $\&$ \ion{Cr}{2} remains 
(see Figure~\ref{xfe_1}). 
The trend of increasing \ion{Cr}{2} with decreasing metallicity is 
due to large line detection/measurement uncertainty; only 1--2 lines were 
used in relatively metal-poor, RHB stars. 
This offset is also present in the detailed Cr transition probability 
study of \citet{Sobeck07}. 
Ionization imbalance or non-LTE effect could be the cause. 

A trend of increasing [Cr I/Fe] with increasing $T_{\rm eff} < 7000~K$
has also been found for RHB stars (see Figure~\ref{xfe_teff1}). 
This is first pointed out by \citet{Lai08} (see their Figure~21). 
Clearly no such trend is apparent in our BHB stars.

Manganese abundances of field and halo metal-poor dwarf and giant stars 
have been shown to be substantially underabundant (see, e.g, 
\citealp{Sobeck06}, \citealp{Lai08}, and references therein). 
Our analysis yields $<$[Mn/Fe]$> \simeq -0.35$. 
The general trend of increasing [Mn/Fe] with at higher [Fe/H] metallicities
in our HB sample is in agreement with those and other previous studies.
We refer the reader to review the extensive discussion of 
\citet{Sobeck06} regarding the production of Mn. 

We derived nickel abundances via spectrum synthesis of the \ion{Ni}{2} 
4067 \AA\ line and the remaining iron-group elements from
EW analysis. 
The reader should be cautious in interpreting the \ion{Co}{1}, 
\ion{Ni}{2}, and \ion{Zn}{1} abundances, as they were determined with 
only 1--2 lines each. 
There are insufficient data to define an abundance pattern of 
\ion{Ni}{2} at this point. 
Our [{Ni}{1}/Fe] values are generally near solar for 
moderately metal-poor stars ([Fe/H] $>2.0$).
The larger star-to-star scatter for very metal-poor stars 
([Fe/H] $< 2.0$) is probably not real, as only one weak \ion{Ni}{1} 
line was used in our analysis, resulting in uncertain Ni abundances
for individual stars.

Zinc has multiple abundant isotopes (\iso{64,66,67,68}{Zn}), but the
isotopic/hyperfine substructure of \ion{Zn}{1} lines are not large
and the observed features are weak \citep{Timmes95}. 
Therefore we treated \ion{Zn}{1} lines as single absorbers. The discussion of 
[Zn/Fe] will be given in \S 8.1.

\subsection{The neutron capture elements: Strontium, Yttrium, Zirconium, 
            Barium, Lanthanum and Europium} 

We derived the strontium abundances using available \ion{Sr}{2} 
4077 \AA, 4161 \AA\ and 4215 \AA\,lines. 
These lines are particularly hard to analyze in RHB stars because 
they are strong and/or partially blended.
For example, the 4077.8 \AA\ resonance line can be affected by 
\ion{Dy}{2} 4078.0 \AA\ and possibly \ion{La}{2} 4077.3 \AA.
We illustrate this in Figure~\ref{Sr}, which shows an example of the 
\ion{Sr}{2} 4077\AA\ synthesis superimposed on the observed spectrum of
an RHB star.
The Dy abundance cannot be determined reliably with the spectra. 
Therefore, the adopted Dy abundance was arbitrarily changed to produce
the best fit to the red wing of the observed \ion{Sr}{2} line profile.

\begin{figure}
\plotone{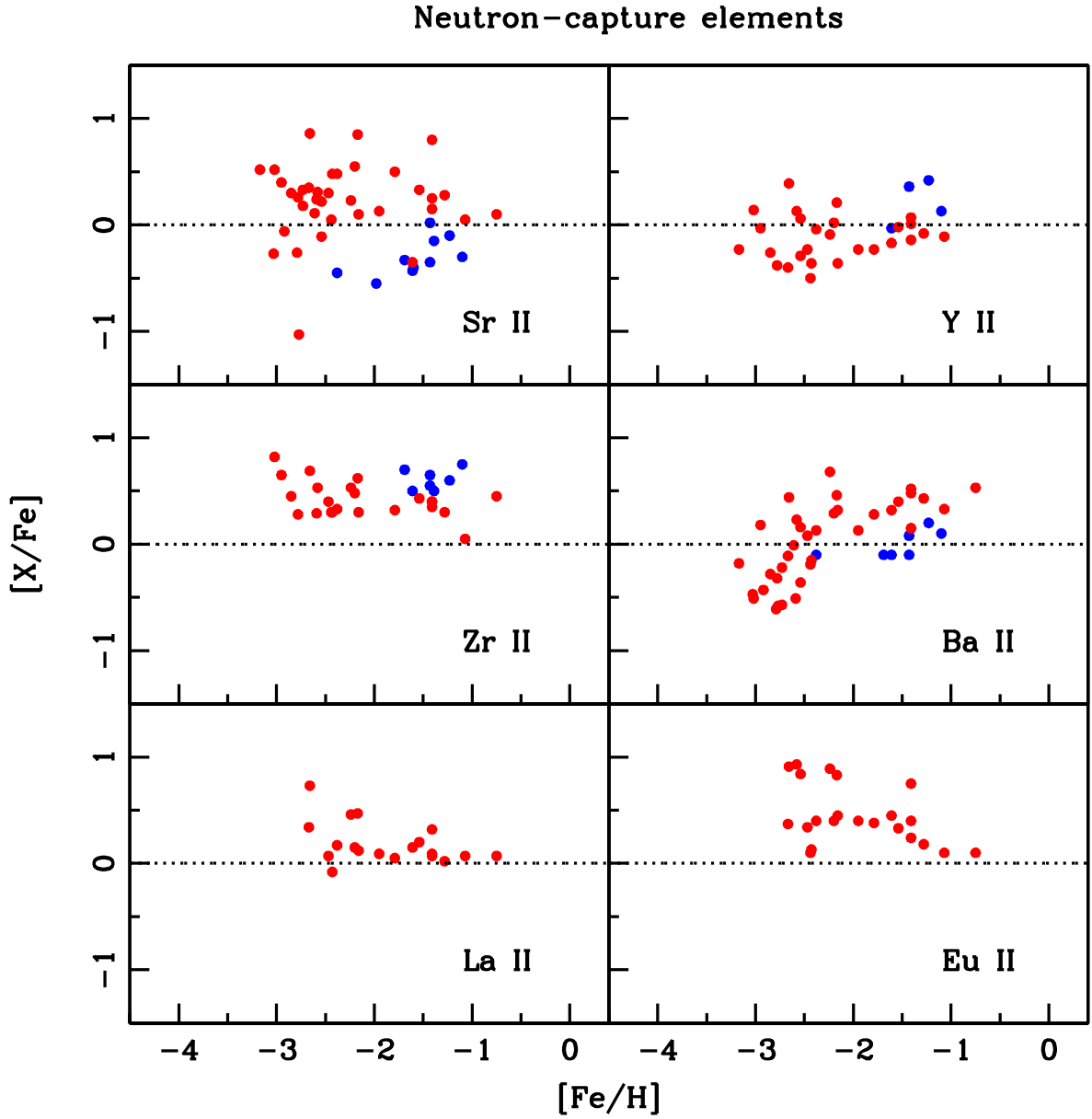}
\caption{Abundance ratios of neutron-capture elements as a function of metallicity.
The red and blue dots represent RHB and BHB stars. \label{xfe_2}}
\end{figure}

\begin{figure}
\plotone{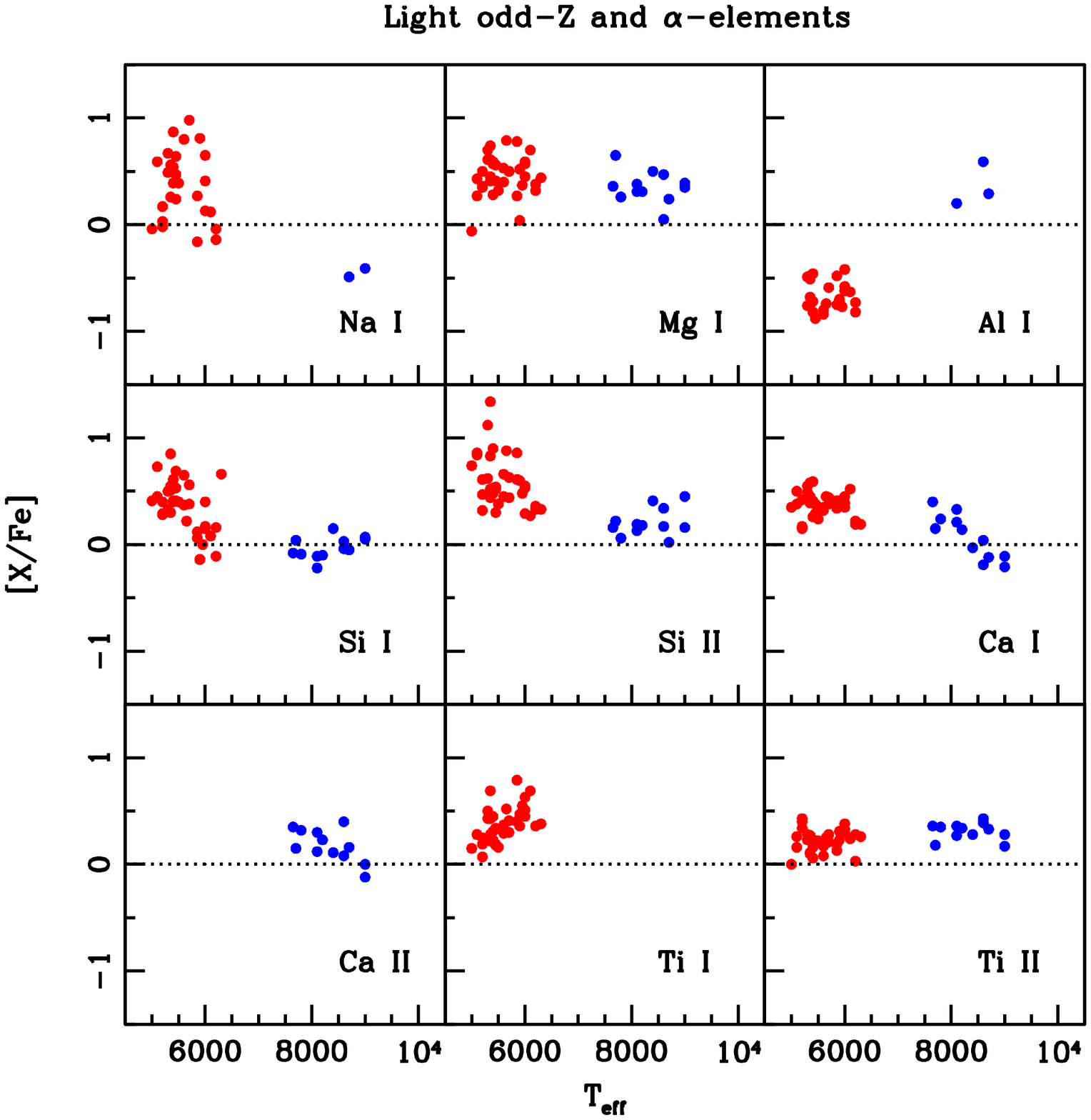}
\caption{Abundance ratios of odd-Z and $\alpha$-elements as a function of spectroscopic $T_{\rm eff}$. 
NLTE corrections applied to \ion{Na}{1}, \ion{Al}{1}, \ion{Si}{1} $\&$ \ion{Si}{2} as described in text.
The red and blue dots represent RHB and BHB stars. \label{xfe_teff}}
\end{figure}

\begin{figure}
\plotone{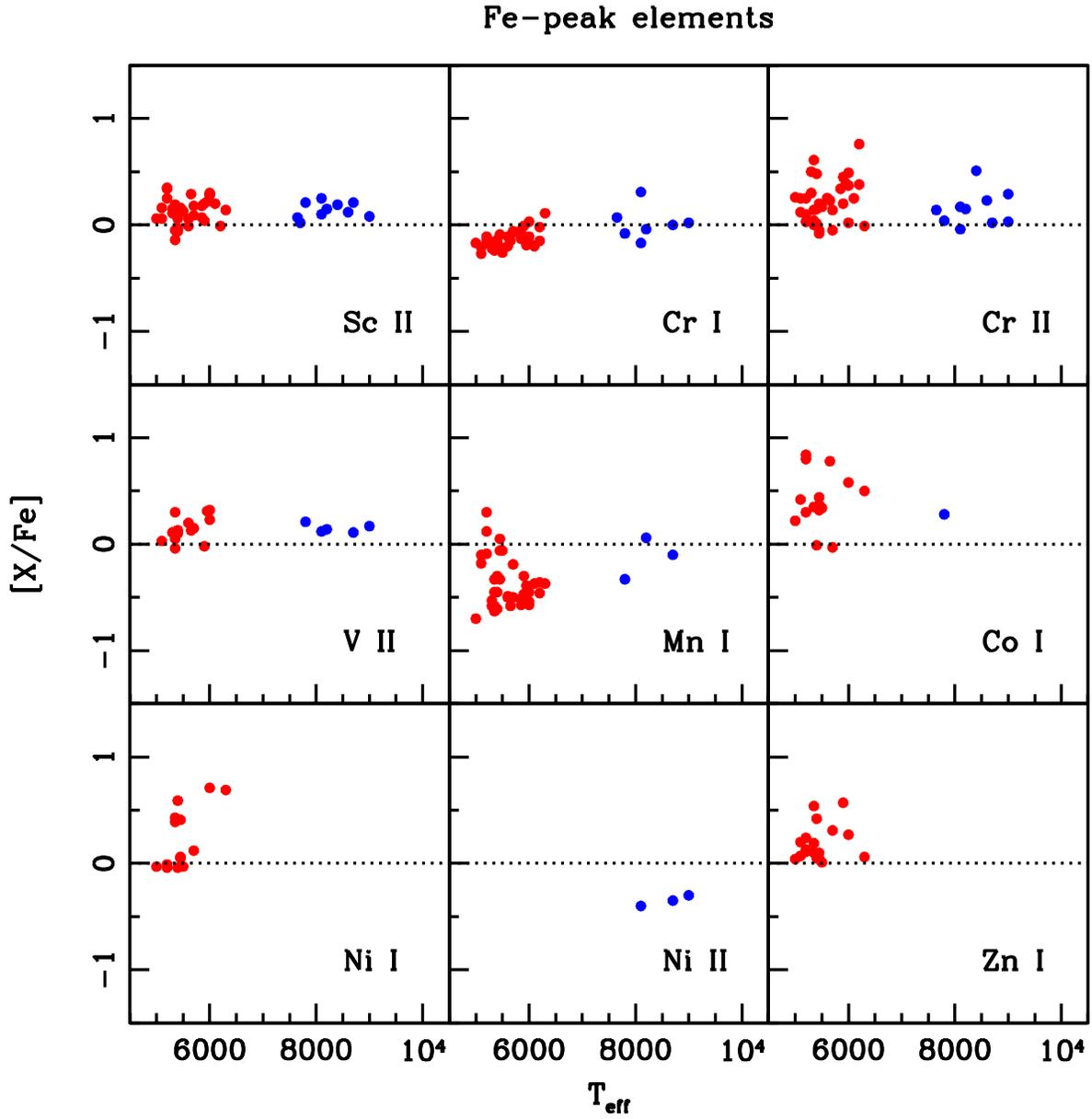}
\caption{Abundance ratios of Fe-peak elements as a function of spectroscopic $T_{\rm eff}$. 
The red and blue dots represent RHB and BHB stars.\label{xfe_teff1}}
\end{figure}

\begin{figure}
\plotone{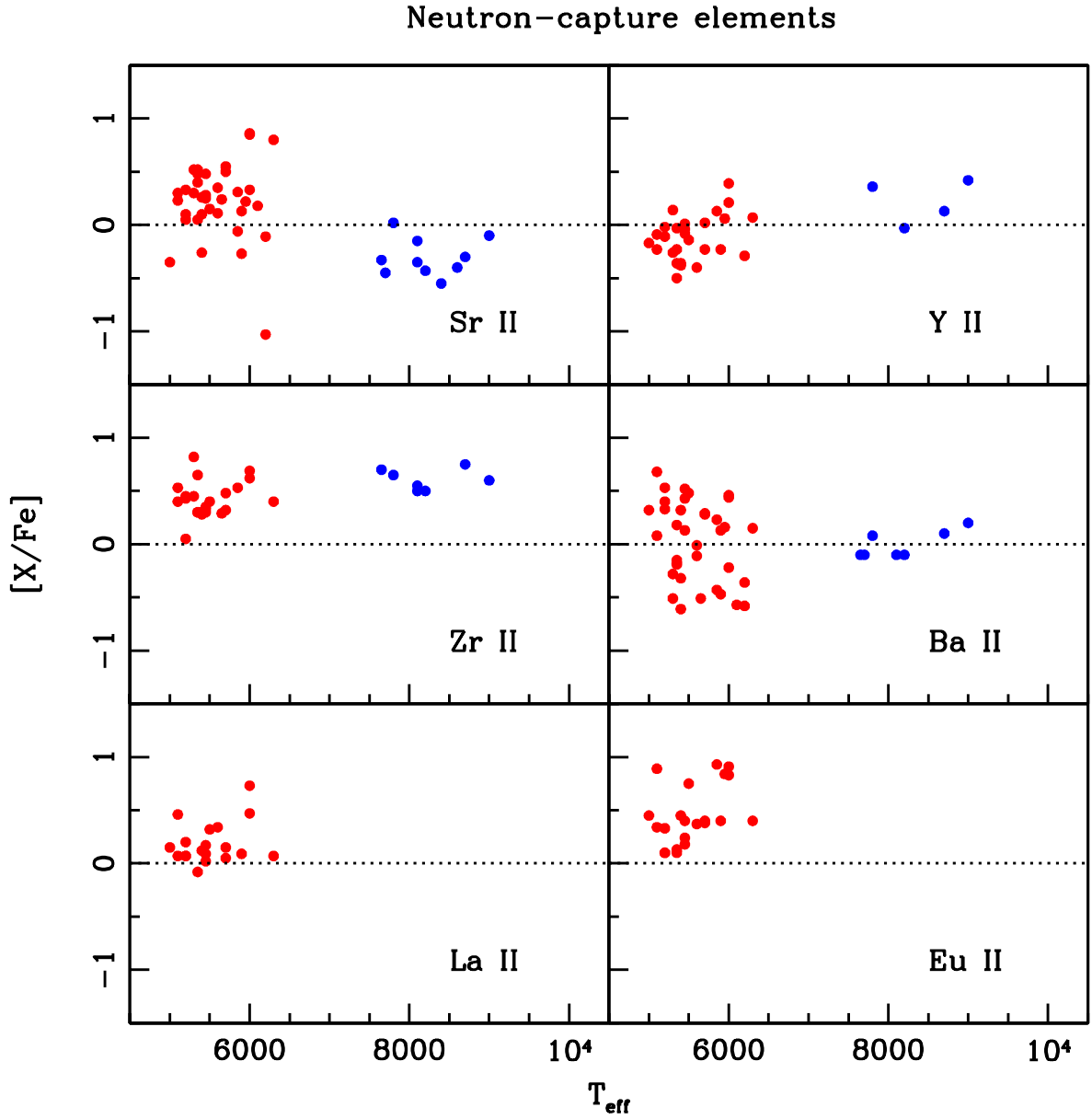}
\caption{Abundance ratios of neutron-capture elements as a function of spectroscopic $T_{\rm eff}$. 
The red and blue dots represent RHB and BHB stars.\label{xfe_teff2}}
\end{figure}

\begin{figure}
\plotone{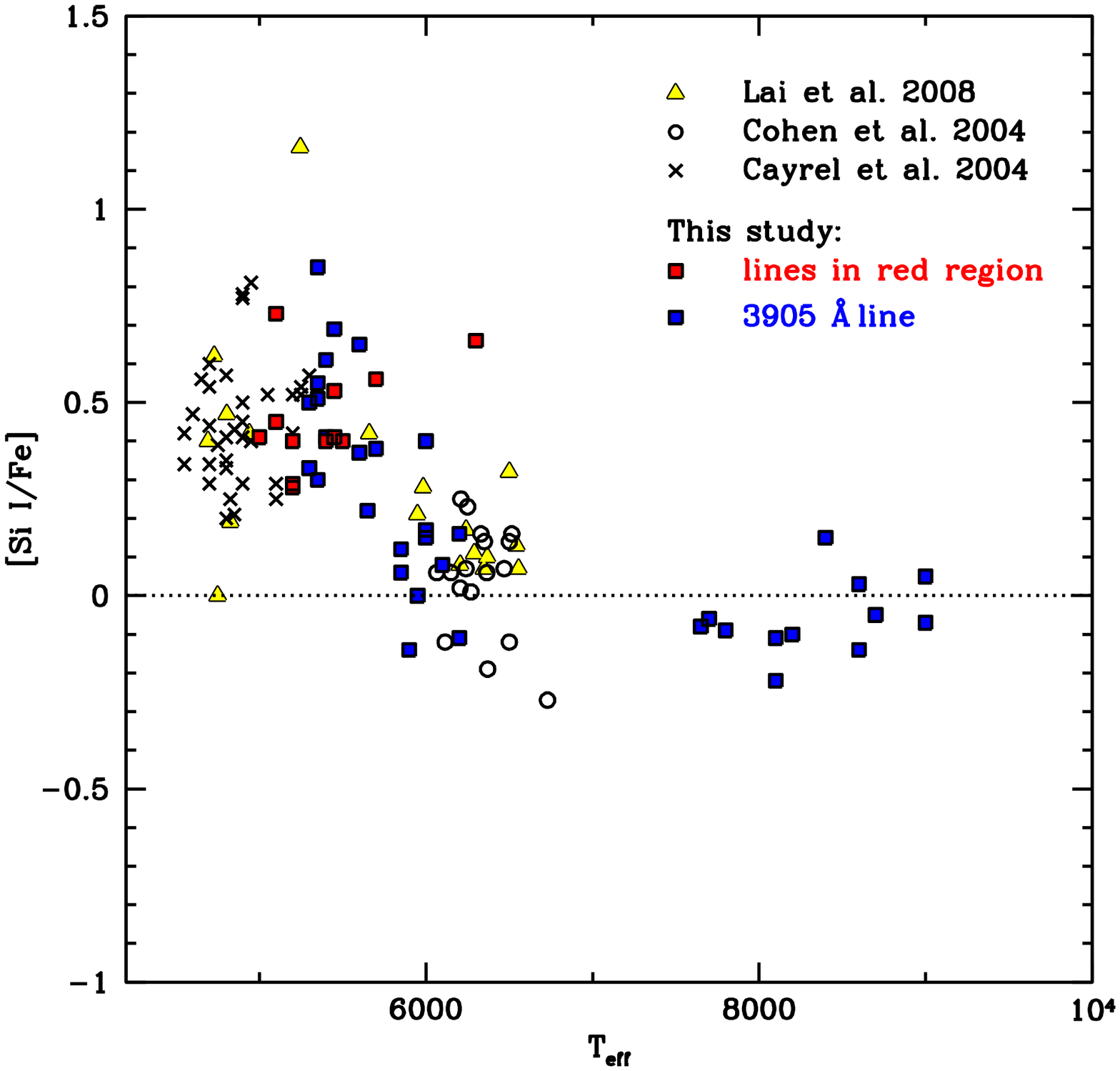}
\caption{Abundance ratios of [Si I/Fe] vs spectroscopic $T_{\rm eff}$, with the addition of data of 
very metal-poor stars giants from \citet{Cayrel04} (crosses), low-luminosity near-turnoff stars 
from \citet{Cohen04} (open circles) and stars in different evolutionary states from \citet{Lai08} 
(yellow triangles). The derived [Si I/Fe] in this study is represented by filled rectangles. NLTE correction applied 
to [Si I/Fe] as described in text. 
The red and blue colors represent Si I lines in red spectral region and 3905 \AA\,line, respectively.     
\label{si_1_teff}}
\end{figure}

\begin{figure}
\plotone{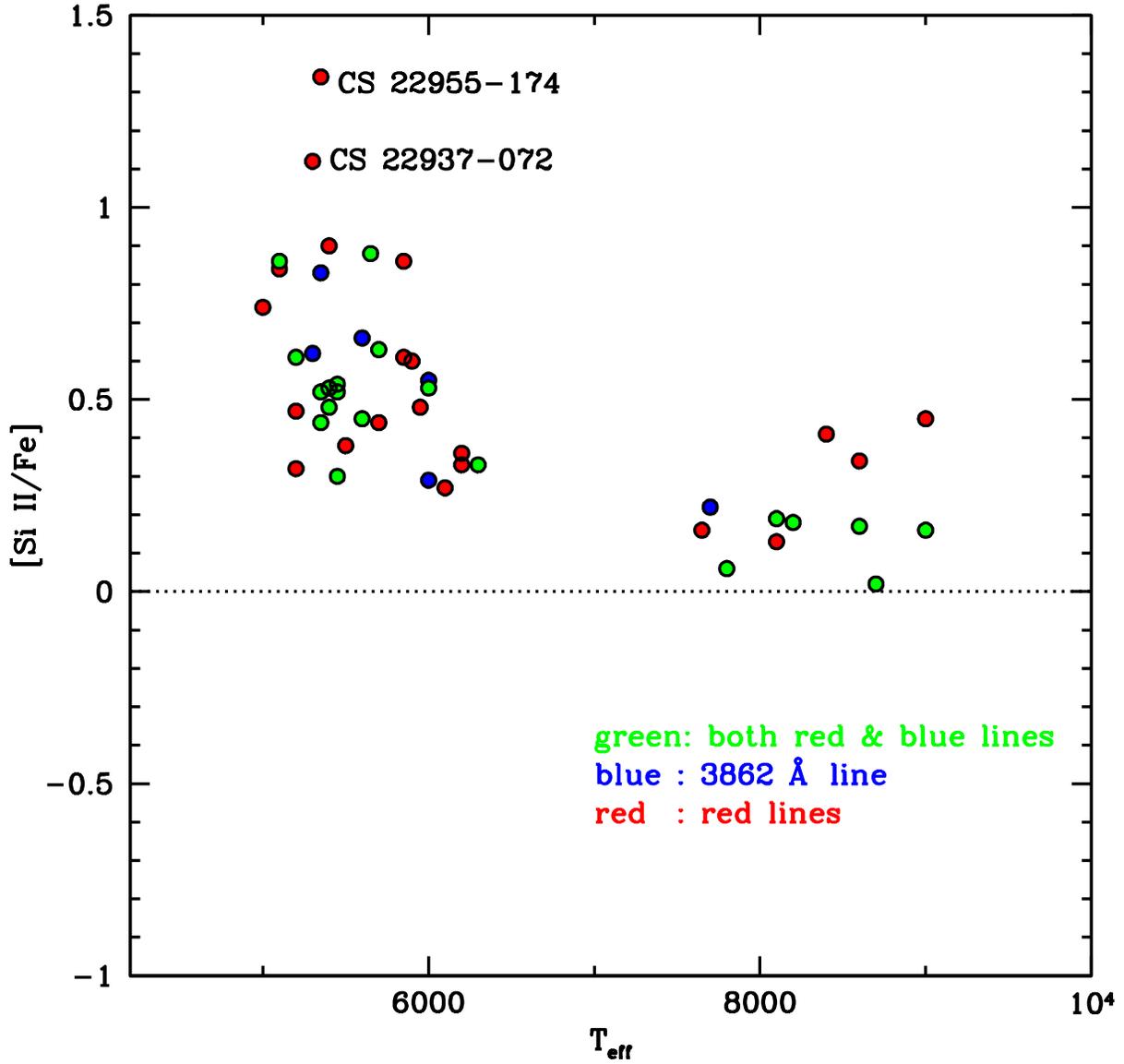}
\caption{Abundance ratios of [Si II/Fe] vs spectroscopic $T_{\rm eff}$. NLTE correction applied 
to [Si II/Fe] as described in text. The colors represent the usage 
of lines in different spectral regions for EW analysis.\label{si_2_teff}}
\end{figure}

\begin{figure}
\plotone{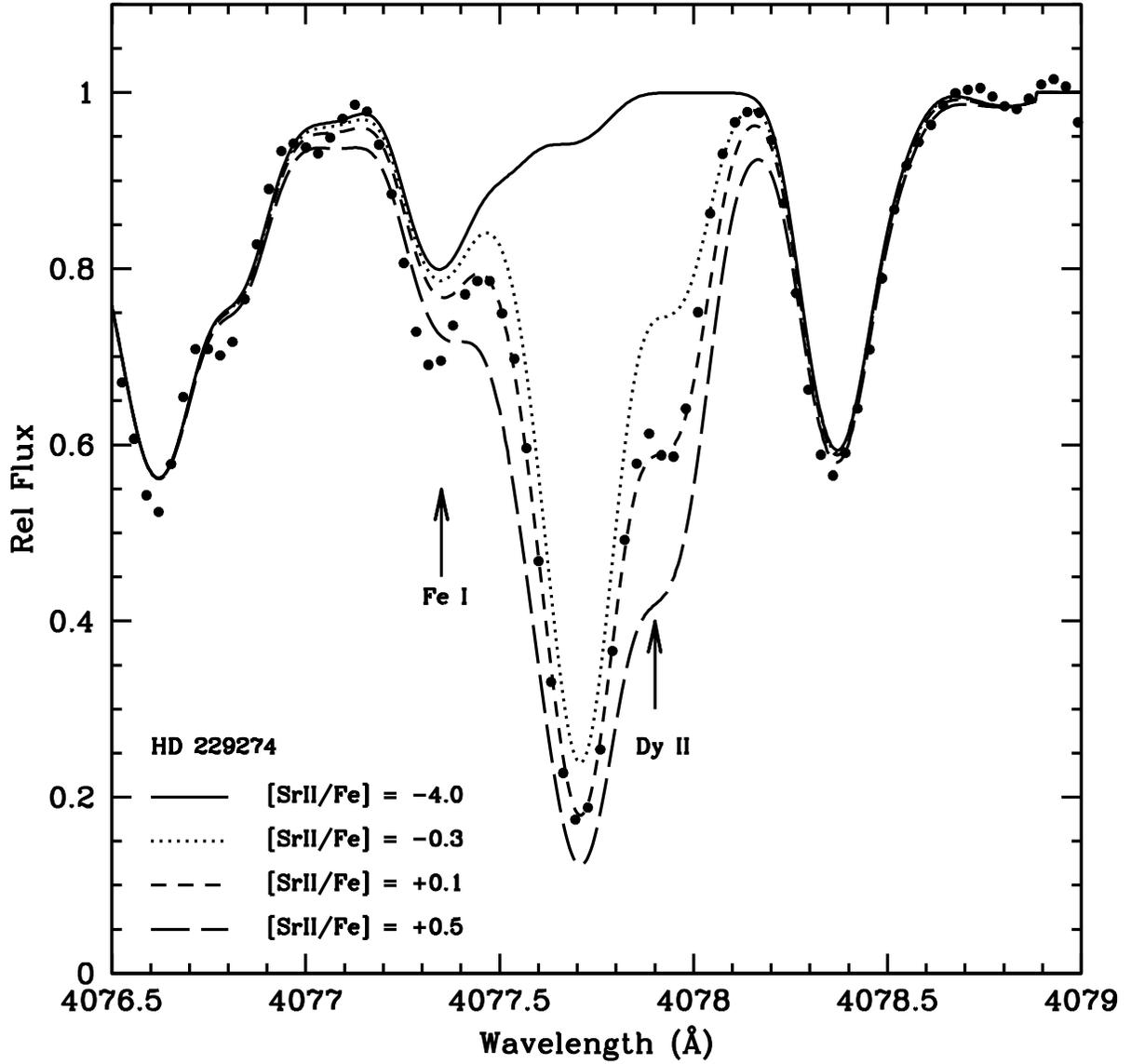}
\caption{An example of synthesized Sr II 4077 \AA\,line superimposed on the observed 
spectrum. The assumed Fe abundance is the same as the metallicity used in the stellar 
parameters. The solid and medium dashed lines represent no Sr contribution and derived 
Sr abundance ratio for this line. The dotted and long dashed lines are $\pm0.4$ dex of 
derived Sr abundance ratio.\label{Sr}}
\end{figure}

\begin{figure}
\plotone{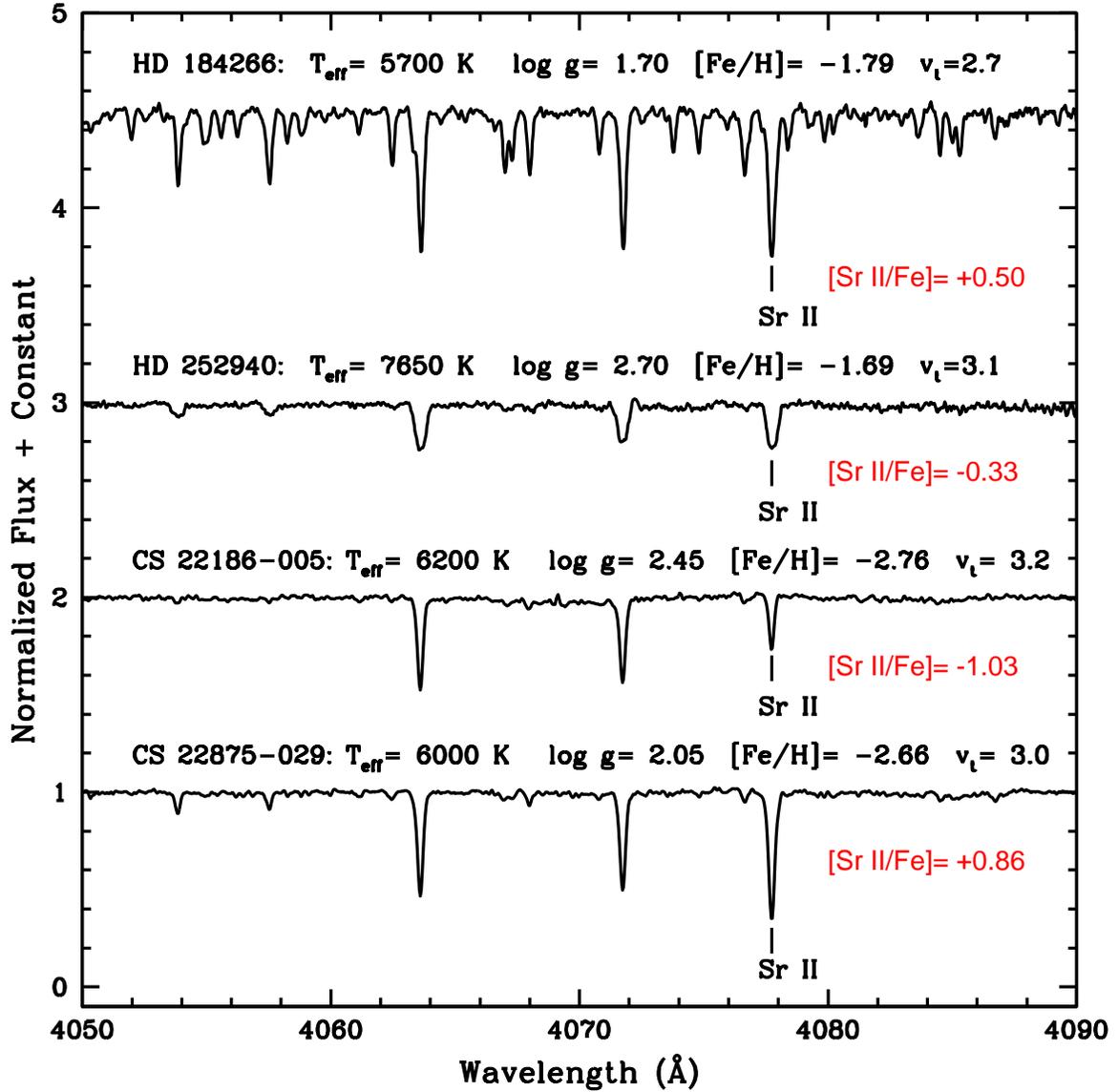}
\caption{The top two spectra show 
the different Sr II line strength between RHB and BHB stars. As shown, Sr II 
line in BHB stars is not as strong as in RHB stars. The bottom 
two stars posses similar stellar parameters but show different line strength in 
Sr II line.\label{comp_srspec}}
\end{figure}

The star-to-star scatter in Sr abundances is large (see Figure~\ref{xfe_2}).  
These variations are intrinsic to the stars, as can be easily seen in
the spectra.
In Figure~\ref{comp_srspec} we show a few examples.
Comparison of stars with similar stellar parameters (\ie, CS\,22186$-$005 
and CS\,22875$-$029 in this figure) shows that the large 
scatter in [Sr/Fe] ratios is real. 
We also note an offset ($\sim 0.5$ dex) of Sr abundance ratios between 
the RHB and BHB stars, which is not present in Yttrium and Zirconium 
abundance ratios (see Figure~\ref{xfe_2} $\&$ \ref{xfe_teff2}). 
This offset may be related to the large \ion{Sr}{2} line strength difference
between the two HB groups. 
Additionally, contamination of the lines by other species, which plagues
the RHB spectra, is not an issue in the BHB stars. 

\begin{figure}
\plotone{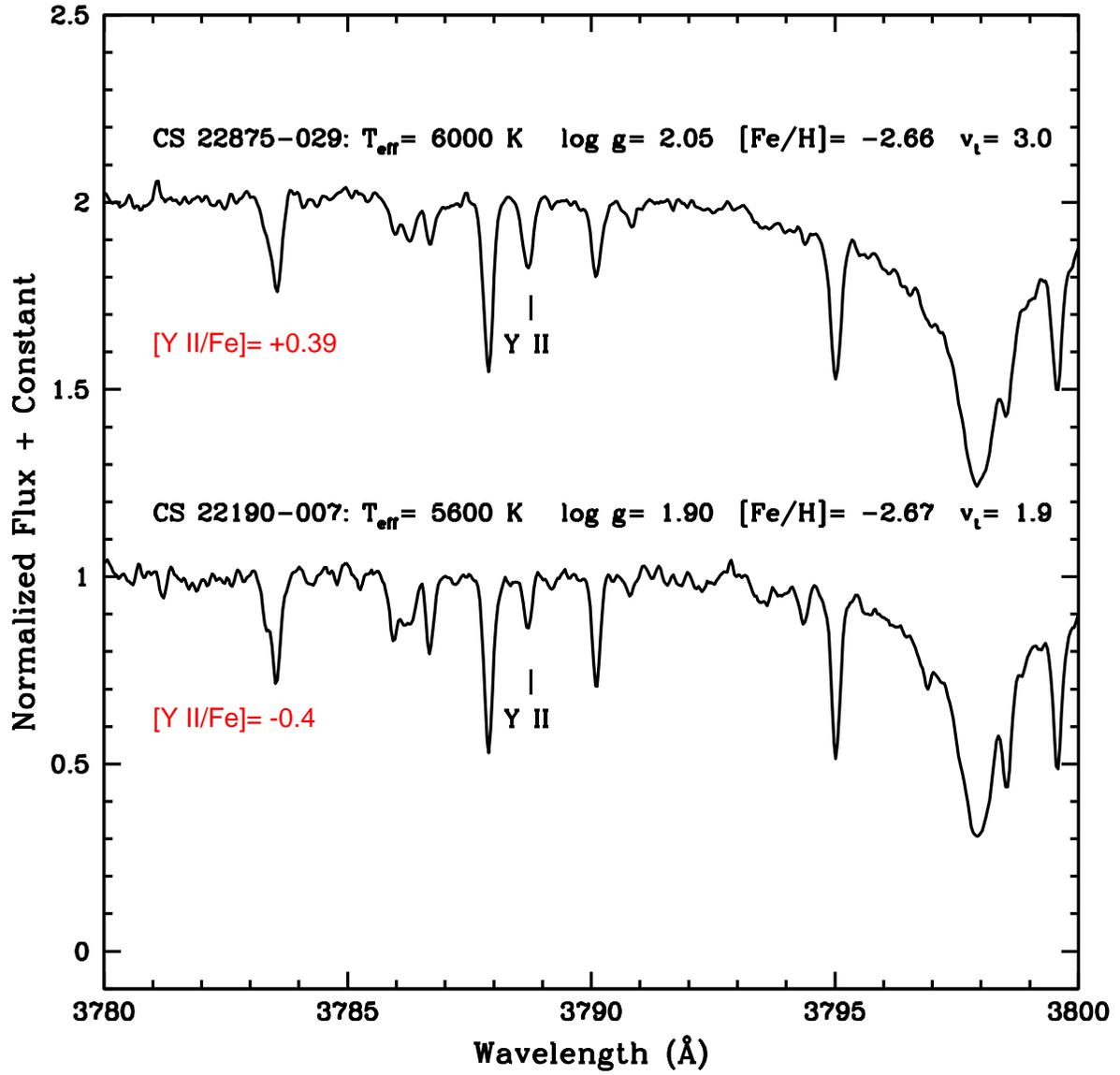}
\caption{Comparison of Y II line strength of stars with similar [Fe/H]. The low 
and high Y II abundance ratios of these two stars contribute to the scatter of 
[Y II/Fe] vs [Fe/H].\label{comp_yspec}}
\end{figure}

We performed EW analysis for Yttrium lines. 
The star-to-star scatter is also large in this element but the analytical
uncertainties are smaller for Y abundances. 
We compare a \ion{Y}{2} line in stars with similar metallicity 
in Figure\ref{comp_yspec}. 
The comparison shows that 
stars with similar metallicity possess different [Y II/Fe]. 

Synthesis were performed for \ion{Zr}{2} 4149 \AA, 4161 \AA, 4090 \AA\ and 4317\AA\ lines,
whenever present in the spectra. 
Generally Zr appears to be overabundant as compared to its neighboring 
light $n$-capture elements Sr and Y. 
We caution that the \ion{Zr}{2} lines are generally very weak, and the
resulting abundance uncertainties are thus large.

Barium is a much-studied member of the heavier $n$-capture element group. 
Its lines are affected by both hyperfine substructure and isotopic splitting. 
A line list with full \ion{Ba}{2} substructure is given in 
\citet{McWilliam98}. 
We adopted the solar abundance ratio distribution among the 
$^{134--138}$Ba isotopes \citep{Lodders03}, and synthesized the
\ion{Ba}{2} lines at 4554 \AA, 5853 \AA, 6141 \AA, and 6496 \AA, whenever present in the spectra.
We note that the 4554 \AA\ line is always substantially stronger than
the other lines, and Ba abundances derived from this line can be severely 
affected by microturbulence and damping.

The spectral lines of La have significant hyperfine substructure, and those of Eu 
have both hyperfine substructure and isotopic substructure. 
There are two natural occurring isotopes, $^{151,153}$Eu, for which we 
adopted the solar abundance ratio distribution \citep{Lodders03}. 
We employed \ion{La}{2} 4086 \AA\ and 4123 \AA\ lines and \ion{Eu}{2} 4129 and
4205 \AA\ lines for abundance analysis. 
In general, Eu and La lines are very weak.
None are detectable in BHB stars, and only 1--2 lines are available in 
RHB stars.

\section{Evolutionary States}

\subsection{$T_{\rm eff}-\log g$ Plane}

We investigated the physical properties of our HB samples, by comparing
our derived temperatures and gravities using the $\alpha$-enhanced, 
HB models of \citet{PCSC06}. 
These models implemented the low $T$-opacities of \citet{Ferguson05} 
and an $\alpha$-enhanced metal distribution that represents typical 
Galactic halo and bulge stars. 
The $\alpha$-enhancement treatment is particularly important because the
$\alpha$-elements are overabundant in metal-poor stellar atmospheres, and
they are major donors of electrons for the for H$^{-}$ continuum opacity. 
We adopted the HB canonical models of various metallicities with $\eta=0.4$. 
The models of \citeauthor{PCSC06} were chosen because they provide a fine 
grid of masses and time steps in contrast to other available HB models. 

\begin{figure}
\plotone{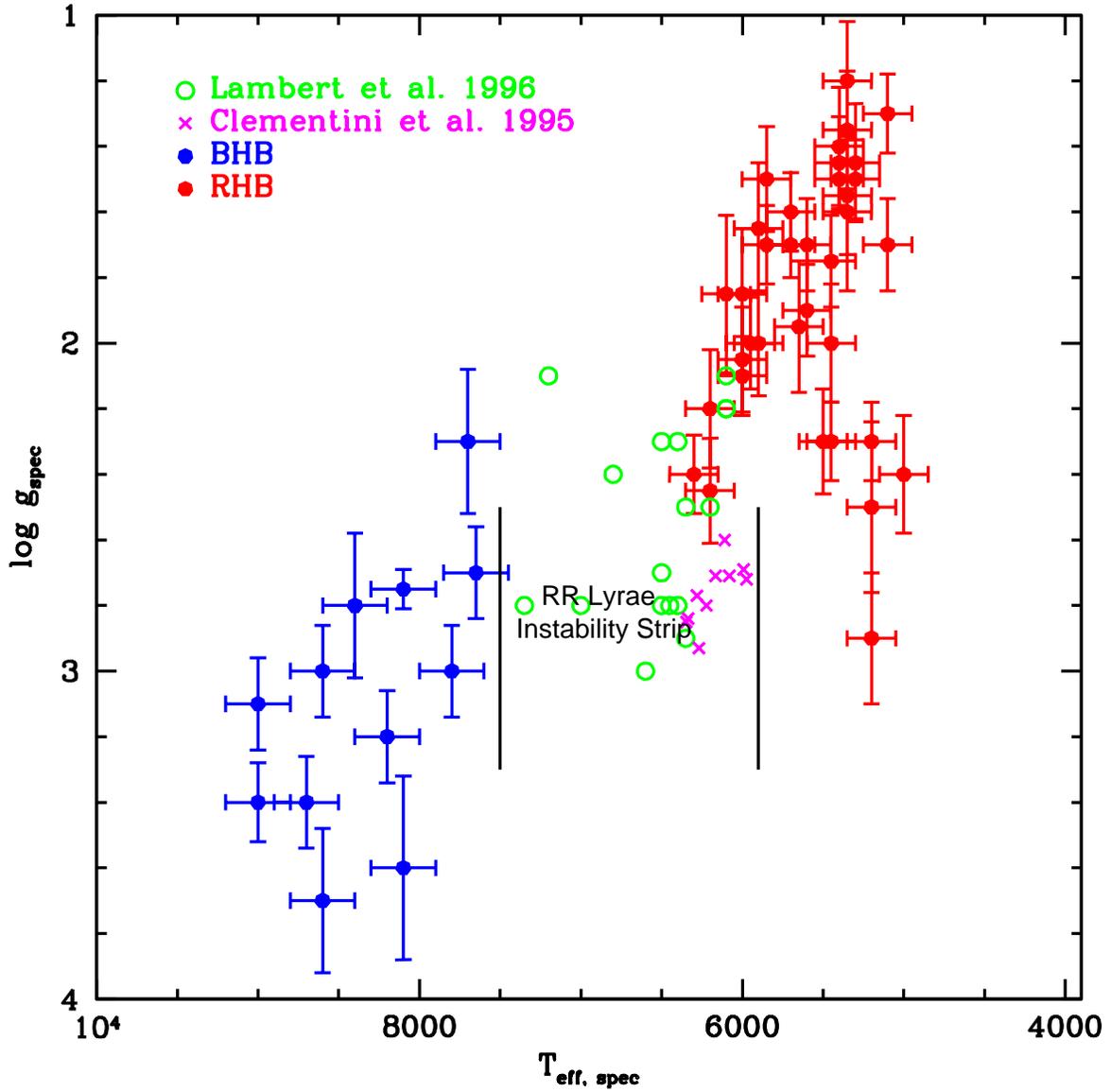}
\caption{The spectroscopic $T_{\rm eff}$ and $\log g$ of our RHB and BHB stars (red and blue dots), 
and $T_{\rm eff}$ and $\log g$ of field RR Lyraes 
from \citealt{Lambert96} and \citealt{Clementini95}) (green open circles $\&$ magenta crosses) on 
the \teff--\logg\ plane. 
\label{hr}}
\end{figure}

In order to convert the bolometric luminosities $L/L_{\sun}$ of the models 
for each mass to \logg\ values, we adopted Eq.~(2) of \citet{Preston06}, 
\begin{equation}
log~g = log(M/M_{\sun}) + 4log~T_{\rm eff} - log (L/L_{\sun})-10.607, 
\end{equation}
in which the constant was evaluated by using the solar \teff\ and \logg\ values.
In Figure~\ref{hr}, we show the spectroscopic \teff\ and \logg\
values of our stars and the field RR Lyraes that are based on spectroscopic 
\teff\ and \logg\ of \citet{Lambert96}, and, photometric \teff\ and 
Baade-Wesselink \logg\ of \citet{Clementini95}, on the \teff--\logg\ plane. 
Both their data and our samples exhibit similar gravity scatter at fixed 
temperature.

To estimate the uncertainties associated with the \citet{PCSC06} HB models,
we compare their luminosities (as translated into \logg) for a given mass 
with \citet{LD90}'s HB model (\ie, [Fe/H]$=-2.26$, $Z=0.0001$, 
$Y=0.23$).\footnote{
\citet{Dorman93} also published HB models with similar parameters, but
their time steps are too large to be useful in this exercise.}
The comparison is summarized in Table~\ref{model_comp}.       
The difference in \logg\ in the two studies is $\lesssim$0.1~dex, much 
smaller than the uncertainties in our spectroscopic \logg\ values. 
Therefore, model choice is not an issue in contributing significant 
error on the mass derivation.

\subsection{Derivation of HB Masses}

Our mass estimation uses HB evolutionary tracks in the 
$T_{\rm eff}-\log g$ plane. As discussed in \S 5.1, 
spectroscopic \logg\ values are generally lower than the photometric ones, 
which would result in deriving more of low mass HB stars. 
Therefore, a correction of the spectroscopic gravities is 
necessary and adopting the photometric gravities is more appropriated to 
represent the physical gravities. 

\citet{Preston06} derived an empirical relation for computing 
photometric gravities (\logg$_{\rm phot}$) by using their spectroscopic 
gravities (\logg$_{\rm spec}$) in conjunction with the 
existing \logg$_{\rm phot}$ of M\,15. 
We adopted this relation, 
\begin{equation}
\logg_{\rm phot}=\logg_{\rm spec}+28.802-7.655{\rm log}\teff_{\rm,spec}
\end{equation}
to obtain the \logg$_{\rm phot}$ for all our RHB stars. 
While there are published \logg$_{\rm phot}$ data for BHB stars in 
other GCs \citep{BBehr03}, there are no useful \logg$_{\rm spec}$ values 
for comparison (\citealp{BBehr03} suggested that their measurements are 
too uncertain to provide any useful information on this issue). 
Additionally, \citeauthor{Preston06} showed that the corrections 
to \logg$_{\rm spec}$ decline with increasing \teff\ and essentially
disappear at the red edge of RR Lyr IS (see their Figure~17). 
This can be understood by noting that the continuous opacity of a hotter 
star is dominated by H$^{-}$, and the dominant electron donor
is hydrogen itself rather than the metals. 
The electron density rises sharply with increasing \teff\ among RHB stars.
Examination of atmosphere models for the M15 RHBs (from 
\citeauthor{Preston06}) suggests that in the line-forming regions,
the electron pressure increases by a factor of more than 30 from the
coolest (\teff\ = 5000~K) to the warmest (\teff\ = 6250~K) stars.
This higher electron pressure helps to enforce LTE in the ionization
equilibria in warmer HB stars. Thus, we assume the spectroscopic \logg\
for our BHB stars is correct and no correction is applied. 
Future spectroscopic investigation of \logg\ for BHB stars in GCs would 
be welcome.

\begin{figure}
\plotone{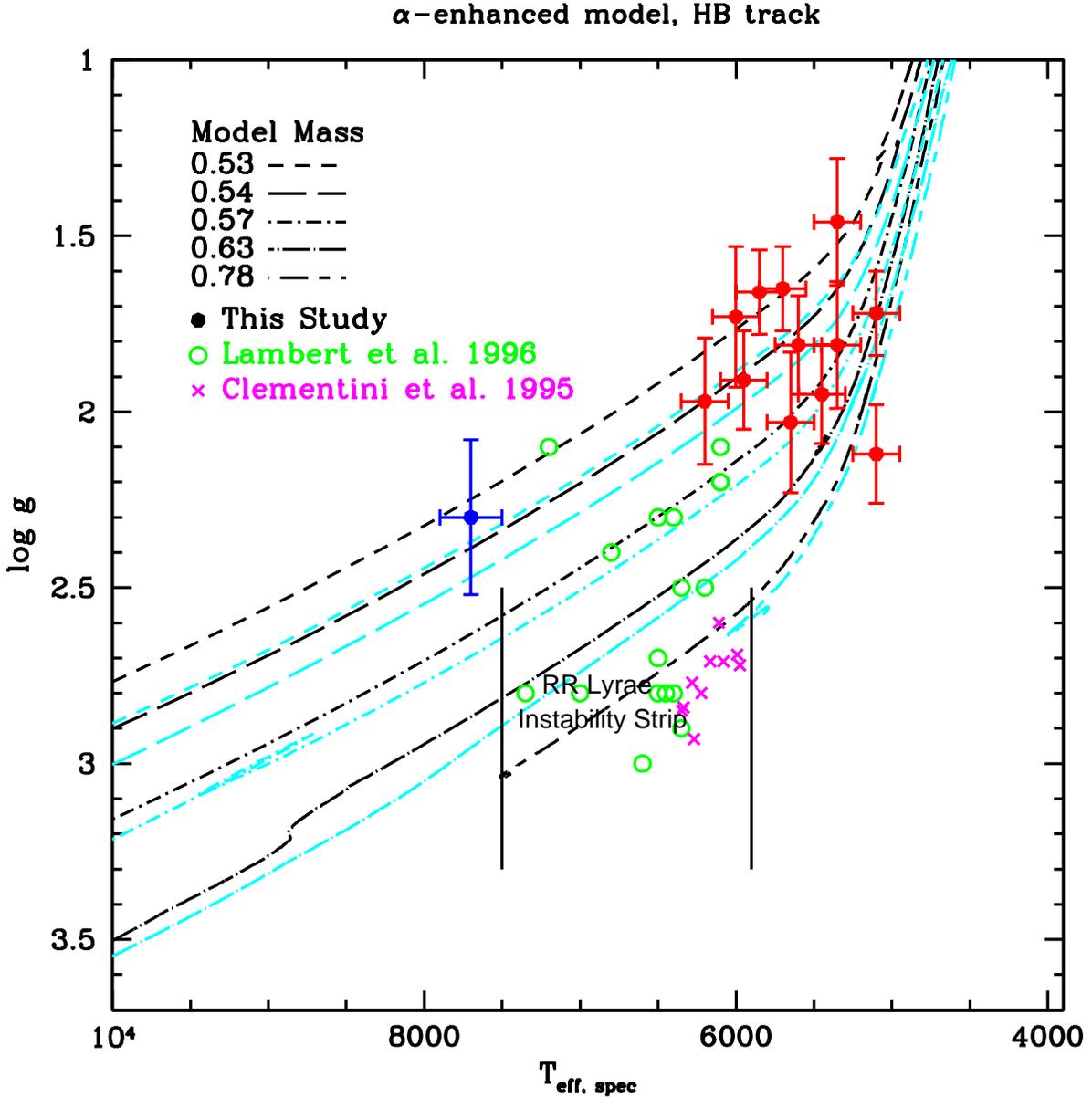}
\caption{The spectroscopic $T_{\rm eff}$ and photometric/spectroscopic $\log g$ of 
a set of our RHB and BHB stars 
(red and blue dots) overlaid on $\alpha$-enhanced HB tracks of [M/H]$=-1.79$, $Z=0.0003$, $Y= 0.245$ (black) and 
[M/H]=$-2.27$, $Z=0.0001$, $Y=0.245$ (cyan). These HB tracks were used to derive the masses of this set of HB stars. 
The $T_{\rm eff}$ and $\log g$ of field RR Lyraes are 
from \citealt{Lambert96} and \citealt{Clementini95} (green open circles $\&$ magenta crosses).
\label{evotrack}}
\end{figure}

\begin{figure}
\plotone{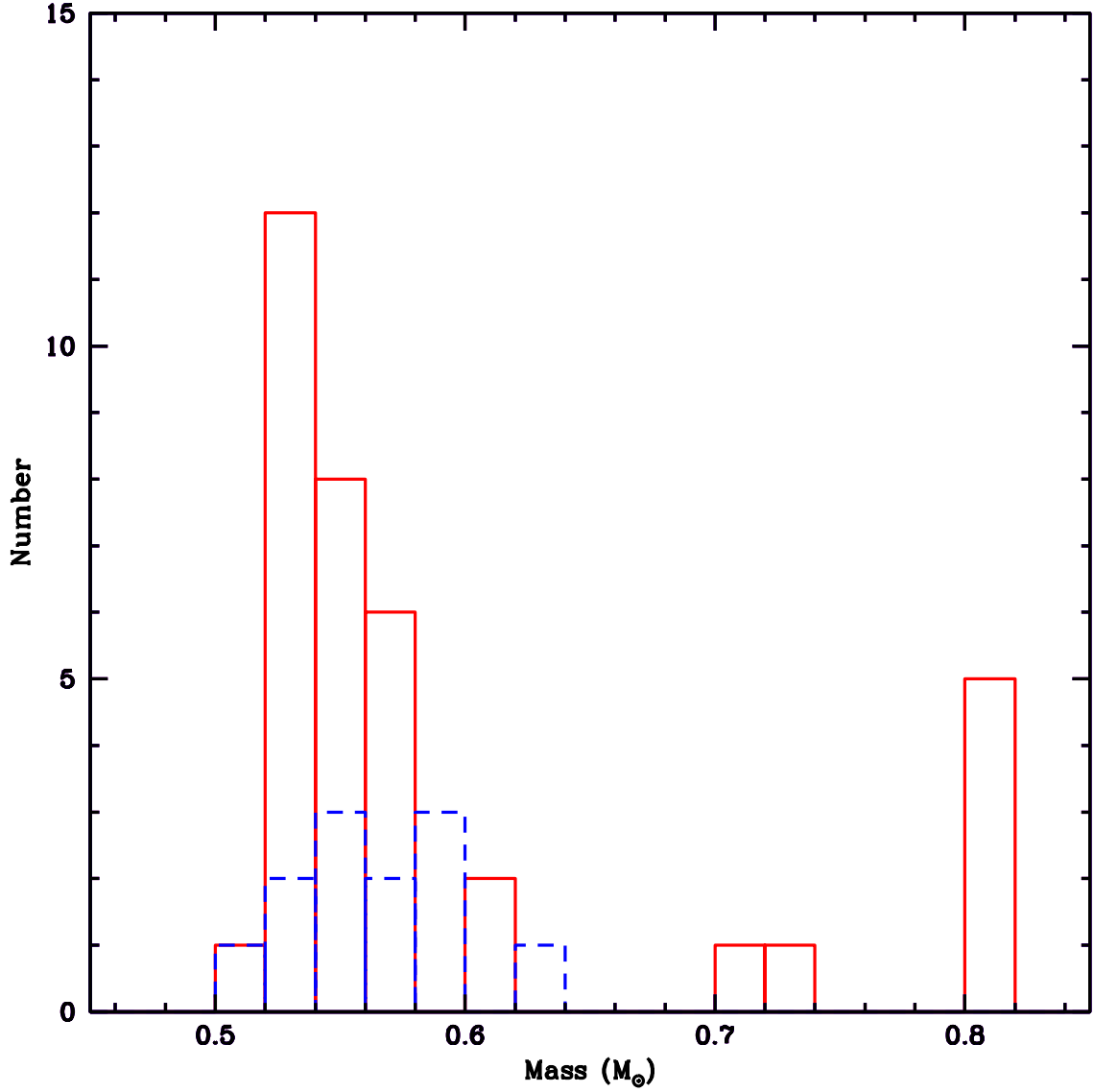}
\caption{The red (solid) and blue (dashed) histograms represent the 
estimated RHB and BHB masses. The mean masses for RHB and BHB stars are 
$0.59~M_\sun$ and $0.56~M_\sun$. Excluding the upper mass limit RHB stars ($M > 0.7~M_\sun$), 
the mean masses are $0.56~M_\sun$ for both RHB and BHB stars. The median masses for RHB and BHB stars 
are $0.54~M_\sun$ and $0.56~M_\sun$, respectively.\label{mass_dist}}
\end{figure}

After calculating RHB \logg$_{\rm phot}$ values, we estimated the masses of 
individual HB star by employing an interpolation scheme. 
To account for different metallicities of our program stars, we first 
chose two models that closely match a star's [Fe/H] and superimposed 
them on the \teff-\logg\ plane along with the \teff$_{\rm ,spec}$ and 
\logg$_{\rm phot}$. 
Then, calculating the linear interpolation between these two 
metallicities and masses:
\begin{equation}
M_{\rm star}=M_{1}+\frac{(M_{2}-M_{1})}{([Fe/H]_{2}-[Fe/H]_{1})}\times([Fe/H]_{\rm star}-[Fe/H]_{1})
\end{equation} 
where $M_{1}$, $M_{2}$ are estimated masses from the two models, 
and [Fe/H]$_{1}$,[Fe/H]$_{2}$ are the two models' iron abundances.  
For stars positioned outside the model mass range 
($0.503 M_\sun \leq M \leq 0.80M_\sun$), we chose the mass that is within 
the $\log g$ and \teff\ errors of the star on \teff--$\log g$ plane.
If there is no mass track lies within the errors, we constrain 
the upper mass limit to be 0.8~$M_{\sun}$, the approximate turnoff mass of 
a old metal-poor main-sequence star. 
In Figure~\ref{evotrack}, we show an example of a set of HB stars superimposed on the 
HB tracks ([M/H]$=-1.79$ and $-2.27$) that were used to derive their masses.  
We summarize the derived masses as a histogram in Figure~\ref{mass_dist} and 
parameters used to derive the masses is listed in Table~\ref{der_masses}. 

The inferred mass distributions have means at $0.59~M_\sun$ 
and $0.56~M_\sun$ for RHB and BHB stars, respectively 
(see Figure~\ref{mass_dist}). 
If we exclude those RHB stars that have masses set to the upper limit 
($M > 0.8 M_\sun$), the mean masses for RHB and BHB 
stars are both  $0.56~M_\sun$, and the median masses are 
$0.54~M_\sun$ and $0.56~M_\sun$.
 
This estimated mean mass is smaller than the HB masses found in some GCs, 
\eg\ M3, for which \citet{VC08} derived mean masses of $0.633~M_{\sun}$ 
and $0.650~M_{\sun}$ for RHB and BHB stars, respectively. 
We also do not find a bimodal or multi-modal HB mass distribution that
appears to exist in many GC's (see \citeauthor{VC08}; \citealp{Catelan04}). 
Several reasons could contribute to these differences.
(1) GC's are mostly mono-metallic, in contrast to the large metallicity 
range of our FHB stars. 
We have needed to multiple evolutionary tracks that correspond most 
closely to the individual metallicities of our FHB stars (refer back to 
the interpolation method as described above). 
(2) Our sample sizes of RHB and BHB stars are too small to clearly indicate
statistically significant mass distributions. 
(3) We have used an empirical correction to spectroscopically-determined
\logg\ values, which directly impacts the derived masses.
(3) Our samples consist more of RHB than BHB stars, where the majority 
agglomerate near the low mass end, resulting in more low mass HB estimates.
(4) Finally, \citeauthor{VC08} cautioned about over-interpretation of masses
derived from the GC CMD method, because they are biased against 
stars in later evolutionary states. 
Thus, it is not clear that our mean masses are substantially different than
those reported for M3.

Additionally, other GC HB mass study have reported mean mass 
in reasonable agreement with ours. 
For example, \citet{Deboer93} obtained $<M_{\rm HB}> = 0.5~M_\sun$ for 
NGC~6397.
Masses of nearby HB stars derived via $Hipparcos$ parallaxes have slightly 
smaller mean masses, $<M_{\rm HB}> = 0.38~M_\sun$, than ours \citep{Deboer97}. 
Finally, the evolutionary and structural models of \citet{Sweigart87} 
suggest a wide range of individual HB masses (0.2 -- 1.2 M$_\sun$). 
We conclude that our derived mean masses for the field HB stars are
reasonable.

\subsection{Blue and Red Edges of the RR Lyrae Instability Strip: [Fe/H]$>-2.5$}

Locations of the blue and red edges (BE and RE) of the 
RR Lyr IS provide powerful constraints on stellar pulsation theory. 
They can be determined directly by examining the color-magnitude 
diagram of GCs that are well populated with RR Lyrs.
Unfortunately, this requirement eliminates most clusters. 

Additionally, accurate cluster reddenings must be known to 
transformation from colors to \teff\ values. 
Determining the blue and red edges from bright field RR Lyr stars via spectroscopic method 
can avoid these complications. 
For the metallicity regime [Fe/H]$< -2.0$, \citet{Preston06} estimated the 
fundamental red edge from the \teff\ distributions of field RHB stars and GC RR Lyrs. 
Since HB colors are affected by metallicity, shifting slightly blueward
with decreasing [Fe/H] (e.g., see Figure~1 of \citealp{Sandage90}), 
we repeated the exercise with our sample. 
We considered only those stars with [Fe/H]$> -2.5$, and compared 
the \teff\ distributions of our field RHB and BHB 
with the distribution for field RR Lyr stars.

\begin{figure}
\plotone{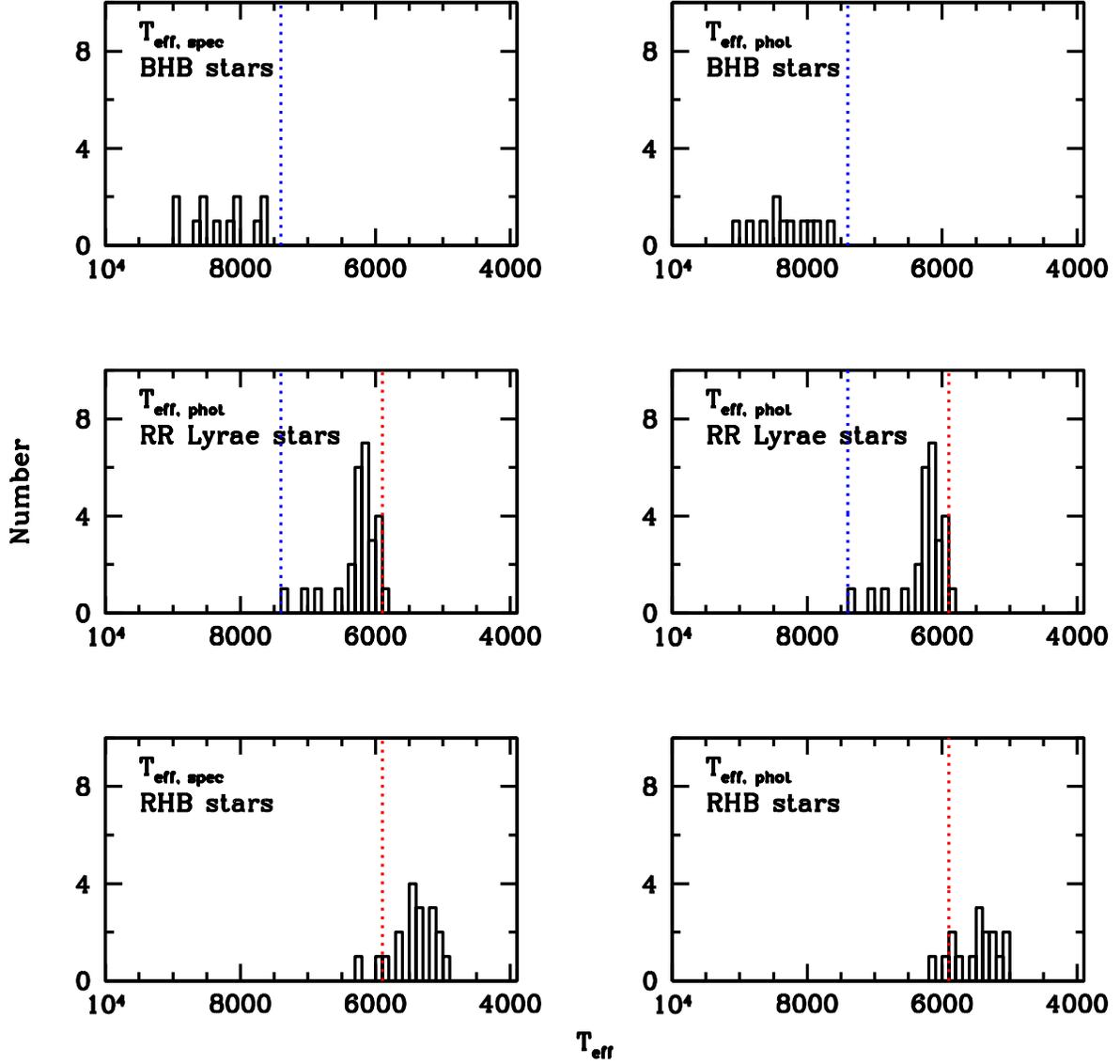}
\caption{The top and bottom panels show the histograms of 
spectroscopic and photometric \teff\ of BHB and RHB stars. The middle panels (same) are the 
photometric \teff\ of field RR Lyr stars extracted from \citet{Lambert96} and \citet{Clementini95}. 
The red and blue dotted lines represent the estimated fundamental red and blue edges of field 
RR Lyr IS for [Fe/H]$>-2.5$.\label{teff_hist}}
\end{figure}

In Figure~\ref{teff_hist}, the top and bottom panels show the distributions 
of spectroscopic and photometric \teff's of BHB and RHB stars 
with [Fe/H]$> -2.5$, respectively. 
The data for field RR Lyr stars (fundamental mode RRab and first 
overtone RRc variables) in both middle panels are extracted from 
\citet{Lambert96} and \citet{Clementini95}. 
It shows the RR Lyr distribution drops at \teff\ = 5900~K and 7000~K. 
Both photometric and spectroscopic \teff\ RHB distributions decline at 
\teff\  $>$ 5700~K and overlap with the RR Lyr distributions (bottom panels). 
We suggest that the weak overlap region, $\simeq$5900~K, is the red edge of 
field HB with [Fe/H]$> -2.5$. 
The \teff's of our BHB sample have no overlap with those of the RR Lyr stars. 
This is expected since RRc type variables, which are bluer than the 
RRab type variables, are generally used for determining the BE, and 
there are only two RRc type variables from \citet{Lambert96} being included 
in the histogram (middle panels). 
Assuming the RRc type variables defined the blue edge in this case, 
we approximated it to be 7400~K.

While field HB stars can be used for deriving red and blue edges, 
we warn that 
the method is not very robust. 
The lack of large BHB samples and uncertainties in \teff\ values of 
field RRc stars are limiting factors on our blue edge estimates. 
The overlapping distributions of field RHB and RRab stars also limit
the red edge accuracy.
Perhaps semi-empirical work (\ie, simulations to map the observed 
distributions) would provide a better constraints on the red and blue edges of FHB stars. 
Before then, deriving \teff's for a large sample of field BHB and RRc 
will be needed.

\section{Discussion}

\begin{figure}
\plotone{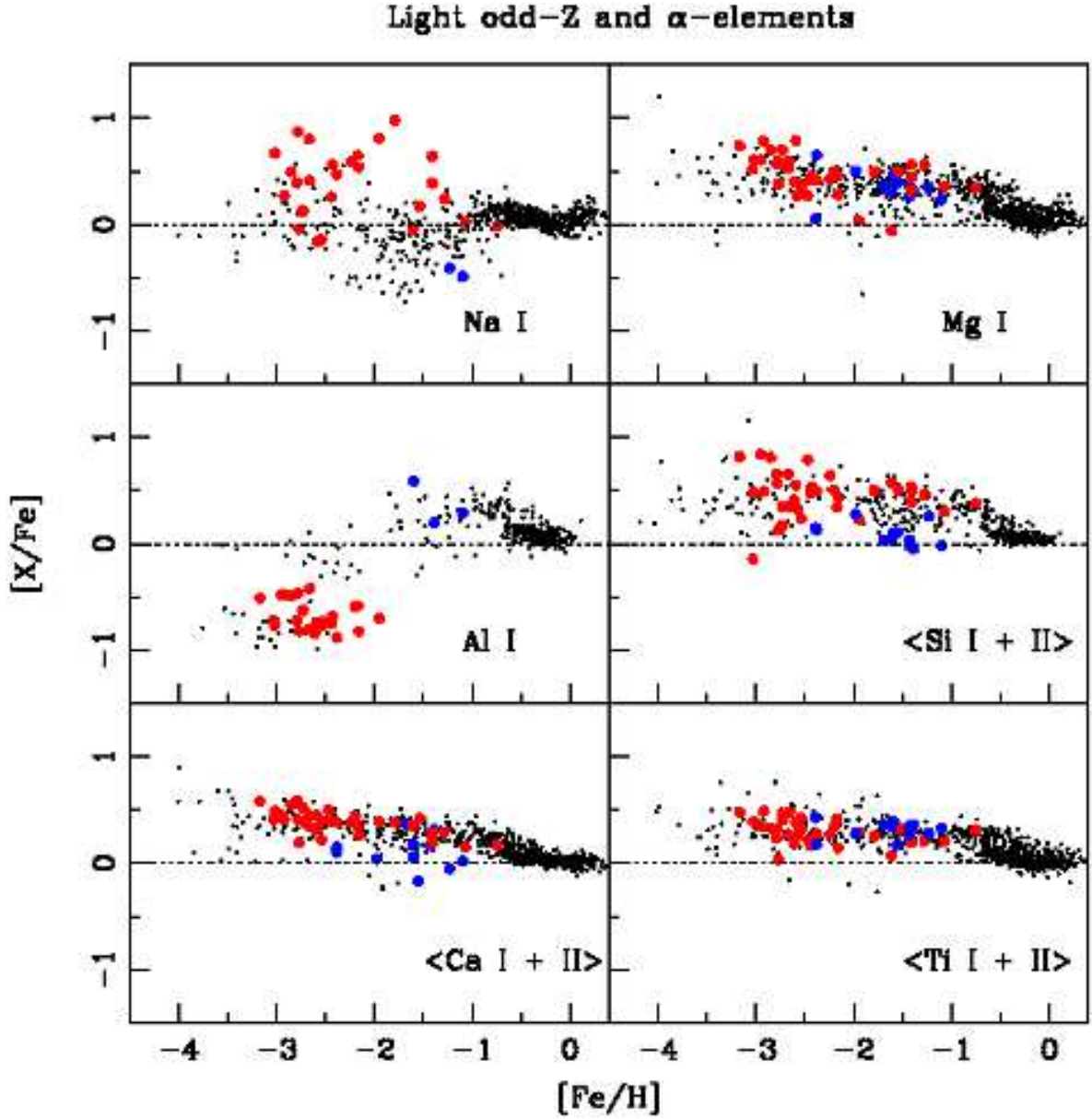}
\caption{Abundance ratios of light odd-Z and $\alpha$-elements in this study superimposed on the data assembled 
by \citet{Venn04} and us. Mean of neutral and ionized species are used for comparisons. NLTE corrections applied to 
\ion{Na}{1}, \ion{Al}{1}, \ion{Si}{1} $\&$ \ion{Si}{2} for our HB stars. The red and blue dots correspond to RHB and BHB stars.\label{x_fe_total}}
\end{figure}

\begin{figure}
\plotone{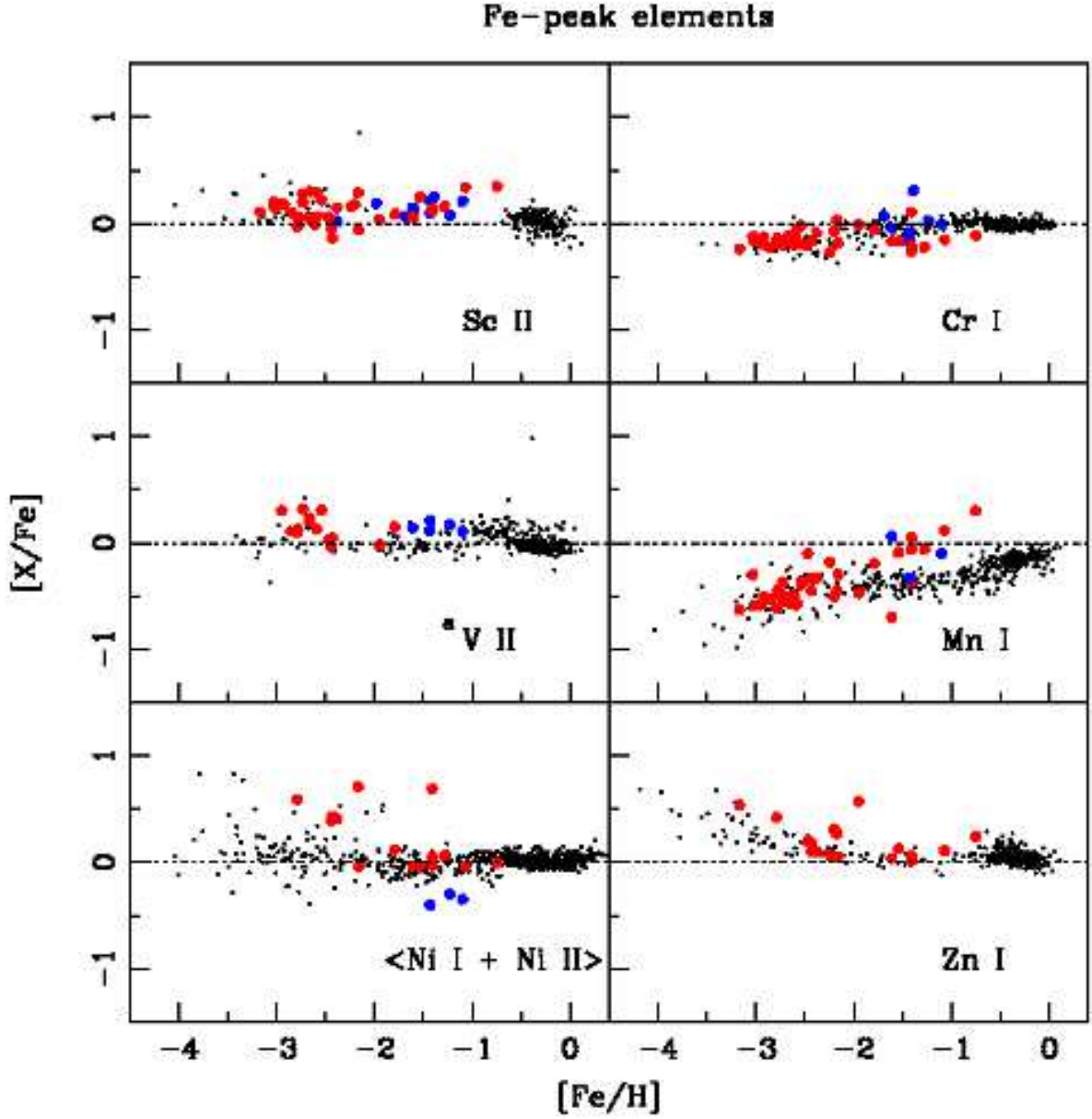}
\caption{Same as Figure~\ref{x_fe_total}, except for Fe-peak elements. a: [\ion{V}{1}/Fe] for stars possess [Fe/H] $> 2.0$ is used 
for comparison. The red and blue dots correspond to RHB and BHB stars.\label{x_fe_total1}}
\end{figure}

In this paper we have explored the chemical compositions of non-variable
RHB and BHB field stars.
Here we will compare our results with abundances in other 
evolutionary groups of halo field stars, and discuss some of the 
possible nucleosynthetic implications.
The comparisons of our [X/Fe] values with those of field stars are 
presented in Figure~\ref{x_fe_total}$-$\ref{x_fe_total2}, 
where neutral and ionized species abundances of several elements have 
been averaged. 
We did not combine \ion{Cr}{1} $\&$ \ion{Cr}{2} abundances, since their
distributions conspicuously diverge at lower metallicities 
(as discussed in \S 6.4). 
Data for field stars were mainly taken from the compilation 
of \citet{Venn04}. 
For those [X/Fe] that are not listed in \citet{Venn04}, we assembled the
comparison samples from several references, which we summarize in
Table~\ref{references}.

\subsection{Light and Iron-peak Elements}

Enrichment of $\alpha$-elements in metal-poor stars has been known for decades. 
The explanation for this behavior presumes predominance of nucleosynthetic
contributions from short-lived massive stars that died in core-collapse 
type II supernovae (SNe II) in early Galactic times. 
The resulting explosions contributed large amounts of light $\alpha$-elements 
(\eg, O, Ne, Mg and Si), smaller amounts of heavier $\alpha$-elements 
(e.g., Ca and Ti) and small amounts of Fe-peak elements to the ISM 
\citep{WW95}. 
Longer-lived, lower-mass stars began to contribute their 
ejecta by adding more Fe-peak elements through Type~Ia supernovae 
(SNe Ia) from lower-mass progenitors which exploded in thermonuclear 
runaway processes at later times. 
When SNe Ia became significant polluters of the ISM, a lowering of the 
[$\alpha$/Fe] values (at higher metallicities) occurred.

In general our HB $\alpha$-element abundances agree with those
of other halo star populations. 
We illustrate this in Figure~\ref{x_fe_total}, where [\ion{Mg}{1}/Fe] 
and [\ion{Ti}{1}/Fe] of our RHB and BHB are in close accord with other 
field stars. 
The $<$[Si I+II/Fe]$>$ and $<$[Ca I+II]$>$ of RHB stars follow the general 
field star trend but these ratios tend to be
lower for BHB stars in the same metallicity range (\ie, $\sim 0.35$ dex lower). 
The offset of mean Ca abundances is mainly due to the lower 
[\ion{Ca}{1}/Fe] of BHB stars (see description in \S 6.1). Similar lines were 
used in both BHB and RHB stars, as such, line selection is probably not the 
cause of the offset.  
As for $<$[Si I+II/Fe]$>$, the star-to-star scatter is large and the 
offset between RHB and BHB stars is dominated by the RHB star 
[\ion{Si}{1}/Fe] dependence on \teff\ (see \S 6.2).

Our BHB and RHB sodium abundance pattern looks quite different than in
other field stars. 
However, little weight should be attached to our results because they
have large uncertainties.
We must rely solely on the Na D lines, and they are very strong in RHB stars.
Aluminum is produced in massive stars, similarly to magnesium, 
but significantly deficient with respect to iron in metal-poor stars. 
The production of Al rises as it reaches the disk-to-halo transition at 
higher metallicity, \ie, [Fe/H] $\gtrsim 1.5$ (\eg, \citealp{Timmes95}). 
Our abundances confirm this, with the caution that our derived trend
with metallicity depends solely on RHB stars at low [Fe/H] and all BHB stars
at high [Fe/H].

Iron-peak elements (with the exception of Ti, discussed above)
are believed to be largely produced during  Type~Ia and Type~II
SNe explosion events. 
In our metallicity regime the iron-peak abundances of main-sequence and
RGB stars generally have their solar values, with the exception 
of Mn and Cu. The derived Fe-peak abundance ratios (\ie, \ion{Sc}{2}, \ion{Cr}{1}, 
and \ion{V}{2}) of our RHB and BHB stars are also in agreement with 
those found in field dwarfs and giants (see Figure~\ref{x_fe_total1}). 
Most of them are expected to be constant in all metallicity regimes. 
Manganese and Zinc are the exceptions. 
In common with previous studies, [Mn/Fe] ratios of our HB stars increase 
as metallicity increases, but the slope of this relation may be larger 
in our sample. 
We do not have a clear physical explanation to this, and caution
that, (a) the trend is based on relatively few points,
and (b) [Mn/Fe] is quite sensitive to stellar parameter choices
(refer to Table~\ref{sensitivity1} $\&$ \ref{sensitivity2}). 
Again, we refer the reader to \citet{Sobeck06} for the production of Mn.   

For nickel abundances we must rely on \ion{Ni}{1} lines for RHB stars 
and \ion{Ni}{2} lines for BHB stars.  
The low \ion{Ni}{2} abundances of BHB stars should not be given
large weight, as they are solely derived from one line. 
The very large [\ion{Ni}{1}/Fe] values of several RHB stars, 
substantially at variance with the general trend of field stars, are 
most likely due to the lack of many detectable lines.
The RHB stars with more than four lines contributing to their Ni abundance 
have ratios in good agreement with the field stars. 

We find [Zn/Fe]~$\simeq$~0.0 throughout the metallicity regime 
of [Fe/H] $> -2.0$, 
which is consistent with the study of \citet{SGC91}. 
Recent work by \citet{Cayrel04} shows increasing 
[Zn/Fe] at decreasing metallicities.
Such a trend could indicate an $\alpha$-rich freezeout process 
contribution to Fe-group element production at low metallicities. 
Our Zn abundance at low metallicity range, \ie, [Fe/H] $< -2.0$, 
perhaps consistent with this recent finding, but
our data points are too sparse for firm conclusions on this point. 
Unfortunately, the comparison can only be made for 
RHB stars since the Zn I lines in BHB stars are too weak to be detected.

\subsection{Neutron-Capture Elements}

Elements heavier than the iron-peak (Z~$>$~30) cannot be efficiently 
synthesized by charged-particle fusion because of Coulomb repulsion and 
the endothermic nature of such reactions.
They are produced in the late stages of stellar evolution via
neutron-capture events, namely the $s$- and $r$-processes (see review by \citealp{Sneden08}).
The $s$-process occurs quiescently in the He-fusion zones of low or 
intermediate mass AGB stars, while the $r$-process is believed to occur 
explosively in neutron rich sites, \eg, Type~II SNe or merging events
of two neutron stars \citep{Rosswog99}.

We have abundances for six $n$-capture elements in HB stars.
Strontium, Yttrium and Zirconium are relatively light $n$-capture elements. 
In the solar system, they are attributed mostly to the ``main'' 
$s$-process \citep{Arlandini99}. 
Barium and Lanthanum are heavier $n$-capture elements also primarily
$s$-process elements in solar-system material.
Europium is our sole representative of solar-system $r$-process elements.

\begin{figure}
\plotone{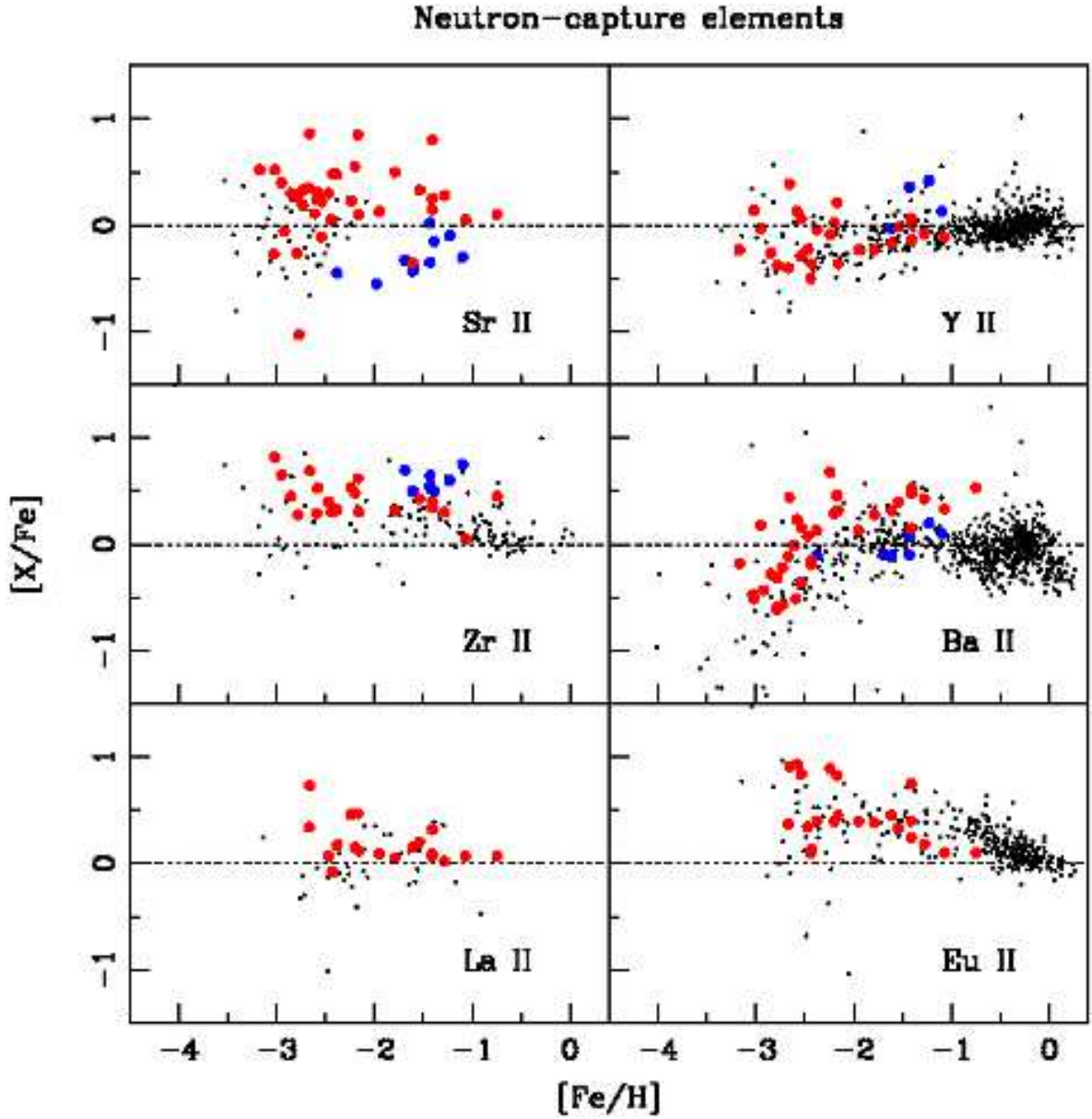}
\caption{Same as Figure~\ref{x_fe_total}, except for $n$-capture elements. The red and blue dots correspond to RHB and BHB stars.\label{x_fe_total2}}
\end{figure}

Our HB $n$-capture abundance ratios are generally in 
accord with field stars studies (see Figure~\ref{x_fe_total2}). 
The offset of [Sr/Fe] between RHB and BHB stars are discussed in \S 6.5. 
Unfortunately, we do not have [Sr/Fe] for field stars with 
[Fe/H] $> -2.0$ for comparison. 
The resonance lines of \ion{Sr}{2} are very strong for moderately metal-poor 
cooler stars and thus Strontium is not well represented in previous
field-star surveys in this metallicity regime.
We conclude that $<$[Sr/Fe]$>$ $\sim$~0 for [Fe/H] $> -2.0$.

Increasing star-to-star scatter with decreasing metallicity is apparent
in the heavier $n$-capture elements Ba, La, and Eu, in accord with trends
seen in other field star samples.
A sharp downward trend of [\ion{Ba}{2}/Fe] with decreasing metallicity 
becomes apparent for [Fe/H]$< -2.0$. 
This pattern is present in field stars studies as well. 
The [La/Fe] should roughly correlate with [Ba/Fe]. 
Unfortunately, we cannot easily detect \ion{La}{2} lines in HB stars below 
[Fe/H] $\simeq$ $-$2.5, where the drop in Ba abundance becomes 
apparent.
The simplest explanation for the rise of [Ba/Fe] at [Fe/H] $> -2.0$ 
is that the $r$-process dominates Ba production at lowest metallicities
while the $s$-process plays a more important role at higher 
metallicities \citep{Busso99}. 

The initial examination of our derived Europium abundances yielded
six RHB stars with [Eu/Fe] $>0.5$, well above the mean trend. 
However, high [Eu/Fe] has also been found in some field stars 
(as shown in Figure~\ref{x_fe_total2}). 
For example, $n$-capture rich star CS\,22892$-$052 has [Eu/Fe] $=+1.64$ 
\citep{Sneden03} and CS\,31082$-$001 has [Eu/Fe] $=+1.63$ \citep{Hill02}.   
The other $n$-capture elements of three of the Eu-rich RHB stars in our samples, \ie, 
CS\,22875$-$029, CS\,22886$-$043 and BD$+$17$\degr$\,3248 
are also high, implying that these three are truly $n$-capture rich stars. 
The overall $n$-capture abundance distributions for the other three
RHB stars with Eu excesses are less certain.
These six RHB stars deserve followup spectroscopic investigation of
the $n$-capture elements.

\subsection{Heavier vs Lighter Neutron-Capture Elements}

\begin{figure}
\plotone{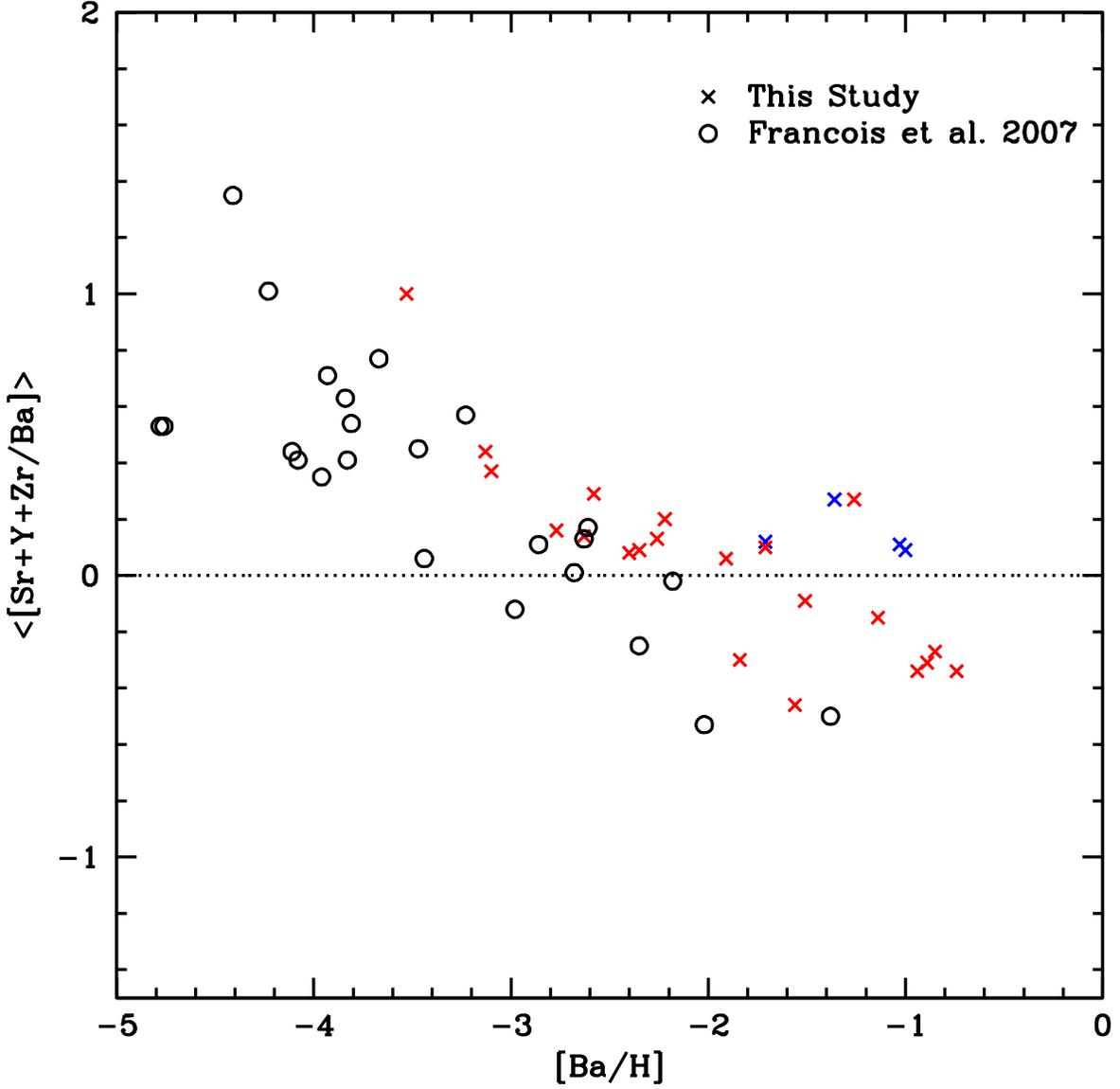}
\caption{Mean abundance ratios of [Sr+Y+Zr/Ba] vs [Ba/H] (red crosses), with the additional data from 
\citet{Francois07} (black open circles). \label{ncapcomp}}
\end{figure}

Abundances of light $n$-capture elements Sr, Y, and Zr appear to be highly 
correlated with each other, and clearly they share a common nucleosynthetic
origin (\eg, \citealp{McWilliam95}; \citealp{Francois07}; \citealp{Aoki05}). 
In Figure~\ref{ncapcomp}, we compare the mean Sr-Y-Zr abundances 
the heavier element Ba for our HB stars, adding in the data of 
\citet{Francois07}.
Only stars with detections of all of these elements are included in this plot. 
The comparison shows a tight correlation 
(\ie, increasing overabundant as decreasing Barium abundances), 
which suggests the correlation exists regardless of metallicity 
regime and evolutionary state. 

\begin{figure}
\plotone{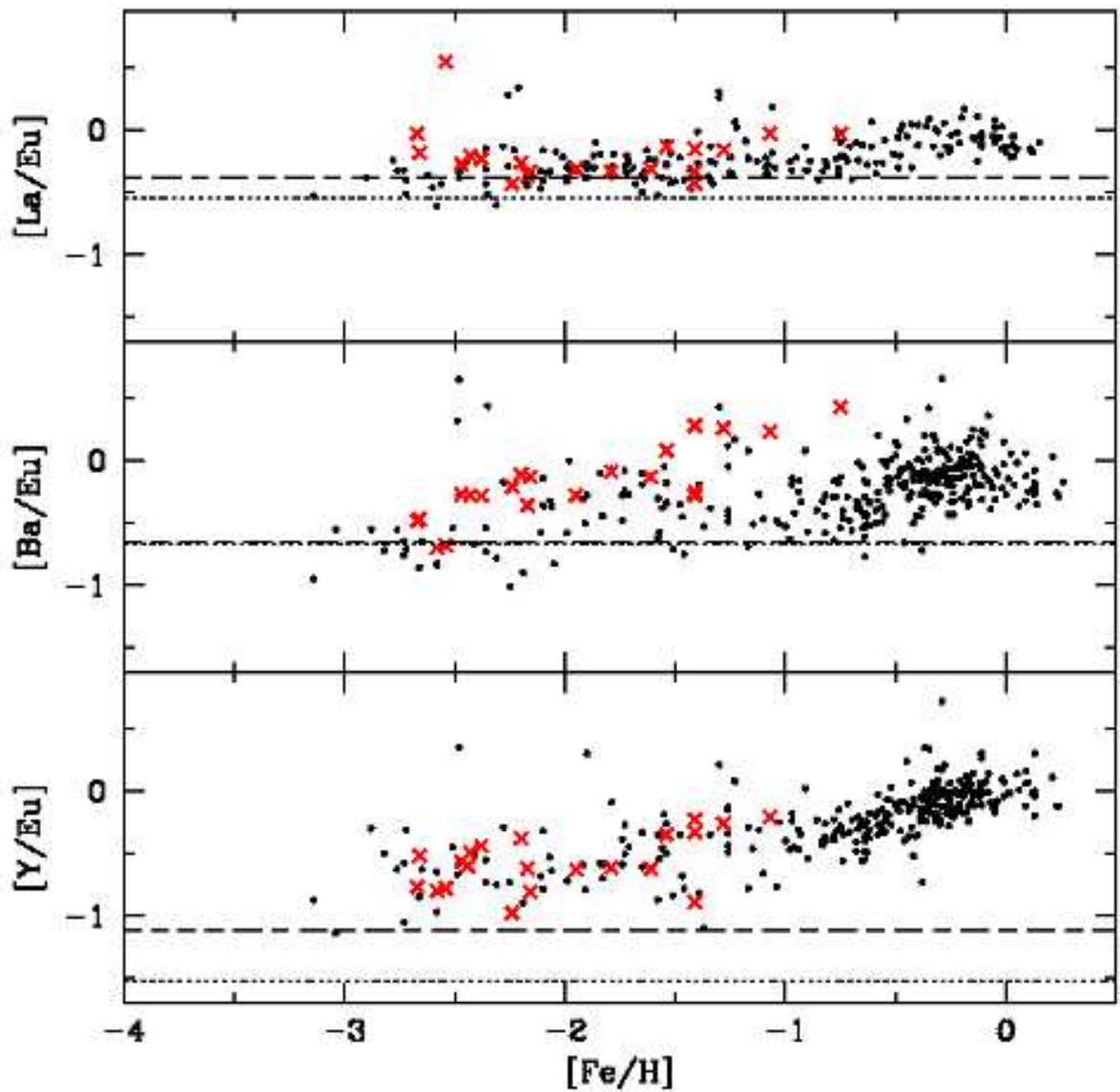}
\caption{Comparison of light vs heavier $n$-capture elemental abundance ratios as a function of metallicity. 
These ratios are used to 
examine $s$ and $r$-process enrichment. The dashed and dotted lines represent 
the estimated pure $r$-process from solar system abundances of \citet{Arlandini99} and 
\citet{Sneden08}, respectively. The red crosses correspond to our RHB stars. The black dots represent 
La, Ba, Y, Eu from \citet{Venn04}, La, Eu from \citet{Simmerer04} and \citet{Woolf95}.\label{xeu_feh}}
\end{figure}

To examine the contributions of the $r$ and $s$-process ratios of 
metal-poor stars, abundances of Y, Ba, La and Eu are 
generally used. 
As discussed above, Y, Ba and La can be formed via $r$ and $s$-processes, 
while Eu is largely formed via the $r$-process. 
In Figure~\ref{xeu_feh}, we plotted the [La/Eu], [Ba/Eu] and [Y/Eu] vs 
[Fe/H] of our HB samples along with those of \citet{Venn04}, \citet{Simmerer04} and 
\citet{Woolf95}, and compare 
them with estimated pure $r$-process solar system abundances 
(\citealp{Arlandini99}; \citealp{Sneden08}). 

The top panel shows the [La/Eu] distribution, which the rise of 
[La/Eu] as metallicity increases progresses slower than [Ba/Eu] and [Y/Eu]. 
The comparison between [La/Eu] and middle panel of [Ba/Eu] demonstrates  
that the larger scatter of [Ba/Eu] is due to the 
Barium not Europium abundances. 
The middle and bottom panels of [Ba/Eu] and [Y/Eu] show large scatter 
in very metal-poor stars regime, which suggests an inhomogeneous mixing 
in early Galactic time. 
We also find a slow increase of [Ba/Eu] and [Y/Eu] as the 
metallicity increases. 
The rise is further evidence of the increasing contribution of the 
$s$-process as metallicity increases (with time in the Galaxy).
The slope of [Ba/Eu] for our HB stars is steeper than the field stars but 
the overall trend is indistinguishable from the large scatter. 
Also, the [Y/Eu] abundances are above the estimated pure $r$-process 
solar-system abundances, which again suggests that the $s$-process (from AGB 
stars) play a significant role in Yttrium production.

\subsection{CS\,22186$-$005}

The RHB star CS\,22186$-$005 has an extremely low Sr abundance, \ie, 
[\ion{Sr}{2}/Fe] $= -1.03$ (see Figures~\ref{comp_srspec} $\&$ \ref{x_fe_total2}).  
As expected, there is no detection of the weaker  \ion{Zr}{2} and \ion{Y}{2} 
in this star.
However, we detected Barium, with an abundance ratio of [\ion{Ba}{2}/Fe] $= -0.58$.
Its Barium abundance follows the general declining trend of metal-poor 
stars that has metallicity below $-2.0$ (see Figure~\ref{x_fe_total2}). 
The resulting abundance ratio, [Ba/Sr] = +0.45, is somewhat surprising 
because in most $n$-capture metal-poor cases, the heavier $n$-capture 
elements are underabundant with respect to lighter ones (as summarized
in see Figure~7 of \citealp{Sneden08}).
Other heavier $n$-capture elements (\ie, Eu and La) were
not detectable with our spectra of CS\,22186$-$005,
This star does not appear to have obvious abundance anomalies among
the lighter elements.

\begin{figure}
\plotone{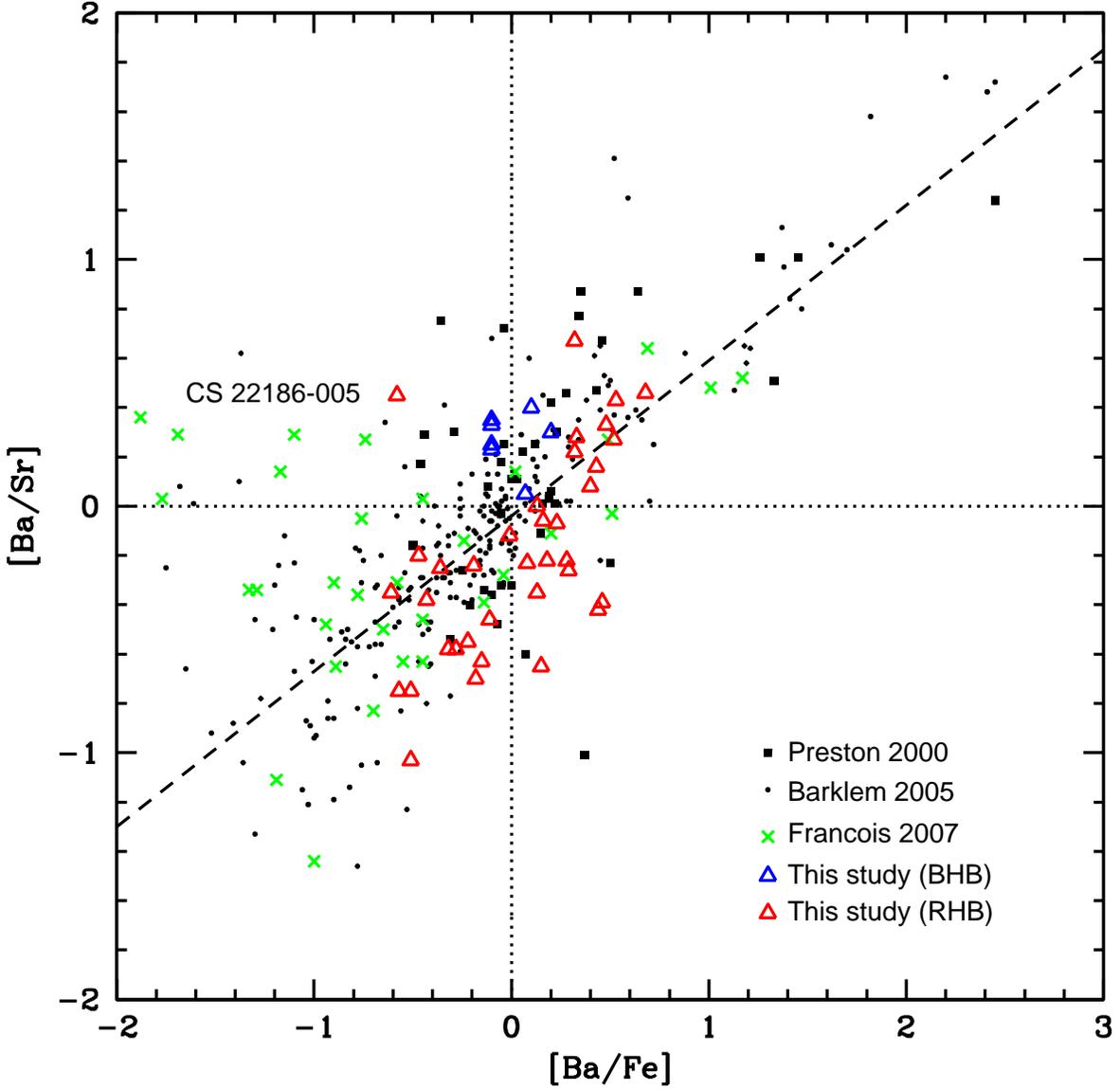}
\caption{Abundance ratios of [Ba/Sr] vs [Ba/Fe]. The long dashed line represent the linear correlation between 
[Ba/Sr] and [Ba/Fe] (see \citealp{Sneden08}). Solid, black rectangulars and dots represent studies of \citet{Preston00} and \citet{Barklem05}, 
respectively. Study by \citet{Francois07} is represented in green crosses. Our RHB and BHB stars are represented by 
red and blue open triangles. \label{srba}}
\end{figure}

In Figure~\ref{srba}, we extend Sneden et al's Figure 7 by adding in 
Sr and Ba abundances of our RHB and BHB stars. 
It is clear that CS\,22186$-$005 is not the only metal-poor star 
that exhibits unusually large [Ba/Sr] ratios at low [Ba/Fe]. 
Such stars have mainly been found among the very metal-poor giant sample
of \citet{Francois07}.
Clearly these stars provide further evidence that $n$-capture synthesis
events cannot easily be characterized by single nucleosynthesis processes.
Followup observations at higher S/N and resolution of this type of star
should be undertaken.

\section{Conclusions}

We present the first large-scample detailed chemical composition study 
of non-variable field RHB and BHB stars. 
The high resolution spectra for our work were obtained with 
the 2.7~m telescope at the McDonald Observatory. 
The sample was selected from the survey of \citet{Behr03}.
Additional RHB spectra from \citet{Preston06} were also added to the analysis.
We derived the model stellar atmospheric parameters, \teff, \logg, [Fe/H], 
and \vturb\ for all program stars based on spectroscopic constraints. 
Of some interest is that the microturbulence of RHB stars increase 
with increasing \teff, in agreement with \citet{Preston06}, while 
microturbulence appears to decline with increasing \teff\ in BHB stars.
More data on BHB stars to solidify this conclusion would be welcome.

Employing these stellar parameters, we derived relative abundance ratios, 
[X/Fe], of the $\alpha$-elements, Fe-peak elements and $n$-capture 
elements for these stars. 
The abundance ratios vs metallicity of our RHB and BHB stars are generally
in accord with other field star studies. 
In particular, the $\alpha$-elements are overabundant, 
[\ion{Al}{1}/Fe] (RHB stars only) 
and [\ion{Mn}{1}/Fe] are underabundant for metal-poor stars. 
Large star-to-star scatter is present in [$n$-capture/Fe] abundance
ratios.

Finally we investigated the physical properties of our RHB and BHB stars 
by locating them in the \teff$-$\logg\ plane, and comparing them 
to HB evolutionary tracks of \citet{PCSC06}, in order to estimate 
individual stellar masses. 
The mass distribution suggests that the majority of our stars have
$M \sim0.56~M_{\sun}$. By comparing the \teff\ 
distribution of our field RHB and BHB stars with the field RR Lyraes of 
\citet{Lambert96} and \citet{Clementini95}, we estimated the temperatures of red and blue edges of 
the RR Lyr IS for stars with [Fe/H]$>-2.5$. 
We derived 5900~K and 7400~K, respectively for these edges.

The general consistency of HB abundance ratios with those of other 
dwarf and giant halo star samples justifies that HB stars can be used 
routinely in the future for Galactic sturcture-metallicity studies
(such as investigations of stellar streams).
More importantly, this work provides a starting point for our future study on 
chemical compositions of RR Lyrs (For et al., in prep). 
Determinations of abundances of these stars throughout their pulsational
cycles will be examined in detail with the same methods as have been
employed in this paper.

\acknowledgments
B.-Q. For acknowledges the invaluable assistance from the mountain 
support staff at the McDonald observatory, and travel assistance from 
a SigmaXi grant-in-aid. 
We are grateful to Bradford Behr and George Preston 
for helpful discussions and advice on their earlier studies. 
This research was supported by U.S. National Science Foundation grants
AST-0607708 and AST-0908978.

\bibliographystyle{apj}
\bibliography{ms}

\clearpage

\clearpage

\clearpage

\clearpage 

\clearpage

\clearpage

\clearpage

\clearpage

\clearpage

\clearpage

\clearpage

\clearpage

\clearpage

\clearpage

\clearpage

\clearpage

\clearpage

\clearpage

\clearpage

\clearpage

\clearpage

\clearpage

\clearpage

\clearpage

\clearpage

\clearpage

\clearpage

\clearpage

\input{table}


\end{document}

%% file: table.tex
\clearpage

\begin{deluxetable}{lccrrrrrrccc}
\tabletypesize{\scriptsize}
\tablewidth{0pt}
\tablecolumns{12}
\tablecaption{Program stars.\label{targets}}
\tablehead{
\colhead{Star} & 
\colhead{R.A.(J2000)}  & 
\colhead{Decl.(J2000)}  & 
\colhead{$B$\tablenotemark{a}}  &
\colhead{$V$\tablenotemark{a,b}}  & 
\colhead{$J$\tablenotemark{c}} & 
\colhead{$H$\tablenotemark{c}}  & 
\colhead{$K_{s}$\tablenotemark{c}}  & 
\colhead{$B-V$} &
\colhead{$V-K$} & 
\colhead{E$(B-V)$\tablenotemark{d}} & 
\colhead{cE$(B-V)$} \\
\colhead{} & 
\colhead{(hr m s)} &
\colhead{($\degr\,\arcmin\,\arcsec$)}   & 
\colhead{(mag)}  &
\colhead{(mag)}  & 
\colhead{(mag)}    & 
\colhead{(mag)}  & 
\colhead{(mag)}  &
\colhead{(mag)} & 
\colhead{(mag)} & 
\colhead{(mag)} & 
\colhead{(mag)} 
}
\startdata
\cutinhead{RHB}
HD\,6229	&	01 03 36.5	&	$+$23 46 06.4	        &	9.31	&	8.60	&	7.088	&	6.646	&	6.575	&	0.71	&	2.025	&	0.034	&	0.021	\\
HD\,6461	&	01 05 25.4	&	$-$12 54 12.1		&	8.4	&	7.65	&	6.149	&	5.676	&	5.587	&	0.75	&	2.063	&	0.025	&	0.013	\\
HD\,25532	&	04 04 11.0	&	$+$23 24 27.1		&	8.85	&	8.24	&	6.688	&	6.327	&	\nodata	&	0.61	&	1.057	&	0.191	&	0.162	\\
HD\,105546	&	12 09 02.7	&	$+$59 01 05.1		&	9.4	&	8.61	&	7.152	&	6.756	&	6.674	&	0.79	&	0.980	&	0.022	&	0.010	\\
HD\,119516	&	13 43 26.7	&	$+$15 34 31.1		&	9.52	&	9.13	&	7.771	&	7.431	&	7.366	&	0.39	&	1.764	&	0.031	&	0.018	\\
BD$+$18$\degr$\,2890	& 14 32 13.5	&	$+$17 25 24.3		&	10.49	&	9.77	&	8.241	&	7.837	&	7.744	&	0.72	&	2.026	&	0.020	&	0.008	\\
BD$+$11$\degr$\,2998 	& 16 30 16.8	&	$+$10 59 51.7		&	9.70	&	9.07	&	7.619	&	7.271	&	7.185	&	0.63	&	1.885	&	0.057	&	0.041	\\
BD$+$09$\degr$\,3223	& 16 33 35.6	&	$+$09 06 16.3		&	9.81	&	9.25	&	7.760	&	7.335	&	7.277	&	0.56	&	1.007	&	0.076	&	0.058	\\
\enddata
\tablenotetext{a}{SIMBAD. http://simbad.u-strasbg.fr/simbad/}
\tablenotetext{b}{\citet{BPS92}.}
\tablenotetext{c}{2MASS All-Sky Point Source Catalog\citep{Skrutskie06}. http://tdc-www.harvard.edu/catalogs/tmpsc.html}
\tablenotetext{d}{Nasa/IPAC extragalactic database.}
\end{deluxetable}

\clearpage

\begin{deluxetable}{lrclrrrr}
\tabletypesize{\scriptsize}
\tablewidth{0pt}
\tablecolumns{8}
\tablecaption{Observation Log.\label{obslog}}
\tablehead{
\colhead{Star} & 
\colhead{UT Date}  & 
\colhead{No. Integration}  & 
\colhead{$t_{exp}$}  &
\colhead{S/N at 7000\AA} & 
\colhead{S/N at 5000\AA} & 
\colhead{S/N at 4000\AA} & 
\colhead{Comments}  \\
\colhead{} & 
\colhead{}  & 
\colhead{}  & 
\colhead{(s)}  &
\colhead{}  & 
\colhead{} & 
\colhead{}  
}
\startdata
BD$+$09$\degr$\,3223	&	30 Jun 2007	&	3	&	1800	                & 223	&	230	&	95	&	1	\\
BD$+$11$\degr$\,2998 	&	01 Jul 2007	&	3	&	1800	                & 230	&      128	&	88	&	1	\\
BD$+$18$\degr$\,2890	&	02 Jul 2007	&	3	&	1800	                & 210	&      124	&	30	&	1	\\
HD\,180903	&	02 Jul 2007	&	3	&	1800	                        & 210	&      88	&	40	&	1,4	\\
HD\,229274	&	02 Jul 2007	&	3	&	1800       	                & 320	&      147	&	100	&	1	\\
HD\,119516	&	03 Jul 2007	&	3	&	1800	                        & 320	&      132	&	60	&	1	\\
HD\,184266	&	04 Jul 2007	&	2	&	900	                        & 360	&      140	&	75	&	1	\\
BD$+$17$\degr$\,3248	&	04 Jul 2007	&	2	&	1800	                & 280	&      108	&	66	&	1	\\
\enddata
\tablenotetext{1}{The echelle grating was blazed to obtain more red portion of the 
spectrum. See text for explanation.}
\tablenotetext{2}{The echelle grating was blazed to obtain optima red 
and blue portion of the spectrum.}
\tablenotetext{3}{Initial analysis was performed. Stellar parameters cannot be obtained due to the 
lack of measurable Fe I or Fe II lines. Excluded from this study.}
\tablenotetext{4}{RR Lyr, excluded from this study.} 
\end{deluxetable}


\begin{deluxetable}{cccrcr}
\tablewidth{0pt}
\tabletypesize{\tiny}
\tablecaption{Equivalent width measurements of program stars. \label{EW}}
\tablecolumns{6}
\tablehead{
\colhead{Wavelength} &
\colhead{Species} &
\colhead{E.P.} &
\colhead{$\log gf$} &
\colhead{Ref.} &
\colhead{EW} \\
\colhead{(\AA)} &
\colhead{} &
\colhead{(eV)} &
\colhead{} &
\colhead{} &
\colhead{(m\AA)}
}
\startdata
\cutinhead{HD\,6229}
5682.63	&	Na I	&	2.102	&	$	-0.71	$	&	1	&	49	\\
5688.19	&	Na I	&	2.104	&	$	-0.46	$	&	1	&	\nodata	\\
5339.93 &       Fe I    &       3.266   &       $       -0.72   $       &       1       &       101     \\
5341.02 &       Fe I    &       1.608   &       $       -1.95   $       &       1       &       141     \\
.       &       .       &       .       &       .                       &       .       &       .       \\
.       &       .       &       .       &       .                       &       .       &       .       \\
.       &       .       &       .       &       .                       &       .       &       .       
\enddata
\tablecomments{Table~\ref{EW} is published in its entirety in the
electronic edition of the {\it Astronomical Journal}.  A portion is
shown here for guidance regarding its form and content.}
\end{deluxetable}

\clearpage

\begin{deluxetable}{lccccccrccr}
\tablewidth{0pt}
\tabletypesize{\scriptsize}
\tablecolumns{11}
\tablecaption{Input stellar atmosphere parameters and derived Fe metallicities.\label{stellarpars}}
\tablehead{
\colhead{Star} & \colhead{$T_{\rm eff}$} & \colhead{$\log g$} & \colhead{[M/H]\tablenotemark{a}} & \colhead{$v_{t}$}
& \colhead{[Fe I/H]} & \colhead{$\sigma$} & \colhead{N} & \colhead{[Fe II/H]} & \colhead{$\sigma$} & \colhead{N} \\
\colhead{} & \colhead{(K)} & \colhead{(dex)} & \colhead{} & \colhead{(km~s$^{-1}$)}
& \colhead{} & \colhead{} & \colhead{} & \colhead{} & \colhead{} & \colhead{} 
}
\startdata
\cutinhead{RHB}
HD\,6229	&	5200	&	2.50	&	$	-1.07	$	&	1.60	&	$	-1.07	$	&	0.13	&	98	&	$	-1.06	$	&	0.13	&	20	\\
HD\,6461	&	5200	&	2.90	&	$	-0.75	$	&	1.40	&	$	-0.75	$	&	0.12	&	94	&	$	-0.74	$	&	0.10	&	13	\\
HD\,25532	&	5450	&	2.00	&	$	-1.41	$	&	2.10	&	$	-1.41	$	&	0.06	&	44	&	$	-1.42	$	&	0.09	&	8	\\
HD\,105546	&	5200	&	2.30	&	$	-1.54	$	&	1.80	&	$	-1.54	$	&	0.08	&	65	&	$	-1.54	$	&	0.06	&	20	\\
HD\,119516	&	5400	&	1.50	&	$	-2.16	$	&	2.20	&	$	-2.16	$	&	0.06	&	49	&	$	-2.16	$	&	0.05	&	15	\\
BD$+$18$\degr$\,2890	&	5000	&	2.40	&	$	-1.61	$	&	1.40	&	$	-1.61	$	&	0.07	&	51	&	$	-1.61	$	&	0.09	&	8	\\
BD$+$11$\degr$\,2998 	&	5450	&	2.30	&	$	-1.28	$	&	1.90	&	$	-1.28	$	&	0.08	&	59	&	$	-1.29	$	&	0.06	&	10	\\
BD$+$09$\degr$\,3223	&	5100	&	1.30	&	$	-2.47	$	&	1.90	&	$	-2.47	$	&	0.05	&	48	&	$	-2.46	$	&	0.06	&	11	\\
\enddata
\tablenotetext{a}{Input model metallicity.}
\end{deluxetable}

\clearpage
\begin{deluxetable}{lrccrccrccrccrccrcc}
\rotate
\tabletypesize{\tiny}
\tablewidth{0pt}
\tablecolumns{19}
\tablecaption{Abundance ratios from Mg, Si and Ca\label{abund1}}
\tablehead{
\colhead{Star} & 
\colhead{[Na I/Fe]}  & 
\colhead{$\sigma$}  & 
\colhead{N}  &
\colhead{[Mg I/Fe]}  & 
\colhead{$\sigma$} & 
\colhead{N}  & 
\colhead{[Si I/Fe]}  & 
\colhead{$\sigma$} & 
\colhead{N} & 
\colhead{[Si II/Fe]} & 
\colhead{$\sigma$} & 
\colhead{N} & 
\colhead{[Ca I/Fe]} & 
\colhead{$\sigma$}   & 
\colhead{N}  &
\colhead{[Ca II/Fe]}  & 
\colhead{$\sigma$} & 
\colhead{N}  
}
\startdata
\cutinhead{RHB}
HD\,6229	&	0.03	&	0.06	&	5	&	0.36	&	0.04	&	3	&	0.28	&	0.06	&	5	&	0.32	&	0.03	&	2	&	0.15	&	0.11	&	12	&	\nodata	&	\nodata	&	\nodata	\\
HD\,6461	&	$-$0.02	&	0.10	&	3	&	0.35	&	0.15	&	2	&	0.29	&	0.02	&	6	&	0.47	&	0.16	&	2	&	0.17	&	0.09	&	13	&	\nodata	&	\nodata	&	\nodata	\\
HD\,25532	&	0.64	&	\nodata	&	1	&	0.56	&	\nodata	&	1	&	0.53	&	0.07	&	5	&	0.54	&	0.18	&	2	&	0.29	&	0.05	&	4	&	\nodata	&	\nodata	&	\nodata	\\
HD\,105546	&	0.17	&	\nodata	&	1	&	0.50	&	0.08	&	3	&	0.40	&	0.10	&	6	&	0.61	&	0.20	&	3	&	0.42	&	0.09	&	12	&	\nodata	&	\nodata	&	\nodata	\\
HD\,119516	&	0.54	&	\nodata	&	1	&	0.28	&	\nodata	&	1	&	0.40	&	\nodata	&	1	&	0.48	&	0.17	&	2	&	0.26	&	0.07	&	7	&	\nodata	&	\nodata	&	\nodata	\\
BD+18\degr\,2890	&	$-$0.04	&	0.02	&	4	&	$-$0.06	&	\nodata	&	1	&	0.41	&	0.08	&	6	&	0.74	&	\nodata	&	1	&	0.35	&	0.07	&	12	&	\nodata	&	\nodata	&	\nodata	\\
BD+11\degr\,2998 	&	0.24	&	\nodata	&	1	&	0.56	&	0.12	&	2	&	0.41	&	0.07	&	5	&	0.52	&	0.07	&	3	&	0.29	&	0.09	&	7	&	\nodata	&	\nodata	&	\nodata	\\
BD+09\degr\,3223	&	\nodata	&	\nodata	&	\nodata	&	0.27	&	\nodata	&	1	&	0.73	&	\nodata	&	1	&	0.86	&	0.16	&	2	&	0.50	&	0.06	&	11	&	\nodata	&	\nodata	&	\nodata	\\
\enddata
\tablenotetext{a}{NLTE correction.}
\end{deluxetable}


\clearpage
\begin{deluxetable}{lrccrccrccrccrccrcc}
\rotate
\tabletypesize{\tiny}
\tablewidth{0pt}
\tablecolumns{19}
\tablecaption{Abundance ratios of Al, Ti, Sc and Cr\label{abund2}}
\tablehead{
\colhead{Star} & 
\colhead{[Al I/Fe]}  & 
\colhead{$\sigma$}  & 
\colhead{N}  &
\colhead{[Ti I/Fe]}  & 
\colhead{$\sigma$} & 
\colhead{N}  & 
\colhead{[Ti II/Fe]}  & 
\colhead{$\sigma$} & 
\colhead{N} & 
\colhead{[Sc II/Fe]} & 
\colhead{$\sigma$} & 
\colhead{N} & 
\colhead{[Cr I/Fe]} & 
\colhead{$\sigma$}   & 
\colhead{N}  &
\colhead{[Cr II/Fe]}  & 
\colhead{$\sigma$} & 
\colhead{N}  
}
\startdata
\cutinhead{RHB}
HD\,6229	&	\nodata	&	\nodata	&	\nodata	&	0.07	&	0.08	&	13	&	0.34	&	0.14	&	10	&	0.34	&	0.12	&	4	&	$-$0.15	&	0.08	&	5	&	0.03	&	0.14	&	5	\\
HD\,6461	&	\nodata	&	\nodata	&	\nodata	&	0.19	&	0.10	&	13	&	0.43	&	0.10	&	9	&	0.35	&	0.11	&	4	&	$-$0.11	&	0.05	&	2	&	0.10	&	0.20	&	5	\\
HD\,25532	&	\nodata	&	\nodata	&	\nodata	&	0.18	&	0.07	&	8	&	0.22	&	0.09	&	7	&	0.12	&	0.06	&	2	&	$-$0.21	&	0.12	&	4	&	$-$0.08	&	0.17	&	5	\\
HD\,105546	&	\nodata	&	\nodata	&	\nodata	&	0.25	&	0.02	&	9	&	0.40	&	0.10	&	8	&	0.25	&	0.08	&	3	&	$-$0.17	&	0.11	&	7	&	0.25	&	0.19	&	6	\\
HD\,119516	&	$-$0.82	&	\nodata	&	1	&	0.23	&	0.06	&	5	&	0.06	&	0.13	&	5	&	$-$0.06	&	\nodata	&	1	&	$-$0.18	&	0.06	&	5	&	0.01	&	0.10	&	5	\\
BD+18\degr\,2890	&	\nodata	&	\nodata	&	\nodata	&	0.15	&	0.09	&	6	&	0.00	&	0.08	&	3	&	0.06	&	0.09	&	2	&	$-$0.17	&	0.01	&	2	&	0.26	&	\nodata	&	1	\\
BD+11\degr\,2998 	&	\nodata	&	\nodata	&	\nodata	&	0.19	&	0.04	&	10	&	0.22	&	0.12	&	6	&	0.16	&	0.05	&	3	&	$-$0.22	&	0.08	&	4	&	$-$0.05	&	0.12	&	3	\\
BD+09\degr\,3223	&	\nodata	&	\nodata	&	\nodata	&	0.28	&	0.08	&	8	&	0.16	&	0.09	&	9	&	0.06	&	0.02	&	3	&	$-$0.21	&	0.07	&	4	&	0.12	&	0.18	&	2	\\
\enddata
\tablenotetext{a}{NLTE correction.}
\end{deluxetable}

\clearpage
\begin{deluxetable}{lrccrccrccrccrccrcc}
\rotate
\tabletypesize{\tiny}
\tablewidth{0pt}
\tablecolumns{19}
\tablecaption{Abundance ratios of Fe-peak elements: V, Mn, Co, Ni and Zn\label{abund3}}
\tablehead{
\colhead{Star} & 
\colhead{[V II/Fe]}  & 
\colhead{$\sigma$}  & 
\colhead{N}  &
\colhead{[Mn I/Fe]}  & 
\colhead{$\sigma$} & 
\colhead{N}  & 
\colhead{[Co I/Fe]}  & 
\colhead{$\sigma$} & 
\colhead{N} & 
\colhead{[Ni I/Fe]} & 
\colhead{$\sigma$} & 
\colhead{N} & 
\colhead{[Ni II/Fe]} & 
\colhead{$\sigma$}   & 
\colhead{N}  &
\colhead{[Zn I/Fe]}  & 
\colhead{$\sigma$} & 
\colhead{N}  
}
\startdata
\cutinhead{RHB}
HD\,6229	&	\nodata	&	\nodata	&	\nodata	&	0.12	&	0.27	&	3	&	0.80	&	\nodata	&	1	&	$-$0.04	&	0.09	&	9	&	\nodata	&	\nodata	&	\nodata	&	0.11	&	0.04	&	2	\\
HD\,6461	&	\nodata	&	\nodata	&	\nodata	&	0.30	&	\nodata	&	1	&	0.84	&	\nodata	&	1	&	$-$0.01	&	0.1	&	9	&	\nodata	&	\nodata	&	\nodata	&	0.24	&	\nodata	&	1	\\
HD\,25532	&	\nodata	&	\nodata	&	\nodata	&	0.05	&	0.07	&	3	&	0.37	&	\nodata	&	1	&	0.05	&	0.12	&	4	&	\nodata	&	\nodata	&	\nodata	&	0.04	&	\nodata	&	1	\\
HD\,105546	&	\nodata	&	\nodata	&	\nodata	&	$-$0.09	&	0.16	&	5	&	0.30	&	0.08	&	2	&	$-$0.03	&	0.13	&	5	&	\nodata	&	\nodata	&	\nodata	&	0.13	&	0.05	&	2	\\
HD\,119516	&	\nodata	&	\nodata	&	\nodata	&	$-$0.30	&	0.08	&	3	&	$-$0.01	&	\nodata	&	1	&	$-$0.04	&	\nodata	&	1	&	\nodata	&	\nodata	&	\nodata	&	0.05	&	0.04	&	2	\\
BD+18\degr\,2890	&	\nodata	&	\nodata	&	\nodata	&	$-$0.70	&	0.08	&	3	&	0.22	&	\nodata	&	1	&	$-$0.03	&	0.09	&	3	&	\nodata	&	\nodata	&	\nodata	&	0.04	&	\nodata	&	1	\\
BD+11\degr\,2998 	&	\nodata	&	\nodata	&	\nodata	&	$-$0.06	&	0.15	&	4	&	0.32	&	0.04	&	2	&	0.06	&	0.03	&	2	&	\nodata	&	\nodata	&	\nodata	&	\nodata	&	\nodata	&	\nodata	\\
BD+09\degr\,3223	&	0.03	&	\nodata	&	1	&	$-$0.10	&	0.11	&	4	&	0.42	&	\nodata	&	1	&	\nodata	&	\nodata	&	\nodata	&	\nodata	&	\nodata	&	\nodata	&	0.20	&	\nodata	&	1	\\
\enddata
\end{deluxetable}

\clearpage
\begin{deluxetable}{lrccrccrccrccrccrcc}
\rotate
\tabletypesize{\tiny}
\tablewidth{0pt}
\tablecolumns{13}
\tablecaption{Abundance ratios of neutron-capture elements: Sr, Y, Zr, Ba, La and Eu\label{abund4}}
\tablehead{
\colhead{Star} & 
\colhead{[Sr II/Fe]} & 
\colhead{$\sigma$}  & 
\colhead{N}  &
\colhead{[Y II/Fe]} & 
\colhead{$\sigma$}  & 
\colhead{N}  &
\colhead{[Zr II/Fe]}  & 
\colhead{$\sigma$}  & 
\colhead{N}  &
\colhead{[Ba II/Fe]}  & 
\colhead{$\sigma$} & 
\colhead{N}  & 
\colhead{[La II/Fe]}  & 
\colhead{$\sigma$} & 
\colhead{N} & 
\colhead{[Eu II/Fe]} & 
\colhead{$\sigma$} & 
\colhead{N} 
}
\startdata
\cutinhead{RHB}
HD\,6229	&	0.05	&	0.05	&	2	&	$-$0.11	&	0.07	&	2	&	0.05	&	0.05	&	2	&	0.33	&	0.09	&	3	&	0.07	&	\nodata	&	1	&	0.10	&	0.15	&	2	\\
HD\,6461	&	0.10	&	0.10	&	2	&	\nodata	&	\nodata	&	\nodata	&	0.45	&	0.15	&	2	&	0.53	&	0.12	&	3	&	0.07	&	\nodata	&	1	&	0.10	&	\nodata	&	1	\\
HD\,25532	&	0.25	&	\nodata	&	1	&	0.01	&	0.10	&	2	&	0.35	&	0.04	&	3	&	0.52	&	0.19	&	3	&	0.09	&	0.08	&	2	&	0.24	&	0.01	&	2	\\
HD\,105546	&	0.33	&	0.02	&	2	&	$-$0.02	&	0.04	&	4	&	0.43	&	0.06	&	3	&	0.40	&	0.16	&	3	&	0.20	&	0.08	&	2	&	0.33	&	0.03	&	2	\\
HD\,119516	&	0.10	&	\nodata	&	1	&	$-$0.36	&	0.06	&	5	&	0.30	&	\nodata	&	1	&	0.32	&	0.22	&	3	&	0.12	&	\nodata	&	1	&	0.45	&	0.05	&	2	\\
BD+18 2890	&	$-$0.35	&	\nodata	&	1	&	$-$0.17	&	\nodata	&	1	&	\nodata	&	\nodata	&	\nodata	&	0.32	&	0.08	&	3	&	0.15	&	0.28	&	2	&	0.45	&	0.10	&	2	\\
BD+11 2998 	&	0.28	&	0.02	&	2	&	$-$0.08	&	0.12	&	2	&	0.30	&	\nodata	&	1	&	0.43	&	0.09	&	3	&	0.02	&	0.02	&	2	&	0.18	&	0.03	&	2	\\
BD+09 3223	&	0.30	&	0.10	&	2	&	$-$0.23	&	0.07	&	2	&	0.40	&	\nodata	&	1	&	0.08	&	0.11	&	4	&	0.07	&	\nodata	&	1	&	0.34	&	0.06	&	2	\\
\enddata
\end{deluxetable}

\clearpage
\begin{deluxetable}{lrrcrr}
\tablewidth{0pt}
\tablecolumns{5}
\tablecaption{Mean abundance ratios of various elements.\label{mabund}}
\tablehead{
\colhead{Element} & 
\colhead{RHB} &
\colhead{$N$} &
\colhead{BHB} &
\colhead{$N$} 
}
\startdata
Na I	&	0.37	&	27	&	$-$0.45	&	2	\\
Mg I	&	0.47	&	36	&	0.36	&	12	\\
Al I	&	$-$0.67	&	25	&	0.36	&	3	\\
Si I	&	0.35	&	36	&	$-$0.03	&	12	\\
Si II	&	0.59	&	35	&	0.21	&	12	\\
Ca I	&	0.37	&	36	&	0.07	&	12	\\
Ca II	&	\nodata	&	\nodata	&	0.18	&	12	\\
Sc II	&	0.13	&	35	&	0.14	&	10	\\
Ti I	&	0.37	&	35	&	\nodata	&	\nodata	\\
Ti II	&	0.23	&	36	&	0.31	&	12	\\
V II	&	0.14	&	14	&	0.15	&	5	\\
Cr I	&	$-$0.14	&	36	&	0.02	&	7	\\
Cr II	&	0.23	&	35	&	0.15	&	10	\\
Mn I	&	$-$0.37	&	36	&	$-$0.13	&	3	\\
Co I	&	0.41	&	15	&	0.28	&	1	\\
Ni I	&	0.22	&	15	&	\nodata	&	\nodata	\\
Ni II	&	\nodata	&	\nodata	&	$-$0.35	&	3	\\
Zn I	&	0.19	&	18	&	\nodata	&	\nodata	\\
Sr II	&	0.23	&	36	&	$-$0.30	&	10	\\
Y II	&	$-$0.12	&	27	&	0.22	&	4	\\
Zr II	&	0.42	&	23	&	0.61	&	7	\\
Ba II	&	0.03	&	36	&	0.00	&	7	\\
La II	&	0.19	&	19	&	\nodata	&	\nodata	\\
Eu II	&	0.45	&	22	&	\nodata	&	\nodata	
\enddata
\end{deluxetable}

\clearpage

\begin{deluxetable}{llccc}
\tabletypesize{\scriptsize}
\tablecaption{Sensitivity of [X/Fe] with stellar parameters.\label{sensitivity1}}
\tablewidth{0pt}
\tablecolumns{6}
\tablehead{
\colhead{Stellar Parameters} & 
\colhead{Species} &  
\colhead{} & 
\colhead{Star} & 
\colhead{} \\
\colhead{} & 
\colhead{$\Delta$[X/Fe]} & 
\colhead{CS\,22898$-$043} & 
\colhead{HS\,25532} & 
\colhead{BD$+$18$\degr$\,2890}
}
\startdata
$T_{\rm eff}+150$   &  Na I    & \nodata       &   +0.16        &    +0.16\\
(K)	&	Mg I	& +0.09	&	+0.08	&	+0.25	\\
\logg+0.15 	&	Na I	&	\nodata	&	$-$0.05	&	$-$0.03	\\
(dex)	&	Mg I	&	+0.01	&	$-$0.02	&	$-$0.01	\\
$[$M/H$]$+0.1&  Na I    &       \nodata &       $-$0.01 &       +0.00   \\
(dex)	&	Mg I	&	\nodata	&	$-$0.01	&	$-$0.01	\\
\vturb+0.2      &	Na I	&	\nodata	&	$-$0.01	&	$-$0.05	\\
(\kmsec)	&	Mg I	&	$-$0.05	&	$-$0.10	&	$-$0.07	
\enddata
\tablecomments{Table~\ref{sensitivity1} is published in its entirety in the
electronic edition of the {\it Astronomical Journal}.  A portion is
shown here for guidance regarding its form and content.}											      
\end{deluxetable}

\clearpage

\begin{deluxetable}{llc}
\tabletypesize{\scriptsize}
\tablecaption{Sensitivity of [X/Fe] with stellar parameters for BHB star.\label{sensitivity2}}
\tablewidth{0pt}
\tablecolumns{3}
\tablehead{
\colhead{Stellar Parameters} & 
\colhead{Species} &  
\colhead{Star} \\
\colhead{} & 
\colhead{$\Delta$[X/Fe]} & 
\colhead{HD\,93329} 
}
\startdata
$T_{\rm eff}+200$	&	Na I	&	+0.18	\\
(K)	&	Mg I	&	+0.14	\\
$\logg+0.15$	&	Na I	&	$-$0.03	\\
(dex)	&	Mg I	&	$-$0.04	\\
$[$M/H$]$+0.1	&	Na I	&	+0.01	\\
(dex)	&	Mg I	&	+0.00	\\
\vturb+0.2	&	Na I	&	$-$0.02	\\
(\kmsec)	&	Mg I	&	$-$0.01	
\enddata
\tablecomments{Table~\ref{sensitivity2} is published in its entirety in the
electronic edition of the {\it Astronomical Journal}.  A portion is
shown here for guidance regarding its form and content.}
\end{deluxetable}

\clearpage
\begin{deluxetable}{lccrr}
\tabletypesize{\scriptsize}
\tablewidth{0pt}
\tablecolumns{5}
\tablecaption{Comparison of HB model\label{model_comp}}
\tablehead{
\colhead{Model} &
\colhead{Mass} &
\colhead{$\log$ \teff} &
\colhead{$\Delta \log g$\tablenotemark{a}} &
\colhead{$\Delta \log L$\tablenotemark{a}} \\
\colhead{} &
\colhead{($M/\Msun$)} &
\colhead{(K)} &
\colhead{} &
\colhead{}
} 
\startdata
\citet{LD90}	&	0.56	&	4.22	&	$+0.02$	&	$-0.02$	\\
\citet{LD90}	&	0.56	&	4.26	&	$+0.11$	&	$-0.11$	\\
\citet{LD90}	&	0.78	&	3.86	&       $-0.01$	&	$+0.01$	\\
\citet{LD90}	&	0.78	&	3.72	&	$+0.09$	&	$-0.09$	
\enddata
\tablenotetext{a}{\citet{PCSC06} minus \citet{LD90} model}
\end{deluxetable}

\clearpage
\begin{deluxetable}{lcccc}
\tabletypesize{\scriptsize}
\tablewidth{0pt}
\tablecolumns{5}
\tablecaption{Estimated HB masses and Parameters Used\label{der_masses}}
\tablehead{
\colhead{Stars} & 
\colhead{$T_{\rm eff, spec}$} & 
\colhead{$\log g$} & \colhead{[Fe/H]} & 
\colhead{Mass} \\
\colhead{} & 
\colhead{(K)} & 
\colhead{(dex)} & 
\colhead{(dex)} & 
\colhead{$M_{\sun}$}
}
\startdata
\cutinhead{RHB}
HD\,6229	&	5200	&	2.86\tablenotemark{a}	&	-1.07	&	0.80	\\
HD\,6461	&	5200	&	3.26\tablenotemark{a}	&	-0.75	&	0.80	\\
HD\,25532	&	5450	&	2.20\tablenotemark{a}	&	-1.41	&	0.60	\\
HD\,105546	&	5200	&	2.66\tablenotemark{a}	&	-1.54	&	0.80	\\
HD\,119516	&	5400	&	1.73\tablenotemark{a}	&	-2.16	&	0.54	\\
BD$+$18$\degr$\,2890	&	5000	&	2.89\tablenotemark{a}	&	-1.61	&	0.80	\\
BD$+$11$\degr$\,2998 	&	5450	&	2.50\tablenotemark{a}	&	-1.28	&	0.72	\\
BD$+$09$\degr$\,3223	&	5100	&	1.72\tablenotemark{a}	&	-2.47	&	0.61	\\
\enddata
\tablenotetext{a}{Photometric $\log g$.}
\tablenotetext{b}{Spectroscopic $\log g$.}
\end{deluxetable}

\clearpage
\begin{deluxetable}{ll}
\tabletypesize{\scriptsize}
\tablewidth{0pt}
\tablecolumns{2}
\tablecaption{Data Sources\label{references}}
\tablehead{
\colhead{References} & \colhead{Element}
}
\startdata
\citet{Venn04} & Na, Mg, Ca, Ti, Ni, Y, Ba, La, Eu \\
\citet{Cohen04} & Si, Al, Sc, Cr, Mn, Sr \\
\citet{Lai08} & Si, Al, Sc, V, Mn, Zn, Sr, Zr \\
\citet{Fulbright00} & Si, Al, Cr, V, Zr \\
\citet{Reddy03} & Al, Sc, Cr, V, Mn, Ni, Zn \\
\citet{Sobeck06} & Mn \\
\citet{Cayrel04} & Si, Zn \\
\citet{SB02} & Si, Ni \\
\citet{Nissen07} & Ni
\enddata
\end{deluxetable}